\documentclass[10pt,aps,prb,twocolumn,english,amsmath,amssymb,floatfix,superscriptaddress,longbibliography,preprintnumbers]{revtex4-2}

\usepackage{graphicx}
\usepackage[dvipsnames]{xcolor}
\usepackage{amsmath,braket}

\definecolor{orange}{rgb}{1,0.5,0}
\definecolor{goodgreen}{rgb}{0.1,0.5,0}
\definecolor{goodred}{rgb}{0.7,0,0}

\usepackage[normalem]{ulem}

\makeatother

\usepackage{babel}
\usepackage{bm}

\usepackage[colorlinks,urlcolor=goodgreen,citecolor=blue,linkcolor=goodred]{hyperref}

\begin{document}
	
	\title{Effects of Electron Form Factor on Quasiparticle Interference in Twisted Bilayer Graphene}
	\author{D.-H.-Minh Nguyen}
	\email{d.h.minh.ng@gmail.com}
	\affiliation{Donostia International Physics Center, 20018 Donostia-San Sebasti\'an, Spain}
	\affiliation{Advanced Polymers and Materials: Physics, Chemistry and Technology, Chemistry Faculty (UPV/EHU), Paseo M. Lardizabal 3, 20018 San Sebastian, Spain}
	
	\author{Francisco Guinea}
	\email{paco.guinea@imdea.org}
	\affiliation{Donostia International Physics Center, 20018 Donostia-San Sebasti\'an, Spain}
	\affiliation{IMDEA Nanoscience, C/ Faraday 9, 28049 Madrid, Spain}
	
	\author{Dario Bercioux}
	\email{dario.bercioux@dipc.org}
	\affiliation{Donostia International Physics Center, 20018 Donostia-San Sebasti\'an, Spain}
	\affiliation{IKERBASQUE, Basque Foundation for Science, Euskadi Plaza, 5, 48009 Bilbao, Spain}
	
	\begin{abstract}
		The overlap matrix of electronic energy eigenstates, sometimes referred to as the form factor, determines the quantum geometric tensor of electrons in solids. Here, we show that the variation in the overlap of two eigenstates with opposite momenta can be directly observed via quasiparticle interference (QPI) imaging. We study the QPI in twisted bilayer graphene using a real-space tight-binding model combined with the kernel polynomial method. The resulting QPI patterns, which are largely independent of whether the two graphene layers are commensurate or incommensurate, reveal all intralayer and interlayer interference processes. While the intralayer interference signals resemble those of monolayer graphene, the interlayer interference --- which vanishes at large twist angles --- displays a chiral structure that reverses between the two layers and between the valence and conduction bands. Furthermore, the QPI patterns explicitly demonstrate the approximate translational symmetries and valley charge conservation in twisted bilayer graphene, validating the topological obstruction to constructing the Wannier orbitals of states at the Dirac cones. Using a continuum model of twisted bilayer graphene, we show that all characteristics of the observed QPI patterns can be explained by the form factor of eigenstates projected onto a single layer. Our results provide fundamental insights into the electronic spectrum and wave functions of twisted bilayer graphene, and establish QPI as an experimental probe for the form factor of back-scattering states.
	\end{abstract}
	\date{\today}
	\maketitle
	
	\section{Introduction}
	Quasiparticle interference (QPI) is a prominent method for probing the spectral properties of electrons in quantum materials~\cite{Simon2011,Chen2017,Avraham2018,Yin2021}. It relies on analyzing the Friedel oscillations in the electronic local density of states (LDOS) near a defect or impurity~\cite{Bena2015,Villain2015}. By taking the Fourier transform (FT) of the variation in LDOS due to the defect, researchers can extract information regarding the material's band structure and quasiparticles' properties. In practice, QPI is performed using scanning tunneling spectroscopy (STS), which provides high-resolution imaging of LDOS at the nanoscale. This approach has been used to visualize the surface states of topological materials~\cite{Guo2010,Zheng2018,Yuan2019,Stuart2022,Xiong2024,Cai2024,Sanchezbarquilla2025,Hoffmann2025,Bagchi2025}, and to identify the pairing symmetry~\cite{Hanaguri2009,Hanke2012,Li2023,Bhattacharyya2023,Nag2025,Wang2025,Christiansen2025} and the dominant orbitals contributing to the superconducting states~\cite{Huang2025} in unconventional superconductors. Recently, QPI has been proposed to serve as a probe for the bulk odd-frequency superconducting pairing~\cite{Chakraborty2022}, an identifier of excitations in quantum spin liquids~\cite{Ruan2021,He2022,Jahin2025}, and a means to demonstrate the interplay between chiral loop-current state and bond-order fluctuations in kagome metals~\cite{Nakazawa2025}. In the context of monolayer and bilayer graphene, QPI has been used extensively to investigate their electronic properties both theoretically~\cite{Kechedzhi2007,Bena2008,Pereg-Barnea2008,Bena2009green,Peres2009local,Lawlor2013,Soule2014,Dutreix2016,Zhang2021robust,Yang2021friedel,Yang2024wavefronts,Li2025local} and experimentally~\cite{chen2005atomic,Ruffieux2005,Brihuega2008,Zhang2009origin,Mallet2012,Gutierrez2016-iv,Jung2016-vd,Tesch2016-jc,Tesch2017,Dombrowski2017,Dutreix2019,Zhang2021quantum,Bao2021,Huempfner2022,Lisi2022,Zhang2023observation,Sun2023determining,Guan2024observation,Mallet2016friedel,Rutter2007scattering,Simon2009symmetry,Yankowitz2014band,Jolie2018,Joucken2020,Zhang2020,Joucken2021direct,Kaladzhyan2021}. For example, QPI has revealed the role of pseudospin~\cite{Brihuega2008,Mallet2012} and Berry phase~\cite{Dutreix2019} in interference patterns of Dirac electrons.
	
	Twisted bilayer graphene (TBG) is a two-dimensional (2D) system of two graphene layers stacked with a relative rotation angle $\theta$. This system possesses a superlattice structure due to the formation of a moir\'{e} pattern~\cite{Pong2005review}, which was commonly observed by scanning tunneling microscopy (STM) on the surface of highly oriented pyrolytic graphite in the 1990s~\cite{Buckley1991large,Yang1992several,Rong1993,Cee1995unusual,Bernhardt1998formation,Ouseph2000scanning}. Following the isolation of graphene via mechanical exfoliation~\cite{Novoselov2004electric}, a great number of studies on the electronic structure of this system were carried out~\cite{LopesdosSantos2007,Latil2007,Shallcross2008,Hass2008tbg,Emtsev2008,Varchon2008,Shallcross2010,Mele2010,Li2010observation,Mele2011,Kindermann2011,Santos2012,Brihuega2012stm,Moon2012,Moon_2013}, leading to the prediction of flat bands at the \textit{magic} twist angle $\theta\approx1.1^{\circ}$ of TBG~\cite{Morell2010,Bistritzer2011-mq,Santos2012}. In 2018, Cao \textit{et al.} brought TBG to the center of intense research by observing superconductivity ($T_{\text{c}}\approx1.7$~K) and other strongly correlated phases at twist angles close to the magic angle~\cite{Cao2018,Lu2019-ga,Yankowitz2019,Saito2020-zp,Nuckolls2020-pe,Cao2020,Oh2021-xr,Jaoui2022-aa}. This breakthrough is particularly significant because the phase diagram of this material, which strongly resembles that of high-$T_\text{c}$ superconductors, can be explored in a single sample by electrostatic gating. A great number of works have investigated the origin of superconductivity in TBG and proposed different possible mechanisms, such as phonon-driven superconductivity~\cite{Lian2019,Zhu2025}, plasmon-mediated superconductivity~\cite{Liu2018,Cea2021,Sharma2020,Peng2024}, and flat band superconductivity enabled by quantum geometry~\cite{Hu2019,Julku2020,Xie2020,Verma2021,Tian2023,Tanaka2025}. As these debates intensify, a profound understanding of the fundamental electronic properties of TBG is essential. With QPI being able to provide experimentally verifiable results, understanding QPI in TBG is of considerable interest~\cite{Phong2020,Rhodes2025}. For instance, by examining TBG with twist angle $\theta\sim1.5^\circ$-$2^\circ$, Phong and Mele~\cite{Phong2020} computed the inverse FT of selected QPI signals for two different Dirac models of TBG: one is Wannier representable and the other is Wannier obstructed. As the predictions of the two models differ, QPI measurements may identify the correct one.
	
	In this work, we investigate QPI in TBG with twist angles $\theta\gtrsim2^{\circ}$, where the many-body effects are negligible at energies close to the charge neutrality point. Using atomistic tight-binding calculations, we demonstrate that the QPI patterns capture the full range of interference processes. We discuss how these patterns reflect the emergent symmetries and topology of TBG. Furthermore, we derive a general analytical relation between the QPI signals associated with the back-scattering processes and the form factor. This relation is then applied to TBG to microscopically explain its QPI patterns based on the form factor of a continuum model. Finally, the Appendix provides additional data on QPI in different cases, such as when the defect location breaks $C_3$ symmetry and when the structure of TBG is incommensurate.
	\section{Theoretical Models}
\begin{figure*}
		\includegraphics[width=\linewidth]{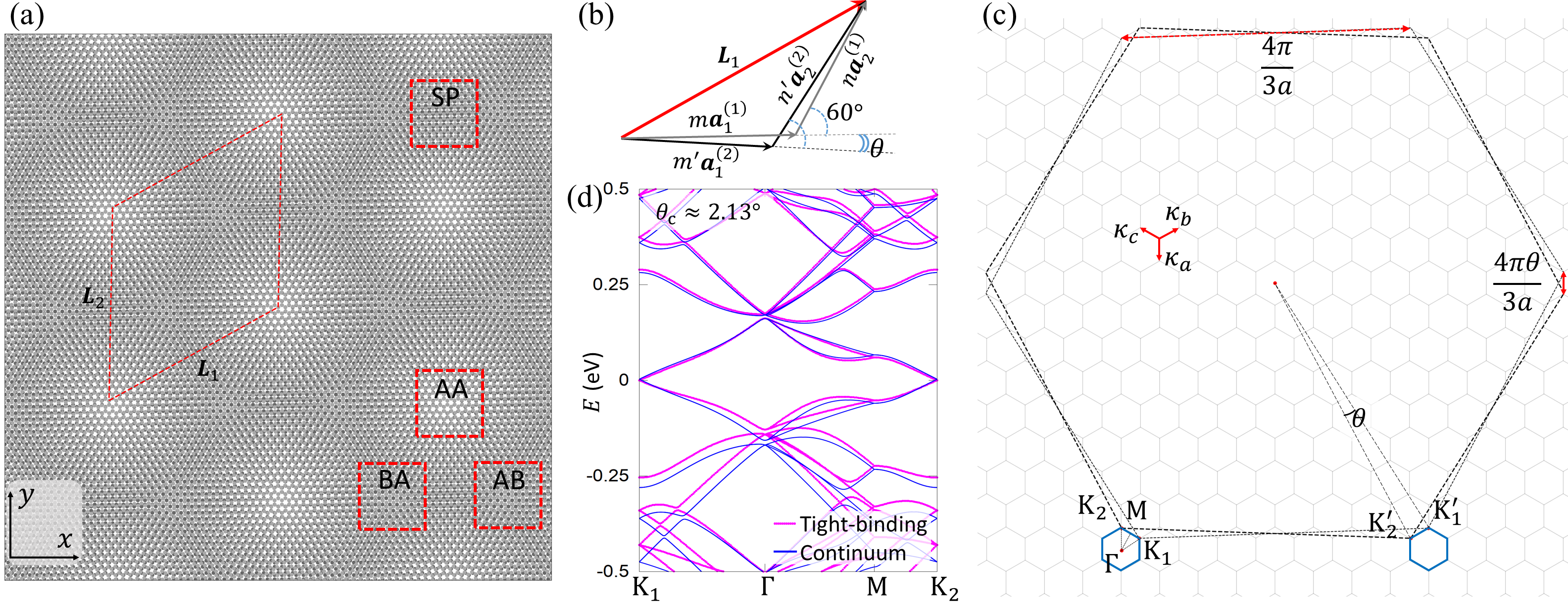}
		\caption{\label{fig:1lattice} \textit{Lattice structure of twisted bilayer graphene}. (a) A commensurate lattice of TBG with the twist angle $\theta_{\text{c}}\approx2.13^\circ$ [$(m,n)=(16,15)$]. The red dashed rhombus represents a supercell and the purple dashed squares indicate different stacking regions within the lattice. (b) Commensurate relation between the superlattice vectors and those of each layer when $|m-n|=1$. (c) The superlattice Brillouin zone of TBG (blue hexagons) with a high-symmetry path K$_1\Gamma$MK$_2$ when $m-n=1$. The Brillouin zones of the two monolayer graphene layers are indicated by the two large dashed hexagons. (d) The electronic spectrum of a commensurate TBG sample with twist angle $\theta_{\text{c}}\approx2.13^{\circ}$ along the high-symmetry path shown in (c), obtained from the tight-binding (magenta lines) and continuum (blue lines) models.}
	\end{figure*}

	In general, TBG is characterized by two structural degrees of freedom: the twist angle $\theta$ and the relative translation vector between the two layers $\boldsymbol{\delta}$. When the lattices of the two layers are commensurate, the combined structure is periodic and a supercell can be defined as its smallest repeating unit, which is independent of the translation vector $\boldsymbol{\delta}$. In this work, the two layers of TBG are denoted as $1$ and $2$, corresponding to the top and bottom layers, where the top layer is rotated counterclockwise by an angle $\theta$. When the twist angle is small ($\theta\lesssim10^{\circ}$), the moir\'{e} pattern of TBG can be seen with the formation of different local regions resembling the regular stacking configurations of bilayer graphene, such as AA, AB, BA, and saddle point (SP) [see Fig.~\ref{fig:1lattice}(a)]. The moir\'{e} beating pattern is characterized by the moir\'{e} lattice vectors~\cite{Hermann2012periodic,Nam2017}, $\bm{L}^{\text{M}}_{1,2}=\left(1 - R^{-1}(\theta)\right)^{-1} \bm{a}_{1,2}$, where $R(\theta)$ is the rotational operator by the twist angle. The moir\'{e} lattice vectors are always present when the two lattices are either commensurate or incommensurate. Meanwhile, when the two graphene lattices are commensurate, there is another type of long-range lattice vectors that define the aforementioned supercells. We define the primitive superlattice vectors $\bm{L}_1$ and $\bm{L}_2$ that translate the supercell to form the entire periodic lattice --- see Fig.~\ref{fig:1lattice}(a). We have
	\begin{align}
		\bm{L}_1 = m\bm{a}_1^{(1)} + n\bm{a}_2^{(1)} &= m'\bm{a}_1^{(2)} + n'\bm{a}_2^{(2)} \label{Eq: coincident vector}\\
		&\text{with }m,n,m',n'\in\mathbb{Z},\nonumber
	\end{align}
	and $\bm{L}_2$ can be obtained by rotating $\bm{L}_1$ by $60^\circ$ or $120^\circ$. Here, $\bm{a}_1^{(\ell)}$ and $\bm{a}_2^{(\ell)}$ are the lattice vectors of layer $\ell=1,2$; $m$ and $n$ are coprime integers. If we choose $\bm{a}_{\nu}^{(2)}=R_{\theta}\bm{a}_{\nu}^{(1)}$ for $\nu=1,2$, it is straightforward to see that $m=n'$ and $n=m'$ [Fig.~\ref{fig:1lattice}(b)]. The twist angle is determined by~\cite{Trambly_de_Laissardiere2010}
	\begin{equation}
		\cos\theta = \frac{1}{2}\frac{m^2+n^2+4mn}{m^2+n^2+mn},\label{Eq: twist angle}
	\end{equation}
	and the superlattice constant is given by
	\begin{equation}
		L = \frac{a\sqrt{m^2+n^2+mn}}{\sqrt{\gcd(m-n,3)}} = \frac{|m-n|a}{2\sin(\theta/2)\sqrt{\gcd(m-n,3)}}.
	\end{equation}
	The factor $\sqrt{\gcd(m-n,3)}$ appears to ensure that $L$ is the length of the primitive superlattice vectors --- see the Appendix for a simplified argument. The value of $L$ is $|m-n|/\sqrt{\gcd(m-n,3)}$ times greater than the moir\'{e} lattice constant, $L_{\text{M}}=\frac{a}{2\sin(\theta/2)}$.
	We note that Eq.~\eqref{Eq: twist angle} yields two values $\pm\theta$ of the twist angle. With our current choice for the lattice vectors of monolayer graphene, $\theta$ is only positive when $m>n$, as can be seen geometrically from Fig.~\ref{fig:1lattice}(b). If we choose $m-n=1$, the primitive superlattice vectors coincide with the moir\'{e} lattice ones.
	
	Angles given by Eq.~\eqref{Eq: twist angle} ensure that the two lattice structures are commensurate and the translational symmetry is present. It is worth noting that the band folding mechanisms are different for different values of $m-n$: Mele~\cite{Mele2010} distinguishes two special classes with $\gcd(m-n,3)=1$ and $\gcd(m-n,3)=3$. For convenience, we denote $\lambda=\gcd(m-n,3)$ hereafter. The relative reciprocal displacement of the Dirac points from the two layers, $\Delta\bm{K}=\bm{K}_2-\bm{K}_1$, is a reciprocal superlattice vector when $\lambda=3$ and is not when $\lambda=1$~\cite{Santos2012}. However, Lopes dos Santos~\cite{Santos2012} showed that the condition $\lambda=1$ worked well for all small twist angles. In the following, we mainly consider the commensurate structures with $\lambda=1$ and only examine structures with $\lambda=3$ in Section~\ref{Sec: Wannier}. For $\lambda=1$, it is straightforward to identify the superlattice Brillouin zone since $\Delta\bm{K}$ connects two adjacent vertexes of this hexagon --- as depicted in Fig.~\ref{fig:1lattice}(c). The notations of the valleys, i.e., the Dirac cones or the $K$ points, are shown in Fig.~\ref{fig:1lattice}(c), where $K_1$ and $K_1'$ belong to the top layer whereas $K_2$ and $K_2'$ are of the other one. The Dirac cones at $K_1$ and $K_2$ have the same winding number and opposite to that of Dirac cones at $K_1'$ and $K_2'$. The adjacent $K$ ($K'$) points are connected via the three vectors $\boldsymbol{\kappa}_a$, $\boldsymbol{\kappa}_b$, and $\boldsymbol{\kappa}_c$, whose magnitude is $\frac{4\pi}{3a}\times2\sin\frac{\theta}{2}\approx\frac{4\pi\theta}{3a}$ for small twist angles.
	\subsection{Tight-binding model}
	In the next section, we use an atomistic tight-binding model~\cite{Trambly_de_Laissardiere2010,Moon_2013} of TBG in the real space representation to simulate its QPI pattern. Within each layer, only the nearest-neighbor hopping terms $t_0=-2.7$~eV~\cite{CastroNeto_2009} are taken into account. The interlayer hopping between two atoms separated by $\bm{d}$ is modeled through
	%
	%
	\begin{align}
		t(\bm{d}) = V_{pp\pi}(d)\left[1-\left(\frac{\bm{d}}{d}\cdot\bm{e}_z\right)^2\right]+V_{pp\sigma}(d)\left(\frac{\bm{d}}{d}\cdot\bm{e}_z\right)^2 
	\end{align}
	%
	%
	with $V_{pp\pi}(d)  = V_{pp\pi}^0 \exp\left(-\frac{d-a_\text{cc}}{r_0}\right)$ and $V_{pp\sigma}(d)  = V_{pp\sigma}^0 \exp\left(-\frac{d-d_0}{r_0}\right)$ for $a_{\text{cc}}=a/\sqrt{3}$, $r_0=0.184a$, and $d_0=1.362a$; $a=0.246$~nm is graphene's lattice constant.
	Here, $V_{pp\pi}^0=-2.7$~eV and $V_{pp\sigma}^0=0.48$~eV are two Slater-Koster parameters~\cite{Trambly_de_Laissardiere2010}. We impose a cutoff value $d_{\text{c}}\approx1.934a$, beyond which the interlayer hopping is set to zero. When the twist angle satisfies Eq.~\eqref{Eq: twist angle}, the structure is periodic with each supercell containing $4(m^2 + n^2 + mn)/\lambda$ carbon atoms, and we can obtain the band structure by writing the Hamiltonian matrix in the momentum space representation. The spinless band structure shown in Fig.~\ref{fig:1lattice}(d), which is independent of the translation vector $\boldsymbol{\delta}$, has two four-fold degenerate Dirac points at $K_1$ and $K_2$ due to the band folding effect: the Dirac points at $K_1'$ and $K_2$ are degenerate, and so are those at $K_1$ and $K_2'$.
	\subsection{Continuum model}
	The continuum model of TBG will be used to explain the QPI patterns since it decouples the degenerate Dirac points. This model is represented in a plane wave basis and is derived based on the electronic decoupling between states at $K$ and $K'$ valleys. The Hamiltonian associated with the $K$ valleys is written as follows
	%
	%
	\begin{align}
		H(\bm{k}) = \sum_{\bm{K}_1}\Big[&\chi^{\dagger}_{\bm{K}_1}h_1(\bm{k}-\bm{K}_1)\chi_{\bm{K}_1} \nonumber\\
		&+ \chi^{\dagger}_{\bm{K}_1+\boldsymbol{\kappa}_a}h_2(\bm{k}-\bm{K}_1-\boldsymbol{\kappa}_a)\chi_{\bm{K}_1+\boldsymbol{\kappa}_a} \nonumber\\
		&+ \sum_{\mu=a,b,c}\left(\chi^{\dagger}_{\bm{K}_1}T_\mu\chi_{\bm{K}_1+\boldsymbol{\kappa}_\mu} + h.c.\right)\Big],\label{Eq: continuum}
	\end{align}
	%
	%
	where $\chi_{\bm{K}_j}$ are two-component vectors, with each component representing a sublattice in a graphene layer, the diagonal blocks
	\begin{equation}
		h_{1/2}(\bm{k}) = \frac{\sqrt{3}t_0a}{2}\begin{pmatrix}
			0 & |\bm{k}|e^{-i(\varphi_{\bm{k}}\mp\theta/2)}\\
			|\bm{k}|e^{i(\varphi_{\bm{k}}\mp\theta/2)} & 0
		\end{pmatrix}
	\end{equation}
	describe the Dirac Hamiltonian at each $K$ valley, and the coupling matrices between the Dirac cones
	\begin{equation}
		T_\mu = \begin{pmatrix}
			u & we^{i\gamma_\mu} \\ we^{-i\gamma_\mu} & u
		\end{pmatrix},\quad \left(\gamma_a,\gamma_b,\gamma_c\right) = \left(0, -\frac{2\pi}{3}, \frac{2\pi}{3}\right).
	\end{equation}
	Here, $\varphi_{\bm{k}}$ is the argument of $k_x + ik_y$, and we choose $u=w=0.105$~eV, which yields good agreement between the two models for the twist angles of interest ($\theta\gtrsim2^{\circ}$) --- see Fig.~\ref{fig:1lattice}(d). The two models agree well in their overall band dispersions, yet the tight-binding model exhibits a stronger particle-hole asymmetry than the continuum model. The continuum Hamiltonian associated with the $K'$ valleys is a time-reversed counterpart of $H(\bm{k})$, given by $H^*(-\bm{k})$. The eigenstate wave function is a superposition of the moir\'{e} plane waves as
	\begin{equation}
		u_{n\bm{k}}(\bm{r}) = \sum_{\bm{K}_1}C_{n\bm{K}_1}(\bm{k})e^{-i\bm{K}_1\cdot\bm{r}} + \sum_{\bm{K}_2}C_{n\bm{K}_2}(\bm{k})e^{-i\bm{K}_2\cdot\bm{r}},\label{Eq: eigenstate expansion}
	\end{equation}
	where $n$ is the band index and $C_{n\bm{K}_{1/2}} = \begin{pmatrix}
		\chi_{n\text{A}\bm{K}_{1/2}}(\bm{k}) & \chi_{n\text{B}\bm{K}_{1/2}}(\bm{k})
	\end{pmatrix}^\text{T}$ are two-component coefficients of the plane wave expansion. Here, $u_{n\bm{k}}(\bm{r})$ is the periodic part of the Bloch wave function $\Psi_{n\bm{k}}(\bm{r}) = e^{i\bm{k}\cdot\bm{r}}u_{n\bm{k}}(\bm{r})$; however, we refer to $u_{n\bm{k}}(\bm{r})$ as the wave function for brevity. The components $\chi_{\text{A}\bm{K}_{1/2}}(\bm{k})$ and $\chi_{\text{B}\bm{K}_{1/2}}(\bm{k})$ correspond to the A and B sublattice of graphene, respectively. Diagonalizing the Hamiltonian matrix requires a sufficient number of plane waves in order for the results to converge. In our calculations, we choose a cutoff radius of $K_\text{cutoff}=1/a$, which works excellently for the twist angles of interest.
	\section{Quasiparticle Interference in Twisted Bilayer Graphene}
	The physics of QPI in a 2D material is given by the spatial FT of the variation in the LDOS due to a defect, which hereafter is referred to as the FT-LDOS. In order to obtain this quantity, we use the tight-binding model of the material in the real space representation. With the TBG sample being a circular disk of radius $100$~nm, the Hamiltonian is represented by a square matrix of order exceeding two million. The variation of the LDOS of TBG due to a defect is obtained by computing the LDOS over a circular area of radius $40$~nm concentric with the sample in two cases: with and without the defect. The defect is modeled by an increase by $\varepsilon_0=0.8|t|$ in the on-site energy of a carbon atom at the center of the sample on layer 1. This value is chosen so that all important scattering processes are present while the signals are clear~\cite{Hong2021-pn}. Here, we neglect the spontaneous time-reversal symmetry breaking due to the defect~\cite{Ulman2014,Lopez-Bezanilla2019,Dietrich2023,Chang2024}.
	\subsection{Kernel polynomial method}
	The spatial LDOS is computed using the kernel polynomial method~\cite{Weisse2006,Le2018,Do_2019,Do_2021,pybinding}. The LDOS at position $\bm{r}$ is given by
	\begin{equation}
		\rho(\bm{r},\tilde{e}) = -\frac{1}{\pi}\Im\left[\braket{\bm{r}|\mathcal{G}^+(\tilde{e},\tilde{H})|\bm{r}}\right],
	\end{equation}
	where $\tilde{e}$ and $\tilde{H}$ are the rescaled energy and Hamiltonian, respectively, and $\mathcal{G}^+(\tilde{e},\tilde{H})$ is the retarded Chebyshev polynomial Green's function~\cite{Braun2014,Ferreira2015,Joao2020kite}. Here, the energy and Hamiltonian are rescaled so that the spectrum becomes dimensionless and varies between $-1$ and $+1$. The retarded Chebyshev polynomial Green's function reads
	\begin{align}
		\mathcal{G}^+(\tilde{e},\tilde{H}) = \frac{1}{\tilde{e}-\tilde{H}+i\eta} &= \sum_{n=0}^{\infty}\mu_n(\tilde{e})\frac{T_n(\tilde{H})}{1 + \delta_{n,0}}\nonumber\\
		&\approx\sum_{n=0}^{N_{\text{c}}}\mu_n(\tilde{e})\frac{T_n(\tilde{H})}{1 + \delta_{n,0}},
	\end{align}
	where the Chebyshev moments are
	\begin{equation}
		\mu_n(\tilde{e}) = -2i\frac{e^{-in\arccos(\tilde{e}+i\eta)}}{\sqrt{1 - (\tilde{e}+i\eta)^2}},
	\end{equation}
	$\eta>0$ is the energy resolution, and $T_n(\tilde{H})$ are the Chebyshev polynomials of $\tilde{H}$. The number of Chebyshev terms, $N_{\text{c}}$, depends on the energy resolution $\eta$. The calculation involves the evaluation of $\braket{\bm{r}|T_n(\tilde{H})|\bm{r}}$. We define the Chebyshev vectors $\ket{\alpha_n} = T_n(\tilde{H})\ket{\bm{r}}$ and have $\braket{\bm{r}|T_n(\tilde{H})|\bm{r}} = \braket{\bm{r}|\alpha_n}$. The Chebyshev vectors are obtained recursively through the relation
	\begin{equation}\label{1.2}
		\begin{split}
			\ket{\alpha_0} &= T_0(\tilde{H})\ket{\bm{r}} = \ket{\bm{r}}\\
			\ket{\alpha_1} &= T_1(\tilde{H})\ket{\bm{r}} = \tilde{H}\ket{\bm{r}}\\
			&\vdots\\
			\ket{\alpha_{n+1}} &= T_{n+1}(\tilde{H})\ket{\bm{r}} = 2\tilde{H}\ket{\alpha_n} - \ket{\alpha_{n-1}}.
		\end{split}
	\end{equation}
	With a sufficient number of Chebyshev moments $\mu_n(\tilde{e})$ and $\braket{\bm{r}|T_n(\tilde{H})|\bm{r}}$, the LDOS can be found accordingly. This calculation is efficiently implemented in the Pybinding~\cite{pybinding} package. 
	\subsection{FT-LDOS}
\begin{figure*}
		\includegraphics[width=\linewidth]{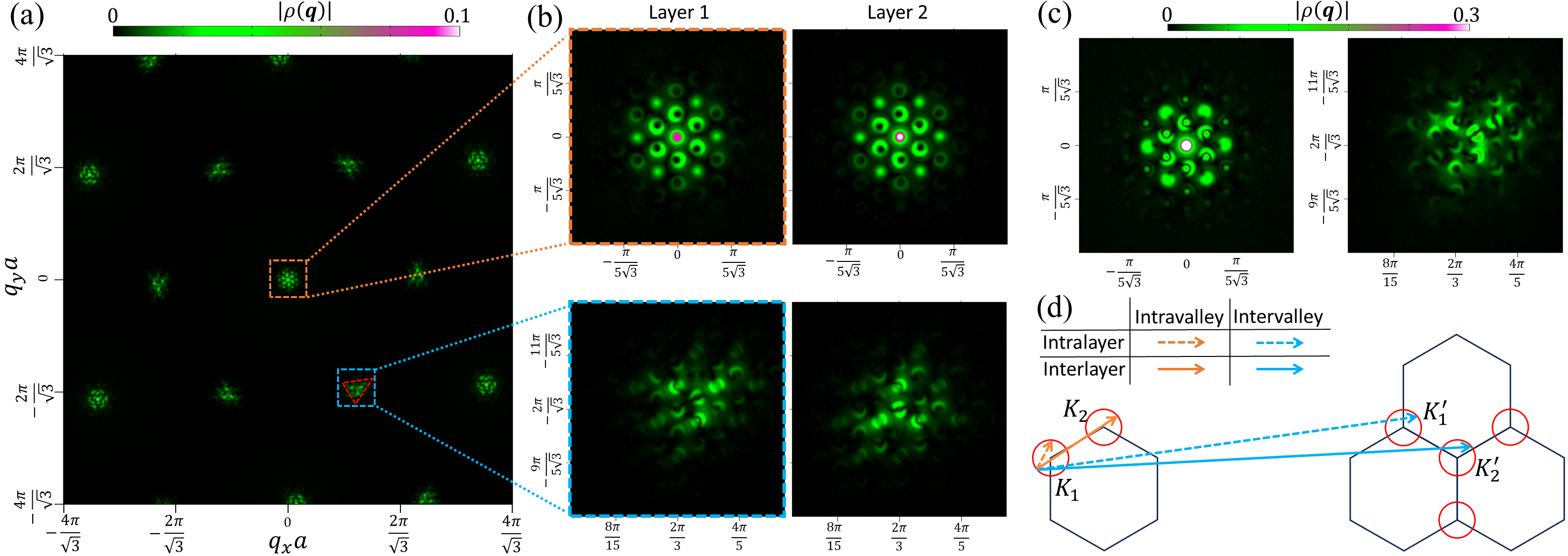}
		\caption{\label{fig:2qpi} \textit{Quasiparticle interference in twisted bilayer graphene}. (a) FT-LDOS of layer 1 of TBG for $\theta_{\text{c}}\approx2.13^{\circ}$ when the defect centers at an AB-stacking region. The Fermi energy is chosen to be $15$~meV and $\eta=5$~meV. The signals include interference between states at valleys of the same winding number (orange box) as well as those at valleys of opposite winding numbers (blue box). (b) The left column shows magnified signals of (a) and the right column includes those of layer 2. (c) The interference signals of layer 1 when the defect centers at an AA-stacking region. (d) Sketch of the interference mechanisms in the reciprocal space of TBG with the red circles depicting the Fermi surface. The orange arrows indicate interference between intravalley valleys while the blue ones indicate those between intervalley valleys.}
	\end{figure*}

	The FT-LDOS of the top layer of our TBG sample for $\theta_{\text{c}}\approx2.13^{\circ}$ is shown in Fig.~\ref{fig:2qpi}(a) at energy $E_{\text{F}}=15$~meV. The defect is placed at the center of the sample, which locally has the AB-stacking configuration and preserves the three-fold ($C_3$) rotational symmetry of the entire sample. The twist angle is given by Eq.~\eqref{Eq: twist angle} with $(m,n)=(16,15)$, leading to a commensurate structure of TBG. Hereafter, we call the central signals enclosed in the orange box ``intravalley", and those within the blue box ``intervalley", which will be explained in the next subsection.
	
	Overall, the interference patterns are constituted by clusters of signals, where the positions of these clusters are the same as those of the QPI patterns of monolayer graphene~\cite{Bena2008,Pereg-Barnea2008,Brihuega2008,Mallet2012}. The signals are filled or empty circles with continuous or disconnected circumferences, whose radii are twice the radius of the Fermi contours. The intravalley cluster is at the same position as the intravalley interference of monolayer graphene, while intervalley clusters are located at positions identical to those of the intervalley signals of monolayer graphene. The distance between the centers of these intervalley clusters is $4\pi/3a$, which is also the distance between the $\bm{K}$ and $\bm{K}'$ points of monolayer graphene. The first interesting feature of the QPI patterns is the rotation of the six intervalley clusters: their individual shapes can be fitted into triangles [red dashed triangle in Fig.~\ref{fig:2qpi}(a)], which are rotated with respect to their adjacent companions by $60^\circ$. This feature creates a chiral structure over a large $\bm{q}$ area.
	
	Next, we take a closer look at the signals of each cluster. Two clusters of signals are magnified in the left column of Fig.~\ref{fig:2qpi}(b). The intravalley signals include a filled circle at the center, six adjacent rings, and higher order signals resembling the central circle and its surrounding rings. The signals are different in geometry and intensity, yet their positions are well defined and they can be arranged into a triangular lattice. The pattern, having a $C_6$-rotational symmetry, is analogous to that of monolayer graphene but occurring on a considerably smaller scale in the reciprocal space; the distance between the centers of two nearest rings is approximately $\frac{4\pi\theta}{3a}$, matching with the side length of the hexagonal moir\'{e} Brillouin zone. Noticeably, the intensity of each ring has a clear minimum value that is rotated by $60^\circ$ between adjacent rings, forming the second chiral structure of the QPI patterns over a smaller scale of $\bm{q}$. On the other hand, the intervalley signals [the bottom left panel of Fig.~\ref{fig:2qpi}(b)] are clearly distinct in shape and geometry, and they can also be arranged into a triangular lattice whose lattice constant is the side length of the moir\'{e} Brillouin zone. We focus on the three brightest signals, which can be seen as filled circles with disconnected circumferences, and the one at their center, which is an empty circle with a disconnected circumference. All these disconnected circumferences consist of two arcs as their intensity vanishes at two points.
	
	Additionally, the right column of Fig.~\ref{fig:2qpi}(b) shows all those signals but obtained from the LDOS of the sample's lower layer --- the chirality is reversed at both the small and large scales of $\bm{q}$. Indeed, if we compute the LDOS of both graphene layers and take the FT, the signals no longer exhibit a clear chiral structure. The chirality change also occurs at the small scale of $\bm{q}$ if we consider electrons in layer 1 with the Fermi energy below the Dirac points. 
	
	Thus far, all the results are obtained with the defect located at the center of an AB-stacking region. When the defect is located at the center of an AA-stacking region and still preserves the $C_3$ rotational symmetry, the FT-LDOS exhibits a similar structure with the same chirality, as shown in Fig.~\ref{fig:2qpi}(c). The main difference is the fillings of the circles: the six rings of the intravalley cluster are now filled whereas the three brightest rings of the intervalley cluster are empty. The FT-LDOS of TBG when the defect is located at a random position and when the structure is incommensurate are presented in the Appendix.
	\subsection{Interference mechanisms}
	The origin of the QPI patterns in the FT-LDOS of TBG can be explained by the interference between the electronic states at the same or different valleys, as depicted in Fig.~\ref{fig:2qpi}(d). When the Fermi energy is slightly above the Dirac band touching points, the Fermi surface of TBG is composed of circular contours centering at the $K$ and $K'$ points. The QPI patterns appear due to the interference between states residing on these contours. For the intravalley cluster, the central signal stems from the interference between two states at the same valley, while the six adjacent rings are present due to the interference between two states at two valleys $K_1$ and $K_2$, or $K_1'$ and $K_2'$. As $K_1$ and $K_2$ (or $K_1'$ and $K_2'$) originate from two different graphene layers, we term the latter interference \textit{interlayer} whereas the former one \textit{intralayer}. Both cases involve interference between states at two valleys with identical winding number; they are thus called \textit{intravalley}.
	Meanwhile, the intervalley clusters correspond to the interference between states at two different valleys with opposite winding numbers. Here, the three brightest signals are associated with the interference between states at two valleys $K_1$ and $K_2'$, or $K_1'$ and $K_2$, which belong to two different layers. The central signals of these clusters stem from the interference between states at two valleys of opposite winding numbers and in the same layer, thus they are referred to as intervalley intralayer interference. Here, the two intralayer interference processes (intravalley and intervalley) are analogous to the QPI patterns of monolayer graphene. The higher-order signals can be explained similarly.
	\section{Symmetry and Topology \label{Sec: Wannier}}
	QPI provides an excellent tool to understand not only the spectrum of a material but also the properties of its wave functions~\cite{Queiroz2018,Dutreix2019,Zhang2019local,Zhang2021robust,Yang2024wavefronts,Wang2025}. In this section, we consider a TBG sample with a commensurate structure satisfying the condition $\lambda=3$ and establish the role of the approximate translational symmetries and valley charge conservation~\cite{Zou2018} in twisted bilayer graphene. Then, by examining the wavefront dislocations in the LDOS, we compare these results with those obtained from two-band models of TBG by Phong and Mele~\cite{Phong2020}, which are related to the Wannier obstruction of the flat bands.
	\subsection{Emergent symmetries}
\begin{figure*}
		\includegraphics[width=\linewidth]{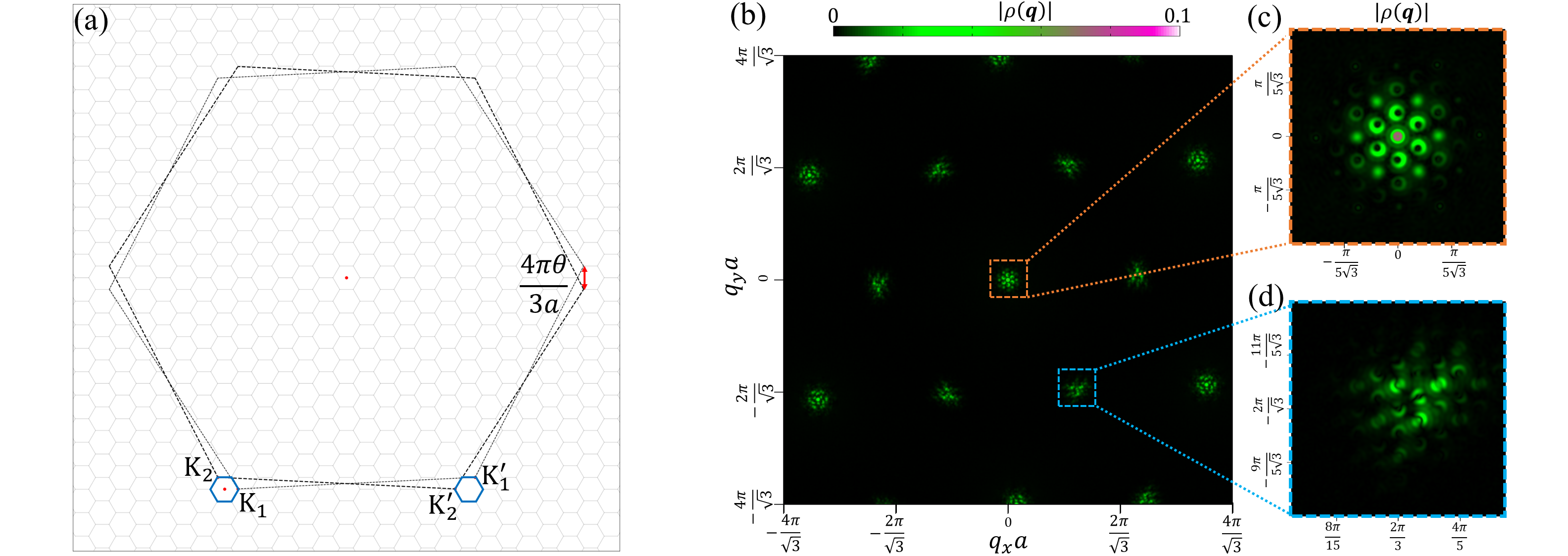}
		\caption{\label{fig:3emergence} \textit{Emergent symmetries in quasiparticle interference}. (a) Sketch of the superlattice Brillouin zone when $\lambda=3$. (b) FT-LDOS of layer 1 of TBG for $\theta_{\text{c}}\approx2.18^{\circ}$ when the defect is located at the center of an AB-stacking region. The Fermi energy is chosen to be $15$~meV and $\eta=5$~meV. (c) Magnified intravalley signals of (b). (d) Magnified intervalley signals of (b).}
	\end{figure*}

	We consider a commensurate TBG sample with twist angle $\theta_{\text{c}}\approx2.18^{\circ}$, given by Eq.~\eqref{Eq: twist angle} for $(m,n)=(47,44)$. With $m-n=3$, the length of the superlattice vectors is $L=\sqrt{3}L_{\text{M}}$, and the relative reciprocal displacement of the Dirac points, $\Delta\bm{K}$, is a reciprocal superlattice vector, as illustrated in Fig.~\ref{fig:3emergence}(a). Here, the length of the reciprocal lattice vectors is approximately $\frac{4\pi\theta}{3a}$, leading to the side of the hexagonal Brillouin zone being $\frac{4\pi\theta}{3\sqrt{3}a}$. The folding of the Dirac points in this case clearly differs from that in the previous section: the $K$ points ($K_1$ and $K_2$) become degenerate and the $K'$ ones ($K_1'$ and $K_2'$) are folded together. Therefore, several works have discussed whether such a difference affects the electronic properties of TBG~\cite{Mele2010,Santos2012,Zou2018}. For large twist angles, the two types of commensurate TBG feature different spectra --- structures with $\lambda=1$ have gapless spectra, similar to monolayer and nontwisted bilayer graphene, whereas those with $\lambda=3$ have a spectral gap at the charge neutrality point~\cite{Mele2010}. However, when the twist angle becomes small, the spectrum always remains gapless for both types of commensurate structures because the Fourier components of the interlayer potential generating the gap vanish linearly as $\theta$ decreases~\cite{Santos2012}.
	
	The FT-LDOS of the sample is shown in Fig.~\ref{fig:3emergence}(b) and its two important clusters of signals are magnified in Figs.~\ref{fig:3emergence}(c) and~\ref{fig:3emergence}(d). The patterns strongly resemble that of Fig.~\ref{fig:2qpi} --- the distance between centers of the circular signals is equal to $\frac{4\pi\theta}{3a}$ instead of $\frac{4\pi\theta}{3\sqrt{3}a}$. The patterns result from the interference between states at Dirac cones of the moir\'{e} Brillouin zone, which differs from the superlattice Brillouin zone. This is strong evidence that the electronic properties of TBG are governed solely by the moir\'{e} length scale $L_{\text{M}}$, regardless of the specific commensurate superlattice. This is referred to as the approximate translational symmetries by Zou \textit{et al.}~\cite{Zou2018}, which suppress the direct coupling between Dirac points of the same winding number and thus prevent the gap opening. Furthermore, as the main features of the FT-LDOS remain unchanged even for incommensurate structures of TBG (see the Appendix), the approximate translational symmetries play a key role in describing the electronic properties of the material, which is one of the premises for constructing the continuum model of TBG~\cite{Bistritzer2011-mq}.
	
	Another emergent symmetry that is reflected in the QPI patterns of TBG is the valley charge conservation. In both Figs.~\ref{fig:2qpi}(a) and~\ref{fig:3emergence}(b), the clusters of signals are clearly separated, distinguishing the intravalley and intervalley signals. This means that the scatterings between Dirac cones of the same winding number only require a small momentum exchange, while those between cones of different winding numbers can only appear due to a large momentum exchange, such as by the atomic-scale defect used in our simulations. Therefore, in the low energy limit, two intravalley Dirac cones are decoupled from the other two cones, preserving the valley pseudospin in TBG.
	\subsection{Wavefront dislocations}
	Thus far, we have only considered the absolute value of the FT-LDOS since this quantity is complex. However, by taking into account the argument of this complex quantity, we can gain insights into the topology of electronic bands of TBG. Studying QPI in monolayer graphene with a hydrogen atom chemisorbed, Dutreix \textit{et al.} showed that the argument of the LDOS exhibits singularities at the intervalley signals, which lead to a singularity in the real space where the wavefront dislocation emerges~\cite{Dutreix2019}. The wavefront dislocation is visualized by taking the inverse FT of the selected signals in the FT-LDOS, which we refer to as the Fourier-filtered LDOS, $\Delta\rho_{\text{filter}}(\omega,\bm{r})$. Importantly, if the defect preserves the rotational symmetry of the lattice~\cite{Liu2024visualizing,Ren2026}, the number of additional wavefronts due to the phase singularity is twice the Berry phase divided by $2\pi$~\cite{Dutreix2019,Zhang2020localBerry}.
	
	Inspired by this property of the QPI, Phong and Mele examined two effective models of the low-energy Dirac cones of TBG~\cite{Phong2020}. One model contains two Dirac cones with opposite chirality, similar to monolayer graphene, and corresponds to a tight-binding model of the flat bands with two Wannier orbitals. These localized orbitals are centered at the AB and BA stacking regions and have three peaks at the AA-stacking regions surrounding the center~\cite{Koshino2018,Kang2018}. Conversely, the other effective model has two Dirac cones of the same chirality and has no equivalent tight-binding model with two orbitals for describing only the flat bands. This is referred to as the topological obstruction of the flat bands of TBG --- to satisfy the symmetries of TBG, tight-binding models have to take into account auxiliary bands separated from the flat bands~\cite{Zou2018,Ahn2019,Carr2019}. Examining the Fourier-filtered LDOS of the two effective models in the presence of a large defect, Phong and Mele found that the Wannier-representable model has a dislocation with two additional wavefronts whereas the Wannier-obstructed one has no dislocation. These results, however, do not exactly match with those obtained by the continuum model of TBG: there is no dislocation when the defect is located at the AA-stacking region and a dislocation with only one additional wavefront when it is centered at the AB-stacking region.
	
\begin{figure*}
		\includegraphics[width=\linewidth]{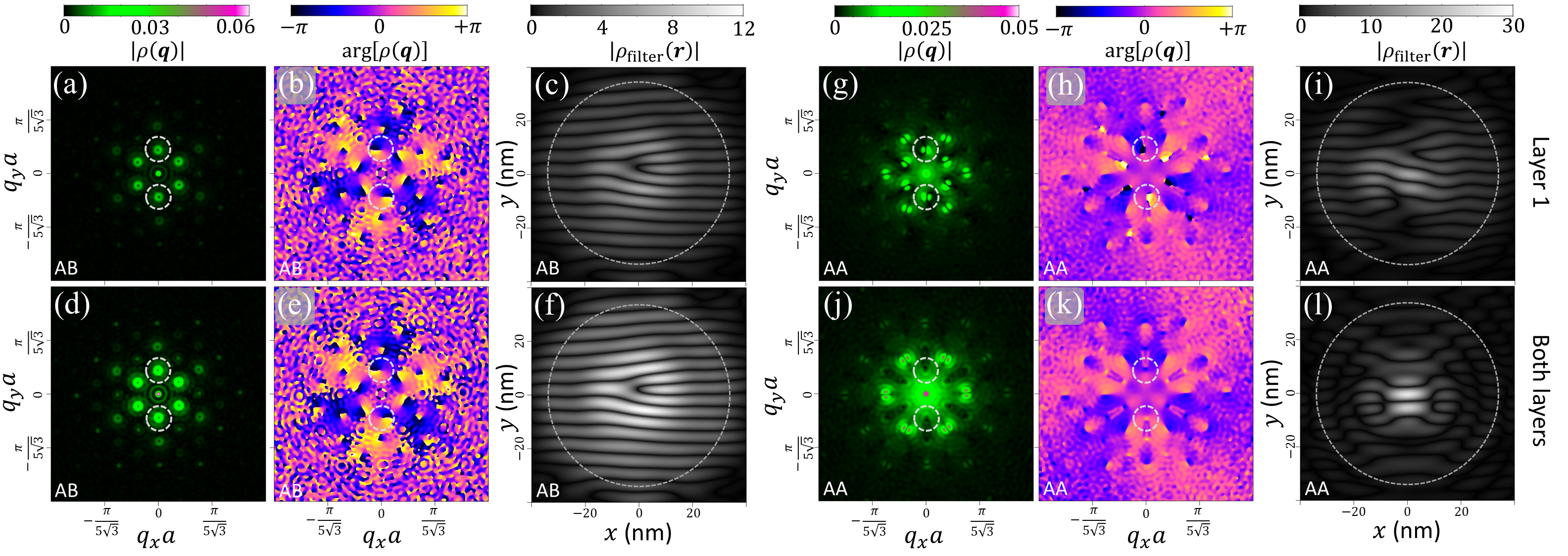}
		\caption{\label{fig:4dislocation} \textit{Wavefront dislocations in twisted bilayer graphene}. The absolute value (a,d,g,j) and argument (b,e,h,k) of the FT-LDOS in a commensurate structure of TBG with twist angle $\theta_{\text{c}}\approx2.18^{\circ}$. The signals within the dashed circles are inverse FT to real space, yielding the Fourier-filtered LDOS shown in (c,f,i,l). The dashed circles in (c,f,i,l) are centered at the defect and indicate the variation of the number of wavefronts. The LDOS is only taken from layer~1 (a,b,c,g,h,i) or from both layers (d,e,f,j,k,l).}
	\end{figure*}

	Using our real-space tight-binding model of TBG, we compute the FT-LDOS and Fourier-filtered LDOS at approximately zero energy\footnote{Here, we choose the value corresponding to the charge neutrality point for twist angle $\theta_{\text{c}}\approx2.13^{\circ}$. However, this point slightly shifts when the twist angle changes.} for twist angle $\theta_{\text{c}}\approx2.18^{\circ}$, i.e., $(m,n)=(47,44)$, when the defect is located at the center of the AB- and AA-stacking regions. The results, including the absolute value and phase of the FT-LDOS and the Fourier-filtered LDOS of two intravalley interlayer signals, are shown in Fig.~\ref{fig:4dislocation} for the LDOS of atoms belonging to only the top layer (top row of Fig.~\ref{fig:4dislocation}) and from both layers (bottom row of Fig.~\ref{fig:4dislocation}). When the defect is at the center of the AB-stacking region, the intravalley interlayer signals are symmetric circles and their phase displays vortices with a winding of $2\pi$. Taking the inverse FT of only those signals around $q_x=0$, we observe a dislocation with two additional wavefronts, which is identical to the result obtained by the Wannier-representable model of Ref.~\cite{Phong2020}. When the defect is at the center of the AA-stacking region, the vortices in the FT-LDOS phase disappear. Consequently, the Fourier-filtered LDOS always has an even number of dislocations so that the total number of wavefronts passing through a circle centered at the defect is conserved. This is similar to the result obtained by the Wannier-obstructed model of Ref.~\cite{Phong2020}. Importantly, the results remain the same when the signals are from only layer~1 and from both layers. From a theoretical point of view, the wave functions of the flat bands reside in both layers of TBG, and the LDOS of both layers is therefore needed to demonstrate the topology of these bands. However, in experiments, the LDOS of the top layer is far stronger than that of the bottom layer as the tunneling probability decays exponentially. Hence, the fact that both cases yield the same number of dislocations indicates that the topology of the flat bands is experimentally accessible.
	
	\section{Quasiparticle Interference and Form Factor}
	In this section, we present a general analytical relation between the form factor and the QPI pattern. We consider a 2D lattice with a single localized defect described by a potential $V(\bm{r})$. The Dyson equation of the retarded Green's function is written in the coordinate representation as follows~\cite{Economou2006-te}
	\begin{align}
		&G^+(\bm{r},\bm{r}';\omega) = G_0^+(\bm{r},\bm{r}';\omega) \nonumber\\
		&+ \int d^2r_1d^2r_2G_0^+(\bm{r},\bm{r}_1;\omega)\braket{\bm{r}_1|V|\bm{r}_2}G^+(\bm{r}_2,\bm{r}';\omega).
	\end{align}
	As the electronic LDOS is proportional to the imaginary part of the diagonal elements of the retarded Green's function, $\rho(\bm{r},\omega) = -\frac{1}{\pi}\Im\left[G^+(\bm{r},\bm{r};\omega)\right]$, we first consider the retarded Green's function within the first Born approximation. Taking the Born series up to the first order of the Dyson equation, we obtain
	\begin{align}
		G^+(\bm{r},\bm{r}'&;\omega) \approx
		G_0^+(\bm{r},\bm{r}';\omega) \nonumber\\
		&+ \int d\bm{}r_1G_0^+(\bm{r},\bm{r}_1;\omega)V(\bm{r}_1)G^+_0(\bm{r}_1,\bm{r}';\omega).
	\end{align}
	Inserting the spectral representation of the Green's function into this expression, we arrive at the following expression
	\begin{align}
		&G^+(\bm{r},\bm{r}',\omega) - G^+_0(\bm{r},\bm{r}',\omega)\nonumber\\
		&\approx \int d^2r_1\sum_{m,\bm{k}}\frac{\psi_{m\bm{k}}(\bm{r})\psi_{m\bm{k}}^*(\bm{r}_1)}{\omega - E_{m\bm{k}} + i\varepsilon}V(\bm{r}_1)\sum_{n,\bm{k}'}\frac{\psi_{n\bm{k}'}(\bm{r}_1)\psi_{n\bm{k}'}^*(\bm{r}')}{\omega - E_{n\bm{k}'} + i\varepsilon}\nonumber\\
		&= \sum_{m,n}\sum_{\bm{k},\bm{k}'}\frac{\psi_{m\bm{k}}(\bm{r})\psi_{n\bm{k}'}^*(\bm{r}')\int d^2r_1\psi_{m\bm{k}}^*(\bm{r}_1)V(\bm{r}_1)\psi_{n\bm{k}'}(\bm{r}_1)}{(\omega - E_{m\bm{k}} + i\varepsilon)(\omega - E_{n\bm{k}'} + i\varepsilon)}.
	\end{align}
	Here, $\psi_{m\bm{k}}(\bm{r})$ and $E_{m\bm{k}}$ are the eigenfunctions and eigenenergies of the pristine lattice, i.e., without the defect. The change in the LDOS due to the defect is given by $\Delta\rho(\bm{r},\omega) = -\frac{1}{\pi}\Im\left[G^+(\bm{r},\bm{r};\omega) - G^+_0(\bm{r},\bm{r};\omega)\right]$. The QPI patterns are given by the FT-LDOS, which reads
	\begin{widetext}    
		\begin{align}
			\Delta\rho(\bm{q},\omega) = \int d^2re^{i\bm{q}\cdot\bm{r}}\Delta\rho(\bm{r},\omega)= \frac{i}{2\pi}\int d^2re^{i\bm{q}\cdot\bm{r}}\sum_{m,n}\sum_{\bm{k},\bm{k}'}\left[\frac{\psi_{m\bm{k}}(\bm{r})\psi_{n\bm{k}'}^*(\bm{r})\int d^2r_1\psi_{m\bm{k}}^*(\bm{r}_1)V(\bm{r}_1)\psi_{n\bm{k}'}(\bm{r}_1)}{(\omega - E_{m\bm{k}} + i\varepsilon)(\omega - E_{n\bm{k}'} + i\varepsilon)} - c.c.\right].
		\end{align}
	\end{widetext}
	For simplicity, we separate the right-hand side into two terms by defining
	so that $\Delta\rho(\bm{q},\omega) = \Delta\rho_1(\bm{q},\omega) - \Delta\rho_2(\bm{q},\omega)$.
	Substituting the Fourier transform of the defect's potential, $V(\bm{r}) = \int d^2pV_{\bm{p}}e^{i\bm{p}\cdot\bm{r}}$, into the first term gives
	\begin{widetext}
		\begin{align}
			\Delta\rho_1(\bm{q},\omega) = \frac{i}{2\pi}\sum_{m,n}\int d^2pV_{\bm{p}}\sum_{\bm{k},\bm{k}'}\frac{\left[\int d^2r\psi_{m\bm{k}}(\bm{r})e^{i\bm{q}\cdot\bm{r}}\psi_{n\bm{k}'}^*(\bm{r})\right]\left[\int d^2r_1\psi_{m\bm{k}}^*(\bm{r}_1)e^{i\bm{p}\cdot\bm{r}_1}\psi_{n\bm{k}'}(\bm{r}_1)\right]}{(\omega - E_{m\bm{k}} + i\varepsilon)(\omega - E_{n\bm{k}'} + i\varepsilon)}.\label{Eq: graphene delta rho 1}
		\end{align}
	\end{widetext}
	The eigenstates of the perfect lattice obey the Bloch theorem and can be written as $\psi_{m\bm{k}}(\bm{r}) = e^{i\bm{k}\cdot\bm{r}}u_{m\bm{k}}(\bm{r})$, where $u_{m\bm{k}}(\bm{r}+\bm{R}_n)=u_{m\bm{k}}(\bm{r})$ is a periodic function and $\bm{R}_n$ are the direct lattice vectors. Letting $\bm{r}=\bm{R}_n+\bm{x}$ with $\bm{x}$ being restricted inside a unit cell, we have
	\begin{align}
		&\int d^2r\psi_{m\bm{k}}(\bm{r})e^{i\bm{q}\cdot\bm{r}}\psi_{n\bm{k}'}^*(\bm{r}) \nonumber\\
		&= \int d^2re^{i(\bm{k}+\bm{q}-\bm{k}')\cdot\bm{r}}u_{m\bm{k}}(\bm{r})u_{n\bm{k}'}^*(\bm{r}) \nonumber\\
		&= \sum_{\bm{R}_n}e^{i(\bm{k}+\bm{q}-\bm{k}')\cdot\bm{R}_n}\int_{\text{u.c.}} d^2xe^{i(\bm{k}+\bm{q}-\bm{k}')\cdot\bm{x}}u_{m\bm{k}}(\bm{x})u_{n\bm{k}'}^*(\bm{x})\nonumber\\
		&= \delta_{\bm{k}+\bm{q},\bm{k}'}\braket{u_{n(\bm{k}+\bm{q})}|u_{m\bm{k}}}
	\end{align}
	and, similarly,
	\begin{align}
		\int d^2r_1\psi_{m\bm{k}}^*(\bm{r}_1)e^{i\bm{p}\cdot\bm{r}_1}\psi_{n\bm{k}'}(\bm{r}_1) = \delta_{\bm{k}-\bm{p},\bm{k}'}\braket{u_{m\bm{k}}|u_{n(\bm{k}-\bm{p})}}.
	\end{align}
	Using these two identities, we can simplify Eq.~\eqref{Eq: graphene delta rho 1} as follows
	\begin{widetext}
		\begin{align}
			&\Delta\rho_1(\bm{q},\omega) \nonumber\\
			&= \frac{i}{2\pi}\sum_{m,n}\int d^2pV_{\bm{p}}\sum_{\bm{k},\bm{k}'}\frac{\delta_{\bm{k}+\bm{q},\bm{k}'}\braket{u_{n(\bm{k}+\bm{q})}|u_{m\bm{k}}}\delta_{\bm{k}-\bm{p},\bm{k}'}\braket{u_{m\bm{k}}|u_{n(\bm{k}-\bm{p})}}}{(\omega - E_{m\bm{k}} + i\varepsilon)(\omega - E_{n\bm{k}'} + i\varepsilon)}= \frac{iV_{-\bm{q}}}{2\pi}\sum_{m,n}\sum_{\bm{k}}\frac{\left|\braket{u_{m\bm{k}}|u_{n(\bm{k}+\bm{q})}}\right|^2}{(\omega - E_{m\bm{k}} + i\varepsilon)(\omega - E_{n(\bm{k}+\bm{q})} + i\varepsilon)}.
		\end{align}
	\end{widetext}
	For $\Delta\rho_2(\bm{q},\omega)$, we follow the same procedure and obtain
	\begin{align}
		&\Delta\rho_2(\bm{q},\omega) \nonumber\\
		&= \frac{iV_{-\bm{q}}}{2\pi}\sum_{m,n}\sum_{\bm{k}}\frac{\left|\braket{u_{m\bm{k}}|u_{n(\bm{k}+\bm{q})}}\right|^2}{(\omega - E_{n(\bm{k}+\bm{q})} - i\varepsilon)(\omega - E_{m\bm{k}} - i\varepsilon)}.
	\end{align}
	With the form factor defined by $\mathcal{M}_{mn}\left(\bm{k},\bm{k}+\bm{q}\right) = \braket{u_{m\bm{k}}|u_{n(\bm{k}+\bm{q})}}$, the FT-LDOS is given by
	\begin{widetext}
		\begin{align}
			\Delta\rho(\bm{q},\omega) = -\frac{V_{-\bm{q}}}{\pi}\sum_{m,n}\sum_{\bm{k}\in\text{FBZ}}\left|\mathcal{M}_{mn}\left(\bm{k},\bm{k}+\bm{q}\right)\right|^2\Im\frac{1}{(\omega - E_{m\bm{k}} + i\varepsilon)(\omega - E_{n(\bm{k}+\bm{q})} + i\varepsilon)}.\label{Eq: graphene1 form factor}
		\end{align}
	\end{widetext}
	The summation over $\bm{k}$ runs over the first Brillouin zone (FBZ). 
	The FT-LDOS is a summation of the norm squared of the form factor, $\big|\mathcal{M}_{mn}\left(\bm{k},\bm{k}+\bm{q}\right)\big|^2$, over momentum $\bm{k}$ and bands, weighted by the factor
	\begin{equation}
		\mathcal{F}_{m\bm{k}}^{n(\bm{k}+\bm{q})}(\omega) = \Im\frac{1}{(\omega - E_{m\bm{k}} + i\varepsilon)(\omega - E_{n(\bm{k}+\bm{q})} + i\varepsilon)}
	\end{equation} 
	that carries the information of the Fermi surface, and the Fourier image $V_{\bm{q}}$ of the defect. Using the Sokhatsky-Weierstrass theorem, we write the factor $\mathcal{F}_{m\bm{k}}^{n(\bm{k}+\bm{q})}(\omega)$ as
	\begin{align}
		\mathcal{F}_{m\bm{k}}^{n(\bm{k}+\bm{q})}(\omega) = -\frac{\pi}{\omega - E_{m\bm{k}}}&\delta\left(\omega - E_{n(\bm{k}+\bm{q})}\right)\nonumber\\
		&- \frac{\pi}{\omega - E_{n(\bm{k}+\bm{q})}}\delta\left(\omega - E_{m\bm{k}}\right).
	\end{align}
	From this expression of $\mathcal{F}_{m\bm{k}}^{n(\bm{k}+\bm{q})}(\omega)$, we can see that  $\Delta\rho(\bm{q},\omega)$ diverges when $\bm{k}$ satisfies the condition $E_{m\bm{k}} = E_{n(\bm{k}+\bm{q})} = \omega$, i.e., both $\bm{k}$ and $\bm{k}+\bm{q}$ reside on the isoenergy contours at energy $\omega$. This can be explained more rigorously by recalling that, for an infinite lattice, the dispersion is a continuously differentiable function of the crystal momentum, which allows us to write $\delta\left(\omega - E_{m\bm{k}}\right) = \sum_{\bm{k}_1\in\{\omega=E_{m\bm{k}}\}}\frac{\delta(\bm{k}-\bm{k}_1)}{\left|\nabla_{\bm{k}}E_{m\bm{k}}\right|}$ and $\delta\left(\omega - E_{n(\bm{k}+\bm{q})}\right) = \sum_{\bm{k}_2\in\{\omega=E_{n(\bm{k}+\bm{q})}\}}\frac{\delta(\bm{k}-\bm{k}_2)}{\left|\nabla_{\bm{k}}E_{n(\bm{k}+\bm{q})}\right|}$. Inserting these expressions into Eq.~\eqref{Eq: graphene1 form factor} and converting the summation over $k$ to an integral in the thermodynamic limit, we obtain
	\begin{align}
		\Delta\rho(\bm{q},\omega) &= V_{-\bm{q}}\sum_{m,n}\frac{\mathcal{A}}{(2\pi)^2}\int_{\text{FBZ}}d^2\bm{k}\left|\mathcal{M}_{mn}\left(\bm{k},\bm{k}+\bm{q}\right)\right|^2 \nonumber\\
		&\times\Bigg[\frac{1}{\omega - E_{m\bm{k}}}\sum_{\bm{k}_2\in\{\omega=E_{n(\bm{k}+\bm{q})}\}}\frac{\delta(\bm{k}-\bm{k}_2)}{\left|\nabla_{\bm{k}}E_{n(\bm{k}+\bm{q})}\right|} \nonumber\\
		&+ \frac{1}{\omega - E_{n(\bm{k}+\bm{q})}}\sum_{\bm{k}_1\in\{\omega=E_{m\bm{k}}\}}\frac{\delta(\bm{k}-\bm{k}_1)}{\left|\nabla_{\bm{k}}E_{m\bm{k}}\right|}\Bigg]
	\end{align}
	with $\mathcal{A}$ being the area of the 2D lattice. This expression reduces to
	\begin{widetext}
		\begin{align}
			\Delta\rho(\bm{q},\omega) = V_{-\bm{q}}\sum_{m,n}\frac{\mathcal{A}}{(2\pi)^2}\Bigg[\sum_{\bm{k}_2\in\{\omega=E_{n(\bm{k}+\bm{q})}\}}&\frac{\left|\mathcal{M}_{mn}(\bm{k}_2,\bm{k}_2+\bm{q})\right|^2}{\omega - E_{m\bm{k}_2}}\frac{1}{\left|\nabla_{\bm{k}}E_{n(\bm{k}+\bm{q})}\right|_{\bm{k}=\bm{k}_2}} \nonumber\\
			&+ \sum_{\bm{k}_1\in\{\omega=E_{m\bm{k}}\}}\frac{\left|\mathcal{M}_{mn}(\bm{k}_1,\bm{k}_1+\bm{q})\right|^2}{\omega - E_{n(\bm{k}_1+\bm{q})}}\frac{1}{\left|\nabla_{\bm{k}}E_{m\bm{k}}\right|_{\bm{k}=\bm{k}_1}}\Bigg].
		\end{align}
	\end{widetext}
	The bracket contains two summations. In the first summation, the wave vector $\bm{k}_2$ runs over the set $\{\omega=E_{n(\bm{k}_2+\bm{q})}\}$ and the summation is singular when $\omega=E_{m\bm{k}_2}$. Meanwhile, the second summation is over the set $\{\omega=E_{m\bm{k}_1}\}$ and becomes singular when $\omega=E_{n(\bm{k}_1+\bm{q})}$. Therefore, 
	we make a rough approximation of $\Delta\rho(\bm{q},\omega)$ by taking only the divergent terms in the two summations. This makes both $\bm{k}_1$ and $\bm{k}_2$ belong to the same set $\Upsilon(\bm{q})=\{\omega=E_{m\bm{k}}=E_{n(\bm{k}+\bm{q})}\}$. With this approximation, the FT-LDOS becomes
	\begin{align}
		\Delta\rho(\bm{q},\omega) \approx V_{-\bm{q}}&\frac{\mathcal{A}}{(2\pi)^2}\sum_{m,n}\sum_{\bm{k}\in\Upsilon(\bm{q})}\frac{\left|\mathcal{M}_{mn}\left(\bm{k},\bm{k}+\bm{q}\right)\right|^2}{\omega - E_{m\bm{k}}}\nonumber\\
		&\times\Bigg(\frac{1}{\left|\nabla_{\bm{k}}E_{n(\bm{k}+\bm{q})}\right|} + \frac{1}{\left|\nabla_{\bm{k}}E_{m\bm{k}}\right|}\Bigg).
	\end{align}
	Assuming that there is only one energy band crossing the level $\omega$, the expression is further simplified into
	\begin{align}
		&\Delta\rho(\bm{q},\omega) \approx V_{-\bm{q}}\frac{\mathcal{A}}{(2\pi)^2}\nonumber\\
		&\times\sum_{\bm{k}\in\Upsilon(\bm{q})}\frac{\left|\mathcal{M}(\bm{k},\bm{k}+\bm{q})\right|^2}{\omega - E_{\bm{k}}}\Bigg(\frac{1}{\left|\nabla_{\bm{k}}E_{\bm{k}+\bm{q}}\right|} + \frac{1}{\left|\nabla_{\bm{k}}E_{\bm{k}}\right|}\Bigg)\label{Eq: graphene1 form factor simple}
	\end{align}
	for $\bm{q}\neq0$. This relation is not a one-to-one mapping since $\Delta\rho$ at a given vector $\bm{q}$ is proportional to a sum of $\big|\mathcal{M}_{mn}\left(\bm{k},\bm{k}+\bm{q}\right)\big|^2$ over all wave vectors $\bm{k}$ in the set $\Upsilon(\bm{q})$. In other words, multiple scattering processes contribute to a single FT-LDOS value. However, for some special cases, the set $\Upsilon(\bm{q})$ may contain only one element $\bm{k}$, making the FT-LDOS proportional to the form factor. In other words, there is only a single vector $\bm{k}$ that satisfies the condition $\omega=E_{m\bm{k}}=E_{n(\bm{k}+\bm{q})}$.
	
	\begin{figure*}
		\includegraphics[width=\linewidth]{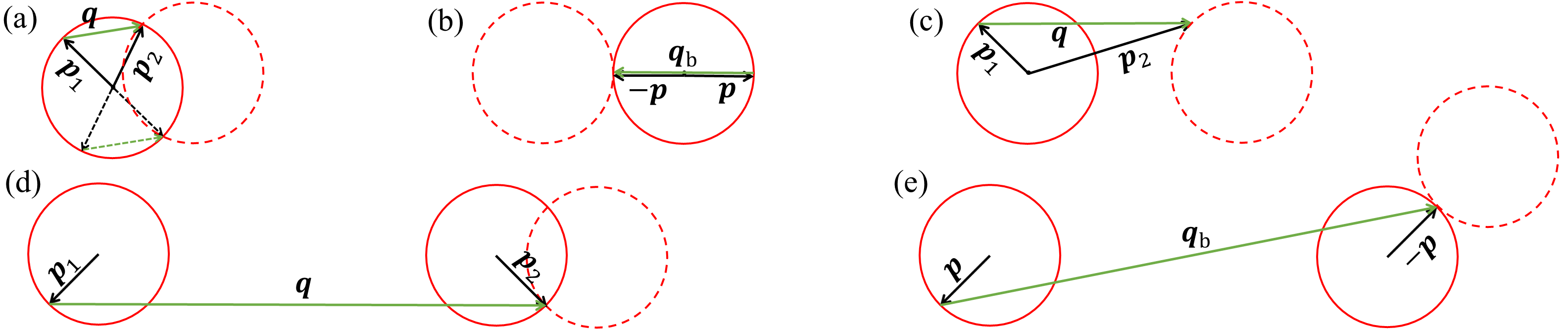}
		\caption{\label{fig:0fermi} \textit{Illustration of the set} $\Upsilon(\bm{q})=\{\omega=E_{\bm{k}}=E_{\bm{k}+\bm{q}}\}$. The Fermi contours, $E_{\bm{k}}=\omega$, are depicted by the red solid circles, while the shifted contours, $E_{\bm{k}+\bm{q}}=\omega$, are depicted by the red dashed circles. The number of crossings between the contours is equal to the number of elements $\bm{k}$ in the set $\Upsilon(\bm{q})$.}
	\end{figure*}
	
	To illustrate this idea, we consider a case where the Fermi surface contains circular Fermi contours of radius $p_\text{F}$. This example can be realized in graphene-based systems with the Fermi energy close to the charge neutrality point. Here, we define the wave vector $\bm{p}=\bm{k}-\bm{K}$, where $\bm{K}$ identifies the center of a circular contour. The set $\Upsilon(\bm{q})$ is thus now defined by the condition $\omega=E_{\bm{p}+\bm{K}}=E_{\bm{p}+\bm{K}+\bm{q}}$. For $\bm{q}$ vectors that can only connect two different states on the same contour [Figs.~\ref{fig:0fermi}(a,b)], the equation $E_{\bm{p}+\bm{K}}=\omega$ yields a set of vectors $\bm{p}$ forming the circle centered at $\bm{K}$. Meanwhile, the equation $E_{\bm{p}+\bm{K}} = E_{\bm{p}+\bm{K}+\bm{q}}$ has two or one solution, corresponding to $0<|\bm{q}|<2p_\text{F}$ or $|\bm{q}|=2p_\text{F}$, respectively. The latter has the greatest magnitude of $\bm{q}$ and involves two states with opposite wave vectors, i.e., $\bm{p}$ and $-\bm{p}$. We refer to this case as the back-scattering process and denote the $\bm{q}$ vector in these cases as $\bm{q}_{\text{b}}$ --- see Fig.~\ref{fig:0fermi}(b). If the $\bm{q}$ vector cannot connect two states on any Fermi contour [Fig.~\ref{fig:0fermi}(c)], the set $\Upsilon(\bm{q})$ has no element. If $\bm{q}$ connects two states on two different contours, there are two scenarios depicted in Figs.~\ref{fig:0fermi}(d) and~\ref{fig:0fermi}(e) that have two and one solutions, respectively. It is straightforward to see that the set $\Upsilon(\bm{q})$ has only one solution $\bm{p}$ in the back-scattering processes where the circle $E_{\bm{p}+\bm{K}/\bm{K}'}=\omega$ touches the circle $E_{\bm{p}+\bm{K}+\bm{q}}=\omega$ at only one point $-\bm{p}$. The FT-LDOS associated with the back-scattering processes is given by
	\begin{align}
		&\Delta\rho(\bm{q}_{\text{b}},\omega) \nonumber\\
		&\approx V_{-\bm{q}_{\text{b}}}\frac{\mathcal{A}}{(2\pi)^2}\frac{\left|\mathcal{M}\left(\bm{k},\bm{k}+\bm{q}_{\text{b}}\right)\right|^2}{\omega - E_{\bm{k}}}\Bigg(\frac{1}{\left|\nabla_{\bm{k}}E_{\bm{k}+\bm{q}_{\text{b}}}\right|} + \frac{1}{\left|\nabla_{\bm{k}}E_{\bm{k}}\right|}\Bigg)\label{Eq: form factor simple}.
	\end{align}
	We can see that the FT-LDOS at $\bm{q}_{\text{b}}$ vectors is approximately proportional to the norm squared of the form factor, weighted by the inverse group velocities of states at the Fermi surface. For circular Fermi contours, the group velocity is isotropic and the norm squared of the form factor is therefore the only factor that determines the variation of $\Delta\rho(\bm{q}_{\text{b}},\omega)$ with respect to $\bm{q}_{\text{b}}$.
	\section{Layer Projected Form Factor and Pseudospin Textures}
	\subsection{Single-layer form factor}
	The connection between the FT-LDOS and the form factor of electrons only holds if the LDOS of all orbitals surrounding the defect is taken into account. However, in QPI experiments, the LDOS is measured by scanning tunneling spectroscopy (STS), which primarily accesses the top layer of the material. Consequently, for multilayer materials, we need to define a different quantity, termed \textit{single-layer form factor} (SLFF), from the projected components of the Bloch states onto that layer. Since the eigenstates of TBG are given by two sets of plane waves and each set is associated with a graphene layer [see Eq.~\eqref{Eq: eigenstate expansion}], the SLFF of the top layer reads
	\begin{equation}
		\bar{\mathcal{M}}(\bm{k},\bm{k}+\bm{q}) = \sum_{\bm{K}_1}C_{\bm{K}_1}^{\dagger}(\bm{k}+\bm{q})C_{\bm{K}_1}(\bm{k})
	\end{equation}
	with $\bm{K}_1$ being the $\bm{K}$ points of the top layer's Brillouin zone.
	\begin{figure*}
		\centering
		\includegraphics[width=\textwidth]{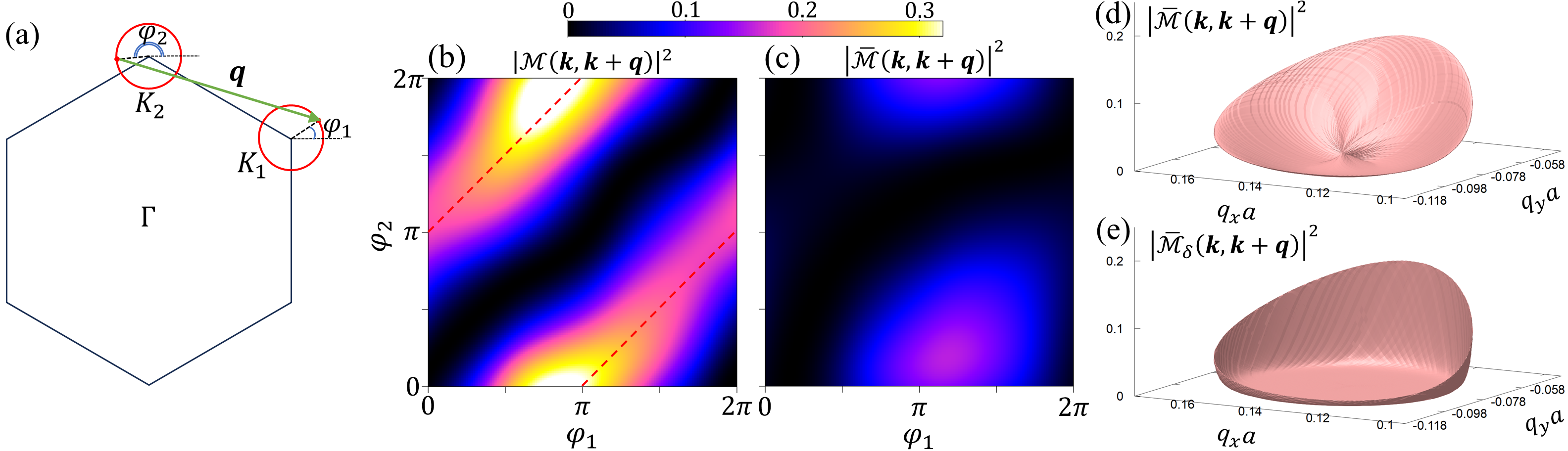}
		\caption{\textit{Single-layer form factor of electrons}. (a) Interference between two states at two adjacent valleys $K_1$ and $K_2$. These states, with momenta $\bm{k}_1$ and $\bm{k}_2$, are represented by the angles $\varphi_1$ and $\varphi_2$. The green arrow indicates the momentum difference $\bm{q}=\bm{k}_1-\bm{k}_2$. (b), (c) The norm squared of the form factor (b) and the SLFF (c) as functions of $\varphi_1$ and $\varphi_2$. (d), (e) The norm squared of the SLFF (d) and SLFFB (e) as functions of $\bm{q}$ for all the values of $\bm{k}=\bm{k}_1$ on a Fermi contour.}
		\label{fig:projected}
	\end{figure*}
	With the Fermi energy below the van Hove singularity~\cite{Brihuega2012stm}, the Fermi surface of TBG consists of isolated closed contours centered at the $\bm{K}$ points. Thus, we divide the form factor into different groups, each of which corresponds to the overlap between states from some specific valleys. As a demonstrative example, we consider the form factor and SLFF of eigenstates at a $K_1$ and $K_2$ valley, as depicted in Fig.~\ref{fig:projected}(a). At each valley $\nu$ ($\nu=1,2$), an eigenstate $\ket{u_{\nu\bm{k}_{\nu}}}$ can be characterized by an angle $\varphi_\nu$, which is the angle between vector $\bm{k}_\nu-\bm{K}_\nu$ and the $k_x$ axis. Using the continuum Hamiltonian~\eqref{Eq: continuum}, we compute the form factor, $\mathcal{M}(\bm{k}_1,\bm{k}_2)$, and show its norm squared in Fig.~\ref{fig:projected}(b) as a function of $\varphi_1$ and $\varphi_2$. The momentum difference between two eigenstates is $\bm{q}=\bm{k}_1-\bm{k}_2$. The domain of $\left|\mathcal{M}(\bm{k}_1,\bm{k}_2)\right|^2$ as a function of $(\varphi_1,\varphi_2)$ has the topology of a torus since it is periodic with respect to these two angles. The highest values of $\left|\mathcal{M}(\bm{k}_1,\bm{k}_2)\right|^2$ distribute along the line $\varphi_2-\varphi_1=\pi$, which corresponds to the back-scattering processes, $\bm{k}_1-\bm{K}_1 = -(\bm{k}_2-\bm{K}_2)$. The maximum value of the form factor's norm squared occurs at $(\varphi_1,\varphi_2)=\left(\frac{5\pi}{6},\frac{11\pi}{6}\right)$: $\bm{q}$ is parallel to $\bm{K}_1-\bm{K}_2$ and is the shortest vector connecting two states at these two valleys. On the other hand, the SLFF's norm squared, $\left|\bar{\mathcal{M}}(\bm{k}_1,\bm{k}_2)\right|^2$, is presented in Fig.~\ref{fig:projected}(c). While the greatest values of the norm squared still distribute along the line of $\varphi_2-\varphi_1=\pi$, the position of the maximum value appears to be located near $(\varphi_1,\varphi_2)=\left(\frac{4\pi}{3},\frac{\pi}{3}\right)$.
	
	To compare the SLFF with the FT-LDOS, we write the SLFF as a function of $\bm{k}=\bm{k}_1$ and $\bm{q}=\bm{k}_1-\bm{k}_2$ and plot it in the $(q_x,q_y)$ plane for the allowed values of $\bm{k}$ and $\bm{q}$ [Fig.~\ref{fig:projected}(d)]. The geometry of the SLFF in this presentation is a complex 3D structure but we can understand some of its basic features. At each $\bm{q}$ point in Fig.~\ref{fig:projected}(d), the number of values of $\left|\bar{\mathcal{M}}(\bm{k},\bm{k}+\bm{q})\right|^2$ is the number of $\bm{k}$ points satisfying the equation $\omega=E_{\bm{k}}=E_{\bm{k}+\bm{q}}$. Therefore, the SLFF is a single-valued function when $\bm{k}-\bm{K}_1 = -(\bm{k}+\bm{q}-\bm{K}_2)$. However, at the center where $\bm{q}=\bm{K}_1-\bm{K}_2$ (corresponding to the line $\varphi_1=\varphi_2$), there is an infinite number of values of $\left|\bar{\mathcal{M}}(\bm{k},\bm{k}+\bm{q})\right|^2$, although the majority of them are zeros. Other $\bm{q}$ points may have two values of $\left|\bar{\mathcal{M}}(\bm{k}_1,\bm{k}_2)\right|^2$ related to two allowed $\bm{k}$ points. Finally, since the FT-LDOS is proportional to the SLFF only at back-scattering wave vectors $\bm{q}_{\text{b}}$, we compute the SLFF at these points by defining
	\begin{equation}
		\bar{\mathcal{M}}_{\delta}(\bm{k},\bm{k}+\bm{q}) = \bar{\mathcal{M}}(\bm{k},\bm{k}+\bm{q}) \delta_{\bm{k}-\bm{K}_j,\bm{K}_l-\bm{k}-\bm{q}}.
	\end{equation}
	We call this quantity the single-layer form factor's boundary (SLFFB) and plot its norm squared in Fig.~\ref{fig:projected}(e). Evidently, all the allowed vectors $\bm{q}$ are restricted to the interior of a circle with a radius twice that of the Fermi contour. Along the circumference of this circle, the SLFFB has a maximum and a minimum value, which indeed agrees with the QPI signals of the intravalley interlayer interference.
	\subsection{Origin of the Chiral Structure}
\begin{figure*}
		\includegraphics[width=\linewidth]{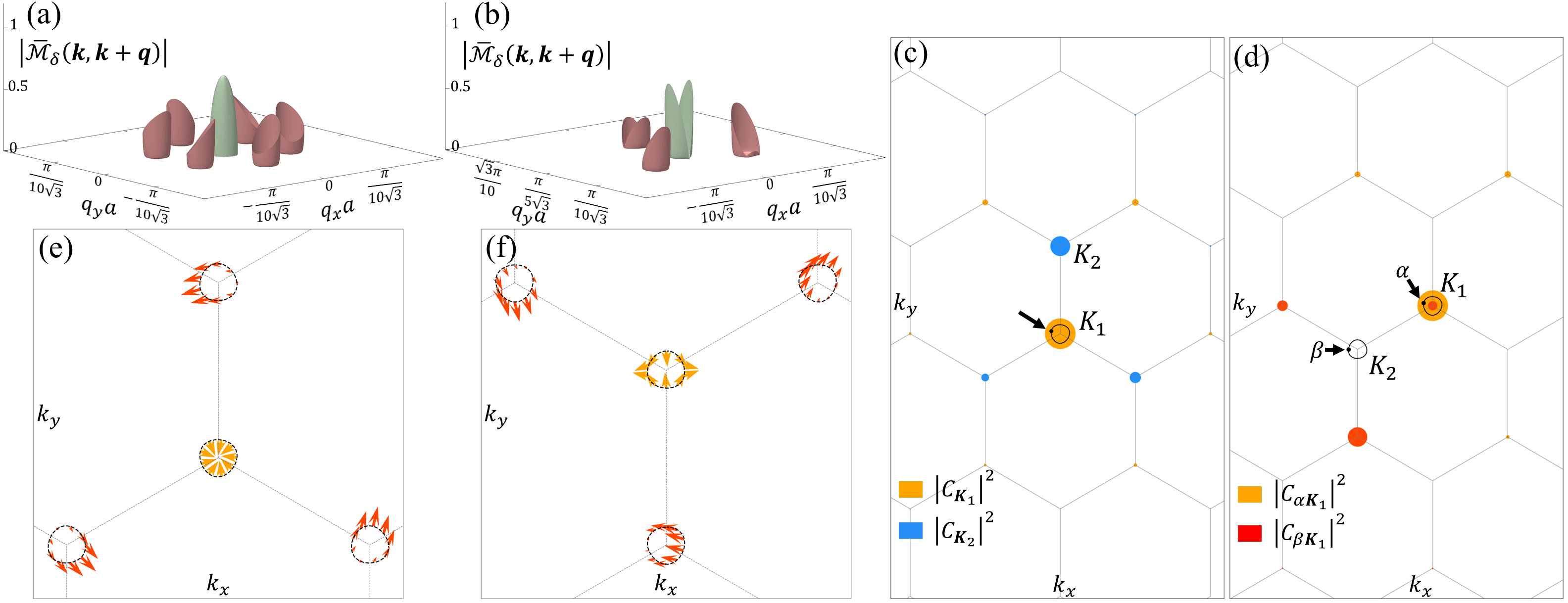}
		\caption{\label{fig:overlap} \textit{Origin of the chiral structure}. (a), (b) The norm of SLFFB associated with some (a) intravalley and (b) intervalley interference processes. The green structures come from intralayer interference while the red structures are interlayer interference. (c) The plane-wave amplitudes of an energy eigenstate with each plane wave being represented by a reciprocal lattice point. The momentum of the eigenstate is depicted by the black dot at a $K_1$ valley whose Fermi contour is indicated by the solid black circle. (d) The plane-wave amplitudes on the top layer of two eigenstates at two adjacent valleys. (e), (f) The pseudospin textures associated with the (e) intravalley and (f) intervalley interference. The orange and red arrows correspond to intralayer and interlayer interference, respectively.}
	\end{figure*}

	The SLFFB of some typical interference processes is plotted in Figs.~\ref{fig:overlap}(a) and~\ref{fig:overlap}(b) using the continuum model with twist angle $\theta=2.13^{\circ}$. Overall, the SLFFB excellently captures the characteristics of the QPI patterns in Fig.~\ref{fig:2qpi}(b). This includes the chiral structure of the intravalley interference and the intensity's minima and maxima. Therefore, we can examine the product of TBG's eigenstates in order to explain the chiral structure of the QPI pattern.
	
	Diagonalizing the continuum Hamiltonian, we obtain the eigenstate at a $\bm{k}$ point as a vector of two-component coefficients, $C_{\bm{K}_{1/2}}$, which are hereafter referred to as plane-wave amplitudes of $\bm{K}_1$/$\bm{K}_2$ points. Their amplitudes squared, $\left|C_{\bm{K}_{1/2}}\right|^2$, are shown in Fig.~\ref{fig:overlap}(c) with the Fermi contour centered at a $\bm{K}_1$ point for a Fermi energy $E_{\text{F}}=20$~meV. The plane-wave amplitudes of the $\bm{K}_1$ point closest to $\bm{k}$ and three $\bm{K}_2$ points surrounding it are significantly greater than other amplitudes. Thus, we neglect the contributions from the other plane waves and write the eigenstate as
	\begin{align}
		u_{K_1\bm{k}}(\bm{r}) \approx C_{\bm{K}_1}&(\bm{k})e^{-i\bm{K}_1\cdot\bm{r}} \nonumber\\
		&+ \sum_{\mu=a,b,c}C_{\bm{K}_1+\boldsymbol{\kappa}_\mu}(\bm{k})e^{-i(\bm{K}_1+\boldsymbol{\kappa}_\mu)\cdot\bm{r}}.
	\end{align}
	When $\bm{k}$ traverses along the Fermi contour of $K_1$ valley, the amplitude of $\bm{K}_1$ plane wave is invariant but those of the three $\bm{K}_2$ points vary following the rotation of $\bm{k}$ around $\bm{K}_1$ --- see Fig.~\ref{fig:overlap}(c).
	
	In general, the overlap of two vectors depends on the phase structures and amplitudes of both vectors. First, we consider how the variation of plane-wave amplitudes of each eigenstate affects the form factor.
	When two eigenstates of two adjacent valleys interfere on the top layer, only their components in that layer influence the interference, for example, see Fig.~\ref{fig:overlap}(d) where we only consider contributions from $\bm{K}_1$ plane waves. When projected onto layer~1, an eigenstate $u_{\alpha\bm{k}_\alpha}(\bm{r})$ of valley $K_1$ is thus represented by a single plane wave, $\bar{u}_{\alpha\bm{k}_\alpha}(\bm{r})\approx C_{\alpha\bm{K}_1}(\bm{k}_\alpha)e^{-i\bm{K}_1\cdot\bm{r}}$, whereas a state $u_{\beta\bm{k}_\beta}(\bm{r})$ of an adjacent $K_2$ valley is a superposition of three plane waves $\bm{K}_2-\boldsymbol{\kappa}_\mu$, $\bar{u}_{\beta\bm{k}_\beta}(\bm{r})\approx\sum_{\mu=a,b,c}C_{\beta\bm{K}_2-\boldsymbol{\kappa}_\mu}(\bm{k}_\beta)e^{-i(\bm{K}_2-\boldsymbol{\kappa}_\mu)\cdot\bm{r}}$. Therefore, the overlap between the two states becomes
	\begin{equation}
		\braket{\bar{u}_\alpha(\bm{k}_\alpha)|\bar{u}_\beta(\bm{k}_\beta)} \approx C_{\alpha\bm{K}_1}^{\dagger}(\bm{k}_\alpha)C_{\beta\bm{K}_1}(\bm{k}_\beta),
	\end{equation}
	which varies as the two states traverse their Fermi contours, leading to the presence of a minimum and maximum in each ring in Figs.~\ref{fig:projected}(e) and~\ref{fig:overlap}(a). For example, in the interference shown in Fig.~\ref{fig:overlap}(d), the overlap of two eigenstates occurs primarily at the $\bm{K}_1$ point closest to $\bm{k}_\alpha$. When $\bm{k}_\alpha$ circles around this $\bm{K}_1$ point, the amplitude of this plane wave in $u_{\alpha\bm{k}_\alpha}(\bm{r})$ remains nearly constant. Meanwhile, when $\bm{k}_\beta$ circles around its nearest $\bm{K}_2$ point, the amplitude of the $\bm{K}_1$ plane wave near $\bm{k}_{\alpha}$ has a minimum and a maximum during the cycle. This variation of the amplitudes contributes to the chiral structure of the form factor.
	
	Next, we demonstrate an approximate picture of the phase structures of the eigenstates when they interfere. Since the single-layer interference of two states is dominated by a single plane wave, each state can be effectively represented by the two-component spinor $C_{\bm{K}_1}(\bm{k})$ of that plane wave. Owing to the sublattice symmetry of TBG, $\left|\chi_{\text{A}\bm{K}_1}(\bm{k})\right| = \left|\chi_{\text{B}\bm{K}_1}(\bm{k})\right|$, we can thus define the pseudospin vector $\bm{s}(\bm{k})=C_{\bm{K}_1}^{\dagger}(\bm{k})\boldsymbol{\sigma}C_{\bm{K}_1}(\bm{k})$ with $\boldsymbol{\sigma}$ being a Pauli vector. The pseudospin textures corresponding to the intravalley and intervalley interference are shown in Figs.~\ref{fig:overlap}(e) and~\ref{fig:overlap}(f), respectively. The intralayer interference has pseudospin textures nearly similar to those of monolayer graphene --- the pseudospin vector rotates clockwise or counterclockwise as we go along a Fermi contour while maintaining a constant magnitude. Meanwhile, the pseudospin vectors associated with interlayer interference almost point in one direction while their lengths vary, which represents the aforementioned variation of amplitudes. The pseudospin textures composed of directions and lengths of the pseudospin vectors explain the chiral structure of the FT-LDOS of TBG.
	
	The pseudospin textures presented in Fig.~\ref{fig:overlap} were obtained by diagonalizing the continuum Hamiltonian with a great number of plane waves. However, an approximate analytical expression can be obtained by limiting the number of plane waves to four. As we have seen in Fig.~\ref{fig:overlap}, the contributions from the second-nearest neighbor $\bm{K}_{1/2}$ points to an eigenstate are small, so we can make an approximation by neglecting them as well as further points. This allows us to use the eight-band model of the continuum Hamiltonian~\cite{Bistritzer2011-mq,Bernevig2021_1} to study the electronic states near a $\bm{K}$ point. In particular, the continuum Hamiltonian governing states near a $K_1$ valley reads
	\begin{widetext}
		\begin{equation}
			H_{\text{tri}}^{(\bm{K}_1)}(\bm{k}) =
			\begin{pmatrix}
				h_1(\bm{k}-\bm{K}_1) & T_a & T_b & T_c\\
				T_a^\dagger & h_2(\bm{k}-\bm{K}_1-\boldsymbol{\kappa}_a) & 0 & 0\\
				T_b^\dagger & 0 & h_2(\bm{k}-\bm{K}_1-\boldsymbol{\kappa}_b) & 0 \\
				T_c^\dagger & 0 & 0 & h_2(\bm{k}-\bm{K}_1-\boldsymbol{\kappa}_c)
			\end{pmatrix}.
		\end{equation}
	\end{widetext}
	At $\bm{k}=\bm{K_1}$, the two eigenvectors at the Dirac point have the form $u_{\text{tri}}(\bm{K_1}) = \begin{pmatrix} C_0 & C_a & C_b & C_c  \end{pmatrix}$ with $C_\mu = -h_2^{-1}(-\boldsymbol{\kappa}_\mu)T_\mu^\dagger C_0$ and $C_0$ being eigenvectors of the monolayer Dirac Hamiltonian. When $\bm{k}=\bm{K_1}+\bm{p}$ ($|\bm{p}|\ll|\boldsymbol{\kappa}_\mu|$), we use perturbation theory by applying a unitary transformation to the Hamiltonian and neglecting the coupling between the Dirac cone and other remote bands. We obtain an effective Hamiltonian for the Dirac cone as~\cite{Bistritzer2011-mq}
	\begin{align}
		H_{\text{Dirac}}(\bm{p}) = \frac{1 - \frac{3w^2}{(\hbar v_\text{D}|\boldsymbol{\kappa}|)^2}}{1 + \frac{3(u^2+w^2)}{(\hbar v_\text{D}|\boldsymbol{\kappa}|)^2}}h_1(\bm{p}),\label{Eq: two-band}
	\end{align}
	from which we obtain the eigenvalues $\mathcal{E}_{0\nu}(\bm{p})$ and eigenvectors $C_{0\nu}(\bm{p})$ ($\nu=1,2$) of the central Dirac cone. In our work, we assume $u\approx w$ since the twist angles of interest are greater than $2^{\circ}$, which allows us to define $\alpha = \frac{w}{\hbar v_{\text{D}}|\boldsymbol{\kappa}|}$ with $\hbar v_{\text{D}} = \frac{\sqrt{3}|t_0|a}{2}$. The eigenvalues and eigenvectors of the Dirac cone at valley $K_1$ read
	\begin{align}
		\begin{split}
			\mathcal{E}_{0\nu}(\bm{p}) &= \pm\frac{1-3\alpha^2}{1+6\alpha^2}\hbar v_{\text{D}}|\bm{p}|,\\
			C_{0\nu}(\bm{p})&=\frac{1}{\sqrt{2}}
			\begin{pmatrix}
				1 \\ -e^{i\left(\varphi_{\bm{p}}-\frac{\theta}{2}\right)}
			\end{pmatrix}.
		\end{split}
	\end{align}
	The pseudospin vectors associated with the intralayer interference form an angle $\varphi_{\bm{p}}+\pi-\theta/2$ with the $x$-axis, as can be seen at the central valleys in Fig.~\ref{fig:overlap}.
	
	To find the two eigenvectors of $H_{\text{tri}}^{(\bm{K}_1)}(\bm{k})$ that are associated with the Dirac cones, we assume that they have the form
	\begin{equation}
		u_{\text{tri}}^{(\nu)}(\bm{p}) = 
		\frac{1}{\sqrt{1+6\alpha^2}}
		\begin{pmatrix}
			C_{0\nu}^\text{T}(\bm{p}) & C_{a\nu}^\text{T}(\bm{p}) & C_{b\nu}^\text{T}(\bm{p}) & C_{c\nu}^\text{T}(\bm{p})
		\end{pmatrix}^\text{T}
	\end{equation}
	and satisfy the eigen-equation $H_{\text{tri}}^{(\bm{K}_1)}(\bm{K}_1+\bm{p})u_\text{tri}^{(\nu)}(\bm{p}) = \mathcal{E}_\nu(\bm{p})u_\text{tri}^{(\nu)}(\bm{p})$. In this case, the components are related by $C_{\mu\nu}(\bm{p}) = \left[\mathcal{E}_{0\nu}(\bm{p}) - h_2(\bm{p}-\boldsymbol{\kappa}_\mu)\right]^{-1}T_\mu^\dagger C_{0\nu}(\bm{p})$. However, as $|\bm{p}|\ll|\boldsymbol{\kappa}_\mu|$, we can approximate this expression with the one used for $\bm{k}=\bm{K}_1$ and get
	\begin{align}
		&C_{\mu\nu}(\bm{p}) \approx -h_2^{-1}(-\boldsymbol{\kappa}_\mu)T_\mu^\dagger C_{0\nu}(\bm{p}) \nonumber\\
		&= -\frac{\alpha}{\sqrt{2}}e^{-i\varphi_{-\boldsymbol{\kappa}_\mu}}\left[e^{-i\left(\gamma_\mu - \frac{\theta}{2}\right)} - e^{i\varphi_{\bm{p}}}\right]
		\begin{pmatrix}
			1 \\ e^{i\left(\gamma_{\mu}+2\varphi_{-\boldsymbol{\kappa}_\mu} + \theta\right)}
		\end{pmatrix},
	\end{align}
	where $\varphi_{\bm{p}}$ ($\varphi_{-\boldsymbol{\kappa}_\mu}$) is the angle between the momentum $\bm{p}$ (vector $-\boldsymbol{\kappa}_\mu$) and the $k_x$ axis. At each $K_2$ valley, the pseudospin vectors point in a single direction that forms an angle $\gamma_{\mu}+2\varphi_{-\boldsymbol{\kappa}_\mu} + \theta$ with the $x$~axis, independent of momentum $\bm{p}$. However, their amplitudes vary with respect to $\bm{p}$ and vanish when $\varphi_{\bm{p}} = - \gamma_\mu + \theta/2$. This analytical description is in agreement with Fig.~\ref{fig:overlap}.
	\section{Conclusion}
	In this work, we investigated QPI patterns in TBG at energies below the van Hove singularity and for twist angles $\theta\gtrsim2^{\circ}$. The QPI patterns reveal all the interference mechanisms, including both intralayer and interlayer interference between states at valleys with either identical or opposite winding numbers. Notably, the signals associated with interlayer interference exhibit a distinct chiral structure. By analyzing two types of commensurate structures of TBG, we demonstrate that QPI patterns validate the approximate translational symmetries and the valley charge conservation in this system. We also examined the Fourier-filtered LDOS, which exhibits varying numbers of additional wavefronts depending on the defect position. While these wavefront dislocations do not unambiguously resolve the topology of the flat bands, the observed approximate translational symmetries provide direct evidence for the topological obstruction to constructing localized Wannier orbitals for these bands~\cite{Zou2018}. 
	
	Furthermore, we derived a general relation between the QPI signals of back-scattering processes and the form factor, which underlies the quantum geometric tensor~\cite{Torma2022}, the topological properties of energy bands~\cite{Song2021}, the many-body ground states~\cite{Bernevig2021_3,Lian2021} and excited states~\cite{Bernevig2021_5}. This relation establishes QPI as an experimental probe for the form factor of back-scattering states. By applying this framework to TBG using the continuum model, we successfully explained the defining characteristics of the observed QPI patterns. We find that their chiral structure stems from the variation in plane-wave amplitudes of the eigenstates when they traverse the Fermi contours. The phenomenon is well described by pseudospin textures, which are in excellent agreement with an analytical model derived from an effective continuum Hamiltonian constructed from four plane waves.
	
	\textit{Notes.|} During the preparation of this manuscript, Mesple \textit{et al.} submitted a preprint in which they experimentally observed the QPI in TBG~\cite{Mesple2025}. Focusing on the intravalley signals, they clearly demonstrated the chiral structure of the interlayer interference at twist angle $4.3^\circ$.
	
	\textit{Acknowledgment.|} We acknowledge fruitful discussions with Florie Mesple, Magdalena Marganska, and Jose Lado. D.H.M.N., and D.B. acknowledge the support from the Transnational Common Laboratory $Quantum-ChemPhys$,  the Department of Education of the Basque Government through
	the project PIBA\_2023\_1\_0007 (STRAINER),  the financial support received from the IKUR Strategy under the collaboration agreement 
	between the Ikerbasque Foundation and DIPC on behalf of the Department of Education of the 
	Basque Government and the Gipuzkoa Provincial Council within the QUAN-000021-01 project, and the support from the Spanish MICINN-AEI through Project No.~PID2024-162933NB-I00~(QUILL). D.H.M.N is supported by Spanish Ministerio de Ciencia, Innovaci\'{o}n y Universidades grant PRE2021-097126. 
	IMDEA Nanociencia acknowledges support from the ‘Severo Ochoa’ Programme for Centres of Excellence in R\&D (CEX2020-001039-S/AEI/10.13039/501100011033). 
	We acknowledge support from NOVMOMAT, project PID2022-142162NB-I00 funded by MICIU/AEI/10.13039/501100011033 and by FEDER, UE as well as financial support through the (MAD2D-CM)-MRR MATERIALES AVANZADOS-IMDEA-NC.

	\bibliography{qpi.bib} 

@article{CastroNeto_2009, 
year = {2009}, 
title = {{The electronic properties of graphene}}, 
author = {Neto, A H Castro and Neto, A H Castro and Guinea, F and Peres, N M R and Novoselov, K S and Geim, A K}, 
journal = {Rev. Mod. Phys.}, 
doi = {10.1103/revmodphys.81.109}, 
url = {https://link.aps.org/doi/10.1103/RevModPhys.81.109}, 
pages = {109 -- 162}, 
number = {1}, 
volume = {81}, 
language = {English},
}

@article{Moon_2013, 
year = {2013}, 
title = {{Optical absorption in twisted bilayer graphene}}, 
author = {Moon, Pilkyung and Koshino, Mikito}, 
journal = {Phys. Rev. B}, 
issn = {1098-0121}, 
doi = {10.1103/physrevb.87.205404}, 
eprint = {1302.5218},
pages = {205404}, 
number = {20}, 
volume = {87},
}

@ARTICLE{Simon2011,
  title     = "Fourier-transform scanning tunnelling spectroscopy: the
               possibility to obtain constant-energy maps and band dispersion
               using a local measurement",
  author    = "Simon, L and Bena, C and Vonau, F and Cranney, M and Aubel, D",
  journal   = "J. Phys. D Appl. Phys.",
  publisher = "IOP Publishing",
  volume    =  {44},
  number    =  {46},
  pages     = "464010",
  month     =  {Nov},
  year      =  {2011},
  url = {https://iopscience.iop.org/article/10.1088/0022-3727/44/46/464010}
}

@article{Chen2017, 
year = {2017}, 
title = {{Quasiparticle interference in unconventional 2D systems}}, 
author = {Chen, Lan and Cheng, Peng and Wu, Kehui}, 
journal = {J. Phys. Condens. Matter}, 
issn = {0953-8984}, 
doi = {10.1088/1361-648x/aa54da}, 
pmid = {27996961},
pages = {103001}, 
number = {10}, 
volume = {29}
}

@article{Avraham2018,
	year = 2018,
	month = {jun},
	publisher = {Wiley},
	volume = {30},
	number = {41},
	pages = {1707628},
	author = {Nurit Avraham and others},
	title = {Quasiparticle Interference Studies of Quantum Materials},
	journal = {Adv. Mater.},
	doi = {10.1002/adma.201707628}
}

@ARTICLE{Yin2021,
  title     = "Probing topological quantum matter with scanning tunnelling
               microscopy",
  author    = "Yin, Jia-Xin and Pan, Shuheng H and Zahid Hasan, M",
  journal   = "Nat. Rev. Phys.",
  publisher = "Springer Science and Business Media LLC",
  volume    =  {3},
  number    =  {4},
  pages     = "249--263",
  month     =  {Mar},
  year      =  {2021},
  copyright = "https://www.springernature.com/gp/researchers/text-and-data-mining",
  language  = "English",
  url = {https://www.nature.com/articles/s42254-021-00293-7#article-info}
}

@ARTICLE{Bena2015,
  title     = "Friedel oscillations: Decoding the hidden physics",
  author    = "Bena, Cristina",
  journal   = "C. R. Phys.",
  publisher = "Cellule MathDoc/Centre Mersenne",
  volume    =  {17},
  number    = "3-4",
  pages     = "302--321",
  month     =  {dec},
  year      =  {2015},
  copyright = "http://creativecommons.org/licenses/by-nc-nd/4.0/",
  language  = "English",
  url = {http://dx.doi.org/10.1016/j.crhy.2015.11.006}
}

@ARTICLE{Villain2015,
  title     = "Jacques Friedel and the physics of metals and alloys",
  author    = "Villain, Jacques and Lavagna, Mireille and Bruno, Patrick",
  journal   = "C. R. Phys.",
  publisher = "Cellule MathDoc/Centre Mersenne",
  volume    =  {17},
  number    = "3-4",
  pages     = "276--290",
  month     =  {Dec},
  year      =  {2015},
  copyright = "http://creativecommons.org/licenses/by-nc-nd/4.0/",
  language  = "English",
  url = {http://dx.doi.org/10.1016/j.crhy.2015.12.010}
}

@article{Guo2010,
  title = {Theory of quasiparticle interference on the surface of a strong topological insulator},
  author = {Guo, H.-M. and Franz, M.},
  journal = {Phys. Rev. B},
  volume = {81},
  issue = {4},
  pages = {041102},
  numpages = {4},
  year = {2010},
  month = Jan,
  publisher = {American Physical Society},
  doi = {10.1103/PhysRevB.81.041102},
  url = {https://link.aps.org/doi/10.1103/PhysRevB.81.041102}
}

@ARTICLE{Zheng2018,
  title     = "Quasiparticle interference on type-I and {type-II} Weyl
               semimetal surfaces: a review",
  author    = "Zheng, Hao and Zahid Hasan, M",
  journal   = "Adv. Phys. X.",
  publisher = "Informa UK Limited",
  volume    =  {3},
  number    =  {1},
  pages     = "1466661",
  month     =  {jan},
  year      =  {2018},
  copyright = "http://creativecommons.org/licenses/by/4.0",
  language  = "English",
  url = {https://www.tandfonline.com/doi/full/10.1080/23746149.2018.1466661}
}

@ARTICLE{Yuan2019,
  title     = "Quasiparticle interference evidence of the topological Fermi arc
               states in chiral fermionic semimetal {CoSi}",
  author    = "Yuan, Qian-Qian and Zhou, Liqin and Rao, Zhi-Cheng and Tian,
               Shangjie and Zhao, Wei-Min and Xue, Cheng-Long and Liu, Yixuan
               and Zhang, Tiantian and Tang, Cen-Yao and Shi, Zhi-Qiang and
               Jia, Zhen-Yu and Weng, Hongming and Ding, Hong and Sun, Yu-Jie
               and Lei, Hechang and Li, Shao-Chun",
  journal   = "Sci. Adv.",
  publisher = "American Association for the Advancement of Science (AAAS)",
  volume    =  {5},
  number    =  {12},
  pages     = "eaaw9485",
  month     =  {dec},
  year      =  {2019},
  language  = "English",
  url = {https://www.science.org/doi/10.1126/sciadv.aaw9485}
}

@article{Stuart2022,
  title = {Quasiparticle interference observation of the topologically nontrivial drumhead surface state in ZrSiTe},
  author = {Stuart, B. A. and Choi, Seokhwan and Kim, Jisun and Muechler, Lukas and Queiroz, Raquel and Oudah, Mohamed and Schoop, L. M. and Bonn, D. A. and Burke, S. A.},
  journal = {Phys. Rev. B},
  volume = {105},
  issue = {12},
  pages = {L121111},
  numpages = {6},
  year = {2022},
  month = Mar,
  publisher = {American Physical Society},
  doi = {10.1103/PhysRevB.105.L121111},
  url = {https://link.aps.org/doi/10.1103/PhysRevB.105.L121111}
}

@article{Xiong2024,
  title = {Distinct quasiparticle interference patterns for surface impurity scattering on various Weyl semimetals},
  author = {Xiong, Feng and He, Chaocheng and Liu, Yong and Black-Schaffer, Annica M. and Nag, Tanay},
  journal = {Phys. Rev. B},
  volume = {109},
  issue = {5},
  pages = {054201},
  numpages = {17},
  year = {2024},
  month = Feb,
  publisher = {American Physical Society},
  doi = {10.1103/PhysRevB.109.054201},
  url = {https://link.aps.org/doi/10.1103/PhysRevB.109.054201}
}

@article{Cai2024,
  title = {Quasiparticle interference of the topological surface state on gray arsenic},
  author = {Cai, Xingfu and Yang, Zhongzheng and Wen, Chenhaoping and Yang, Yang and Zhu, Xiangde and Ning, Wei and Yan, Shichao and Xue, Jiamin},
  journal = {Phys. Rev. B},
  volume = {109},
  issue = {3},
  pages = {035414},
  numpages = {5},
  year = {2024},
  month = Jan,
  publisher = {American Physical Society},
  doi = {10.1103/PhysRevB.109.035414},
  url = {https://link.aps.org/doi/10.1103/PhysRevB.109.035414}
}

@article{Sanchezbarquilla2025,
  title = {Electronic band structure from quasiparticle interference and Landau quantization in ${\mathrm{WTe}}_{2}$},
  author = {S\'anchez-Barquilla, Raquel and Vega, Francisco Mart\'{\i}n and Ruiz, Alberto M. and Jo, Na Hyun and Herrera, Edwin and Baldov\'{\i}, Jos\'e J. and Ochi, Masayuki and Arita, Ryotaro and Bud'ko, Sergey L. and Canfield, Paul C. and Guillam\'on, Isabel and Suderow, Hermann},
  journal = {Phys. Rev. B},
  volume = {112},
  issue = {16},
  pages = {165401},
  numpages = {16},
  year = {2025},
  month = Oct,
  publisher = {American Physical Society},
  doi = {10.1103/7xnv-8wvt},
  url = {https://link.aps.org/doi/10.1103/7xnv-8wvt}
}

@article{Bagchi2025,
  title = {Quasiparticle interference on the surface of ${\mathrm{Bi}}_{2}{\mathrm{Se}}_{3}$-terminated ${\mathrm{(PbSe)}}_{5}$(${\mathrm{Bi}}_{2}{\mathrm{Se}}_{3}{)}_{6}$},
  author = {Bagchi, Mahasweta and R\"u\ss{}mann, Philipp and Bihlmayer, Gustav and Bl\"ugel, Stefan and Ando, Yoichi and Brede, Jens},
  journal = {Phys. Rev. B},
  volume = {112},
  issue = {4},
  pages = {045411},
  numpages = {12},
  year = {2025},
  month = Jul,
  publisher = {American Physical Society},
  doi = {10.1103/h8v9-sl53},
  url = {https://link.aps.org/doi/10.1103/h8v9-sl53}
}

@ARTICLE{Hoffmann2025,
  title     = {{Fermi arcs dominating the electronic surface properties of
               trigonal PtBi$_{2}$}},
  author    = {Hoffmann, Sven and Schimmel, Sebastian and Vocaturo, Riccardo
               and Puig, Joaquin and Shipunov, Grigory and Janson, Oleg and
               Aswartham, Saicharan and Baumann, Danny and B{\"u}chner, Bernd
               and van den Brink, Jeroen and Fasano, Yanina and Facio, Jorge I
               and Hess, Christian},
  journal   = "Advanced Physics Research",
  publisher = "Wiley",
  volume    =  {4},
  number    =  {5},
  month     =  {May},
  year      =  {2025},
  language  = {English},
  url = {https://advanced.onlinelibrary.wiley.com/doi/full/10.1002/apxr.202400150}
}

@article{Chakraborty2022,
  title = {Quasiparticle Interference as a Direct Experimental Probe of Bulk Odd-Frequency Superconducting Pairing},
  author = {Chakraborty, Debmalya and Black-Schaffer, Annica M.},
  journal = {Phys. Rev. Lett.},
  volume = {129},
  issue = {24},
  pages = {247001},
  numpages = {6},
  year = {2022},
  month = {Dec},
  publisher = {American Physical Society},
  doi = {10.1103/PhysRevLett.129.247001},
  url = {https://link.aps.org/doi/10.1103/PhysRevLett.129.247001}
}

@article{Bhattacharyya2023,
  title = {Superconducting gap symmetry from Bogoliubov quasiparticle interference analysis on ${\mathrm{Sr}}_{2}{\mathrm{RuO}}_{4}$},
  author = {Bhattacharyya, Shinibali and Kreisel, Andreas and Kong, X. and Berlijn, T. and R\o{}mer, Astrid T. and Andersen, Brian M. and Hirschfeld, P. J.},
  journal = {Phys. Rev. B},
  volume = {107},
  issue = {14},
  pages = {144505},
  numpages = {13},
  year = {2023},
  month = {Apr},
  publisher = {American Physical Society},
  doi = {10.1103/PhysRevB.107.144505},
  url = {https://link.aps.org/doi/10.1103/PhysRevB.107.144505}
}

@article{Huang2025,
  title = {Revealing the Orbital Origins of Exotic Electronic States with Ti Substitution in Kagome Superconductor ${\mathrm{CsV}}_{3}{\mathrm{Sb}}_{5}$},
  author = {Huang, Zihao and Chen, Hui and Tan, Hengxin and Han, Xianghe and Ye, Yuhan and Hu, Bin and Zhao, Zhen and Shen, Chengmin and Yang, Haitao and Yan, Binghai and Wang, Ziqiang and Liu, Feng and Gao, Hong-Jun},
  journal = {Phys. Rev. Lett.},
  volume = {134},
  issue = {5},
  pages = {056001},
  numpages = {7},
  year = {2025},
  month = {Feb},
  publisher = {American Physical Society},
  doi = {10.1103/PhysRevLett.134.056001},
  url = {https://link.aps.org/doi/10.1103/PhysRevLett.134.056001}
}

@ARTICLE{Nag2025,
  title     = {{Highly anisotropic superconducting gap near the nematic quantum
               critical point of FeSe$_{1-x}$S$_x$}},
  author    = {Nag, Pranab Kumar and Scott, Kirsty and de Carvalho, Vanuildo S
               and Byland, Journey K and Yang, Xinze and Walker, Morgan and
               Greenberg, Aaron G and Klavins, Peter and Miranda, Eduardo and
               Gozar, Adrian and Taufour, Valentin and Fernandes, Rafael M and
               da Silva Neto, Eduardo H},
  journal   = {Nat. Phys.},
  publisher = {Springer Science and Business Media LLC},
  volume    =  {21},
  number    =  {1},
  pages     = {89--96},
  month     =  {Jan},
  year      =  {2025},
  language  = {English},
  url = {https://www.nature.com/articles/s41567-024-02683-x}
}

@ARTICLE{Hanaguri2009,
  title     = "Coherence factors in a high-tc cuprate probed by quasi-particle
               scattering off vortices",
  author    = "Hanaguri, T and Kohsaka, Y and Ono, M and Maltseva, M and
               Coleman, P and Yamada, I and Azuma, M and Takano, M and Ohishi,
               K and Takagi, H",
  journal   = "Science",
  publisher = "American Association for the Advancement of Science (AAAS)",
  volume    =  {323},
  number    =  {5916},
  pages     = "923--926",
  month     =  {feb},
  year      =  {2009},
  language  = "English",
  url = {https://www.science.org/doi/10.1126/science.1166138}
}

@article{Hanke2012,
  title = {Probing the Unconventional Superconducting State of LiFeAs by Quasiparticle Interference},
  author = {H\"anke, Torben and Sykora, Steffen and Schlegel, Ronny and Baumann, Danny and Harnagea, Luminita and Wurmehl, Sabine and Daghofer, Maria and B\"uchner, Bernd and van den Brink, Jeroen and Hess, Christian},
  journal = {Phys. Rev. Lett.},
  volume = {108},
  issue = {12},
  pages = {127001},
  numpages = {5},
  year = {2012},
  month = {Mar},
  publisher = {American Physical Society},
  doi = {10.1103/PhysRevLett.108.127001},
  url = {https://link.aps.org/doi/10.1103/PhysRevLett.108.127001}
}

@ARTICLE{Li2023,
  title     = "Anisotropic gap structure and sign reversal symmetry in
               monolayer {Fe(Se,Te})",
  author    = "Li, Yu and Shen, Dingyu and Kreisel, Andreas and Chen, Cheng and
               Wei, Tianheng and Xu, Xiaotong and Wang, Jian",
  journal   = "Nano Lett.",
  publisher = "American Chemical Society (ACS)",
  volume    =  {23},
  number    =  {1},
  pages     = "140--147",
  month     =  {jan},
  year      =  {2023},
  language  = "English",
  url = {https://pubs.acs.org/doi/10.1021/acs.nanolett.2c03735}
}

@ARTICLE{Wang2025,
  title     = "Odd-parity quasiparticle interference in the superconductive
               surface state of {UTe2}",
  author    = "Wang, Shuqiu and Zhussupbekov, Kuanysh and Carroll, Joseph P and
               Hu, Bin and Liu, Xiaolong and Pangburn, Emile and Crepieux,
               Adeline and Pepin, Catherine and Broyles, Christopher and Ran,
               Sheng and Butch, Nicholas P and Saha, Shanta and Paglione,
               Johnpierre and Bena, Cristina and Davis, J C S{\'e}amus and Gu,
               Qiangqiang",
  journal   = "Nat. Phys.",
  publisher = "Springer Science and Business Media LLC",
  volume    =  21,
  number    =  10,
  pages     = "1555--1562",
  month     =  sep,
  year      =  2025,
  copyright = "https://creativecommons.org/licenses/by/4.0",
  doi = {10.1038/s41567-025-03000-w},
  language  = "English"
}

@article{Christiansen2025,
  title = {Quasiparticle Interference of Spin-Triplet Superconductors: Application to ${\mathrm{UTe}}_{2}$},
  author = {Christiansen, Hans and Andersen, Brian M. and Hirschfeld, P. J. and Kreisel, Andreas},
  journal = {Phys. Rev. Lett.},
  volume = {135},
  issue = {21},
  pages = {216001},
  numpages = {7},
  year = {2025},
  month = {Nov},
  publisher = {American Physical Society},
  doi = {10.1103/pttl-8m71},
  url = {https://link.aps.org/doi/10.1103/pttl-8m71}
}

@ARTICLE{Nakazawa2025,
  title     = "Origin of switchable quasiparticle-interference chirality in
               loop-current phase of kagome metals measured by
               scanning-tunneling-microscopy",
  author    = "Nakazawa, Seigo and Tazai, Rina and Yamakawa, Youichi and Onari,
               Seiichiro and Kontani, Hiroshi",
  journal   = "Nat. Commun.",
  publisher = "Springer Science and Business Media LLC",
  volume    =  16,
  number    =  1,
  pages     = "9545",
  month     =  oct,
  year      =  2025,
  copyright = "https://creativecommons.org/licenses/by-nc-nd/4.0",
  doi = {10.1038/s41467-025-64588-4},
  language  = "English"
}

@article{Ruffieux2005,
  title = {Charge-density oscillation on graphite induced by the interference of electron waves},
  author = {Ruffieux, P. and Melle-Franco, M. and Gr\"oning, O. and Bielmann, M. and Zerbetto, F. and Gr\"oning, P.},
  journal = {Phys. Rev. B},
  volume = {71},
  issue = {15},
  pages = {153403},
  numpages = {4},
  year = {2005},
  month = {Apr},
  publisher = {American Physical Society},
  doi = {10.1103/PhysRevB.71.153403},
  url = {https://link.aps.org/doi/10.1103/PhysRevB.71.153403}
}

@article{chen2005atomic,
  title={Atomic structure of the 6H--SiC (0 0 0 1) nanomesh},
  author={Chen, Wei and Xu, Hai and Liu, Lei and Gao, Xingyu and Qi, Dongchen and Peng, Guowen and Tan, Swee Ching and Feng, Yuanping and Loh, Kian Ping and Wee, Andrew Thye Shen},
  journal={Surface science},
  volume={596},
  number={1-3},
  pages={176--186},
  year={2005},
  doi={10.1016/j.susc.2005.09.013},
  publisher={Elsevier}
}

@article{Dombrowski2017,
  title = {Energy-Dependent Chirality Effects in Quasifree-Standing Graphene},
  author = {Dombrowski, Daniela and Jolie, Wouter and Petrovi\ifmmode \acute{c}\else \'{c}\fi{}, Marin and Runte, Sven and Craes, Fabian and Klinkhammer, J\"urgen and Kralj, Marko and Lazi\ifmmode \acute{c}\else \'{c}\fi{}, Predrag and Sela, Eran and Busse, Carsten},
  journal = {Phys. Rev. Lett.},
  volume = {118},
  issue = {11},
  pages = {116401},
  numpages = {6},
  year = {2017},
  month = {Mar},
  publisher = {American Physical Society},
  doi = {10.1103/PhysRevLett.118.116401},
  url = {https://link.aps.org/doi/10.1103/PhysRevLett.118.116401}
}

@ARTICLE{Gutierrez2016-iv,
  title     = "Imaging chiral symmetry breaking from Kekul{\'e} bond order in
               graphene",
  author    = "Guti{\'e}rrez, Christopher and Kim, Cheol-Joo and Brown, Lola
               and Schiros, Theanne and Nordlund, Dennis and Lochocki, Edward B
               and Shen, Kyle M and Park, Jiwoong and Pasupathy, Abhay N",
  journal   = "Nat. Phys.",
  publisher = "Springer Science and Business Media LLC",
  volume    =  12,
  number    =  10,
  pages     = "950--958",
  month     =  oct,
  year      =  2016,
  doi = {10.1038/nphys3776},
  language  = "English"
}

@ARTICLE{Jung2016-vd,
  title     = "Fingerprints of multiple electron scatterings in single-layer
               graphene",
  author    = "Jung, Minbok and Sohn, So-Dam and Park, Jonghyun and Lee, Keun-U
               and Shin, Hyung-Joon",
  journal   = "Sci. Rep.",
  publisher = "Springer Science and Business Media LLC",
  volume    =  6,
  number    =  1,
  pages     = "22570",
  month     =  mar,
  year      =  2016,
  doi = {10.1038/srep22570},
  copyright = "https://creativecommons.org/licenses/by/4.0",
  language  = "English"
}

@ARTICLE{Tesch2016-jc,
  title     = "Structural and electronic properties of graphene nanoflakes on
               Au(111) and Ag(111)",
  author    = "Tesch, Julia and Leicht, Philipp and Blumenschein, Felix and
               Gragnaniello, Luca and Fonin, Mikhail and Marsoner
               Steinkasserer, Lukas Eugen and Paulus, Beate and Voloshina,
               Elena and Dedkov, Yuriy",
  journal   = "Sci. Rep.",
  publisher = "Springer Science and Business Media LLC",
  volume    =  6,
  number    =  1,
  pages     = "23439",
  month     =  mar,
  year      =  2016,
  doi = {10.1038/srep23439},
  copyright = "https://creativecommons.org/licenses/by/4.0",
  language  = "English"
}

@article{Tesch2017,
  title = {Impurity scattering and size quantization effects in a single graphene nanoflake},
  author = {Tesch, Julia and Leicht, Philipp and Blumenschein, Felix and Gragnaniello, Luca and Bergvall, Anders and L\"ofwander, Tomas and Fonin, Mikhail},
  journal = {Phys. Rev. B},
  volume = {95},
  issue = {7},
  pages = {075429},
  numpages = {5},
  year = {2017},
  month = Feb,
  publisher = {American Physical Society},
  doi = {10.1103/PhysRevB.95.075429},
  url = {https://link.aps.org/doi/10.1103/PhysRevB.95.075429}
}

@article{Dutreix2019, 
year = {2019}, 
title = {{Measuring the Berry phase of graphene from wavefront dislocations in Friedel oscillations}}, 
author = {{Dutreix, C. and Gonz\'alez-Herrero, H. and Brihuega, I. and Katsnelson, M. I. and Chapelier, C. and Renard, V. T.}}, 
journal = {Nature}, 
issn = {0028-0836}, 
doi = {10.1038/s41586-019-1613-5}, 
pages = {219--222}, 
number = {7777}, 
volume = {574} 
}

@article{Brihuega2008,
  title = {Quasiparticle Chirality in Epitaxial Graphene Probed at the Nanometer Scale},
  author = {Brihuega, I. and Mallet, P. and Bena, C. and Bose, S. and Michaelis, C. and Vitali, L. and Varchon, F. and Magaud, L. and Kern, K. and Veuillen, J. Y.},
  journal = {Phys. Rev. Lett.},
  volume = {101},
  issue = {20},
  pages = {206802},
  numpages = {4},
  year = {2008},
  month = {Nov},
  publisher = {American Physical Society},
  doi = {10.1103/PhysRevLett.101.206802},
  url = {https://link.aps.org/doi/10.1103/PhysRevLett.101.206802}
}

@article{Zhang2009origin,
  title={Origin of spatial charge inhomogeneity in graphene},
  author={Zhang, Yuanbo and Brar, Victor W and Girit, Caglar and Zettl, Alex and Crommie, Michael F},
  journal={Nature Physics},
  volume={5},
  number={10},
  pages={722--726},
  year={2009},
  doi={10.1038/nphys1365},
  publisher={Nature Publishing Group UK London}
}

@article{Mallet2012,
  title = {Role of pseudospin in quasiparticle interferences in epitaxial graphene probed by high-resolution scanning tunneling microscopy},
  author = {Mallet, P. and Brihuega, I. and Bose, S. and Ugeda, M. M. and G\'omez-Rodr\'{\i}guez, J. M. and Kern, K. and Veuillen, J. Y.},
  journal = {Phys. Rev. B},
  volume = {86},
  issue = {4},
  pages = {045444},
  numpages = {14},
  year = {2012},
  month = {Jul},
  publisher = {American Physical Society},
  doi = {10.1103/PhysRevB.86.045444},
  url = {https://link.aps.org/doi/10.1103/PhysRevB.86.045444}
}

@article{Huempfner2022,
  title={$\pi$ band folding and interlayer band filling of graphene upon interface potassium intercalation},
  author={Huempfner, Tobias and Otto, Felix and Fritz, Torsten and Forker, Roman},
  journal={Advanced Materials Interfaces},
  volume={9},
  number={25},
  pages={2200585},
  year={2022},
  doi = {10.1002/admi.202200585},
  publisher={Wiley Online Library}
}

@article{Lisi2022,
  title = {Two-Way Twisting of a Confined Monolayer: Orientational Ordering within the van der Waals Gap between Graphene and Its Crystalline Substrate},
  author = {Lisi, Simone and Guisset, Val\'erie and David, Philippe and Mazaleyrat, Estelle and G\'omez Herrero, Ana Cristina and Coraux, Johann},
  journal = {Phys. Rev. Lett.},
  volume = {129},
  issue = {9},
  pages = {096101},
  numpages = {6},
  year = {2022},
  month = {Aug},
  publisher = {American Physical Society},
  doi = {10.1103/PhysRevLett.129.096101},
  url = {https://link.aps.org/doi/10.1103/PhysRevLett.129.096101}
}

@article{Zhang2021quantum,
  title={Quantum interferences of pseudospin-mediated atomic-scale vortices in monolayer graphene},
  author={Zhang, Yu and Su, Ying and He, Lin},
  journal={Nano Letters},
  volume={21},
  number={6},
  pages={2526--2531},
  year={2021},
  doi={10.1021/acs.nanolett.0c05066},
  publisher={ACS Publications}
}

@article{Zhang2023observation,
  title={Observation of robust and long-ranged superperiodicity of electronic density induced by intervalley scattering in graphene/transition metal dichalcogenide heterostructures},
  author={Zhang, Mo-Han and Ren, Ya-Ning and Zheng, Qi and Zhou, Xiao-Feng and He, Lin},
  journal={Nano Letters},
  volume={23},
  number={7},
  pages={2630--2635},
  year={2023},
  doi={10.1021/acs.nanolett.2c04957},
  publisher={ACS Publications}
}

@article{Sun2023determining,
  title={Determining spin-orbit coupling in graphene by quasiparticle interference imaging},
  author={Sun, Lihuan and Rademaker, Louk and Mauro, Diego and Scarfato, Alessandro and P{\'a}sztor, {\'A}rp{\'a}d and Guti{\'e}rrez-Lezama, Ignacio and Wang, Zhe and Martinez-Castro, Jose and Morpurgo, Alberto F and Renner, Christoph},
  journal={Nature communications},
  volume={14},
  number={1},
  pages={3771},
  year={2023},
  doi={10.1038/s41467-023-39453-x},
  publisher={Nature Publishing Group UK London}
}

@article{Bao2021,
  title = {Experimental Evidence of Chiral Symmetry Breaking in Kekul\'e-Ordered Graphene},
  author = {Bao, Changhua and Zhang, Hongyun and Zhang, Teng and Wu, Xi and Luo, Laipeng and Zhou, Shaohua and Li, Qian and Hou, Yanhui and Yao, Wei and Liu, Liwei and Yu, Pu and Li, Jia and Duan, Wenhui and Yao, Hong and Wang, Yeliang and Zhou, Shuyun},
  journal = {Phys. Rev. Lett.},
  volume = {126},
  issue = {20},
  pages = {206804},
  numpages = {7},
  year = {2021},
  month = May,
  publisher = {American Physical Society},
  doi = {10.1103/PhysRevLett.126.206804},
  url = {https://link.aps.org/doi/10.1103/PhysRevLett.126.206804}
}

@article{Guan2024observation,
  title={Observation of Kekul{\'e} vortices around hydrogen adatoms in graphene},
  author={Guan, Yifei and Dutreix, Clement and Gonz{\'a}lez-Herrero, H{\'e}ctor and Ugeda, Miguel M and Brihuega, Ivan and Katsnelson, Mikhail I and Yazyev, Oleg V and Renard, Vincent T},
  journal={Nature Communications},
  volume={15},
  number={1},
  pages={2927},
  year={2024},
  doi={10.1038/s41467-024-47267-8},
  publisher={Nature Publishing Group UK London}
}

@article{Mallet2016friedel,
  title={Friedel oscillations in graphene-based systems probed by Scanning Tunneling Microscopy},
  author={Mallet, Pierre and Brihuega, Ivan and Cherkez, Vladimir and G{\'o}mez-Rodr{\'\i}guez, Jose Mar{\`\i}a and Veuillen, Jean-Yves},
  journal={Comptes Rendus Physique},
  volume={17},
  number={3-4},
  pages={294--301},
  year={2016},
  doi={10.1016/j.crhy.2015.12.013},
  publisher={Elsevier}
}

@article{Rutter2007scattering,
  title={Scattering and interference in epitaxial graphene},
  author={Rutter, Gregory M and Crain, JN and Guisinger, NP and Li, T and First, PN and Stroscio, JA},
  journal={Science},
  volume={317},
  number={5835},
  pages={219--222},
  year={2007},
  doi={10.1126/science.1142882},
  publisher={American Association for the Advancement of Science}
}

@article{Simon2009symmetry,
  title={Symmetry of standing waves generated by a point defect in epitaxial graphene},
  author={Simon, L and Bena, Cristina and Vonau, F and Aubel, D and Nasrallah, H and Habar, M and Peruchetti, JC},
  journal={The European Physical Journal B},
  volume={69},
  number={3},
  pages={351--355},
  year={2009},
  url={https://link.springer.com/article/10.1140/epjb/e2009-00142-3},
  publisher={Springer}
}

@article{Yankowitz2014band,
  title={Band structure mapping of bilayer graphene via quasiparticle scattering},
  author={Yankowitz, Matthew and Wang, Joel I-Jan and Li, Suchun and Birdwell, A Glen and Chen, Yu-An and Watanabe, Kenji and Taniguchi, Takashi and Quek, Su Ying and Jarillo-Herrero, Pablo and LeRoy, Brian J},
  journal={Apl Materials},
  volume={2},
  number={9},
  year={2014},
  doi={https://doi.org/10.1063/1.4890543},
  publisher={AIP Publishing}
}

@article{Jolie2018,
  title = {Suppression of Quasiparticle Scattering Signals in Bilayer Graphene Due to Layer Polarization and Destructive Interference},
  author = {Jolie, Wouter and Lux, Jonathan and P\"ortner, Mathias and Dombrowski, Daniela and Herbig, Charlotte and Knispel, Timo and Simon, Sabina and Michely, Thomas and Rosch, Achim and Busse, Carsten},
  journal = {Phys. Rev. Lett.},
  volume = {120},
  issue = {10},
  pages = {106801},
  numpages = {6},
  year = {2018},
  month = Mar,
  publisher = {American Physical Society},
  doi = {10.1103/PhysRevLett.120.106801},
  url = {https://link.aps.org/doi/10.1103/PhysRevLett.120.106801}
}

@article{Joucken2021direct,
  title={Direct visualization of native defects in graphite and their effect on the electronic properties of bernal-stacked bilayer graphene},
  author={Joucken, Fr{\'e}d{\'e}ric and Bena, Cristina and Ge, Zhehao and Quezada-Lopez, Eberth and Pinon, Sarah and Kaladzhyan, Vardan and Taniguchi, Takashi and Watanabe, Kenji and Ferreira, Aires and Velasco Jr, Jairo},
  journal={Nano Letters},
  volume={21},
  number={17},
  pages={7100--7108},
  year={2021},
  doi={10.1021/acs.nanolett.1c01442},
  publisher={ACS Publications}
}

@article{Joucken2020,
  title = {Determination of the trigonal warping orientation in Bernal-stacked bilayer graphene via scanning tunneling microscopy},
  author = {Joucken, Fr\'ed\'eric and Ge, Zhehao and Quezada-L\'opez, Eberth A. and Davenport, John L. and Watanabe, Kenji and Taniguchi, Takashi and Velasco, Jairo},
  journal = {Phys. Rev. B},
  volume = {101},
  issue = {16},
  pages = {161103},
  numpages = {6},
  year = {2020},
  month = Apr,
  publisher = {American Physical Society},
  doi = {10.1103/PhysRevB.101.161103},
  url = {https://link.aps.org/doi/10.1103/PhysRevB.101.161103}
}

@article{Zhang2020,
  title = {Local Berry Phase Signatures of Bilayer Graphene in Intervalley Quantum Interference},
  author = {Zhang, Yu and Su, Ying and He, Lin},
  journal = {Phys. Rev. Lett.},
  volume = {125},
  issue = {11},
  pages = {116804},
  numpages = {6},
  year = {2020},
  month = Sep,
  publisher = {American Physical Society},
  doi = {10.1103/PhysRevLett.125.116804},
  url = {https://link.aps.org/doi/10.1103/PhysRevLett.125.116804}
}

@article{Kaladzhyan2021,
  title = {Quasiparticle interference patterns in bilayer graphene with trigonal warping},
  author = {Kaladzhyan, Vardan and Joucken, Fr\'ed\'eric and Ge, Zhehao and Quezada-Lopez, Eberth A. and Taniguchi, Takashi and Watanabe, Kenji and Velasco, Jairo and Bena, Cristina},
  journal = {Phys. Rev. B},
  volume = {104},
  issue = {23},
  pages = {235425},
  numpages = {13},
  year = {2021},
  month = Dec,
  publisher = {American Physical Society},
  doi = {10.1103/PhysRevB.104.235425},
  url = {https://link.aps.org/doi/10.1103/PhysRevB.104.235425}
}

@article{Kechedzhi2007,
  title = {Influence of Trigonal Warping on Interference Effects in Bilayer Graphene},
  author = {Kechedzhi, K. and Fal'ko, Vladimir I. and McCann, E. and Altshuler, B. L.},
  journal = {Phys. Rev. Lett.},
  volume = {98},
  issue = {17},
  pages = {176806},
  numpages = {4},
  year = {2007},
  month = Apr,
  publisher = {American Physical Society},
  doi = {10.1103/PhysRevLett.98.176806},
  url = {https://link.aps.org/doi/10.1103/PhysRevLett.98.176806}
}

@article{Bena2008,
  title = {Effect of a Single Localized Impurity on the Local Density of States in Monolayer and Bilayer Graphene},
  author = {Bena, Cristina},
  journal = {Phys. Rev. Lett.},
  volume = {100},
  issue = {7},
  pages = {076601},
  numpages = {4},
  year = {2008},
  month = Feb,
  publisher = {American Physical Society},
  doi = {10.1103/PhysRevLett.100.076601},
  url = {https://link.aps.org/doi/10.1103/PhysRevLett.100.076601}
}

@article{Bena2009green,
  title = {Green's functions and impurity scattering in graphene},
  author = {Bena, Cristina},
  journal = {Phys. Rev. B},
  volume = {79},
  issue = {12},
  pages = {125427},
  numpages = {7},
  year = {2009},
  month = Mar,
  publisher = {American Physical Society},
  doi = {10.1103/PhysRevB.79.125427},
  url = {https://link.aps.org/doi/10.1103/PhysRevB.79.125427}
}

@article{Pereg-Barnea2008,
  title = {Chiral quasiparticle local density of states maps in graphene},
  author = {Pereg-Barnea, T. and MacDonald, A. H.},
  journal = {Phys. Rev. B},
  volume = {78},
  issue = {1},
  pages = {014201},
  numpages = {8},
  year = {2008},
  month = Jul,
  publisher = {American Physical Society},
  doi = {10.1103/PhysRevB.78.014201},
  url = {https://link.aps.org/doi/10.1103/PhysRevB.78.014201}
}

@article{Peres2009local,
  title={Local density of states and scanning tunneling currents in graphene},
  author={Peres, NMR and Yang, Ling and Tsai, Shan-Wen},
  journal={New Journal of Physics},
  volume={11},
  number={9},
  pages={095007},
  doi={10.1088/1367-2630/11/9/095007},
  year={2009}
}

@article{Lawlor2013,
  title = {Friedel oscillations in graphene: Sublattice asymmetry in doping},
  author = {Lawlor, J. A. and Power, S. R. and Ferreira, M. S.},
  journal = {Phys. Rev. B},
  volume = {88},
  issue = {20},
  pages = {205416},
  numpages = {11},
  year = {2013},
  month = Nov,
  publisher = {American Physical Society},
  doi = {10.1103/PhysRevB.88.205416},
  url = {https://link.aps.org/doi/10.1103/PhysRevB.88.205416}
}

@article{Soule2014,
  title = {Quasiparticle spectroscopy as a probe of the topological phase in graphene with heavy adatoms},
  author = {Soul\'e, Paul and Franz, M.},
  journal = {Phys. Rev. B},
  volume = {89},
  issue = {20},
  pages = {201410},
  numpages = {4},
  year = {2014},
  month = May,
  publisher = {American Physical Society},
  doi = {10.1103/PhysRevB.89.201410},
  url = {https://link.aps.org/doi/10.1103/PhysRevB.89.201410}
}

@article{Dutreix2016,
  title = {Friedel oscillations at the surfaces of rhombohedral $N$-layer graphene},
  author = {Dutreix, C. and Katsnelson, M. I.},
  journal = {Phys. Rev. B},
  volume = {93},
  issue = {3},
  pages = {035413},
  numpages = {20},
  year = {2016},
  month = Jan,
  publisher = {American Physical Society},
  doi = {10.1103/PhysRevB.93.035413},
  url = {https://link.aps.org/doi/10.1103/PhysRevB.93.035413}
}

@article{Zhang2021robust,
  title = {Robust wavefront dislocations of Friedel oscillations in gapped graphene},
  author = {Zhang, Shu-Hui and Yang, Jin and Shao, Ding-Fu and Wu, Zhenhua and Yang, Wen},
  journal = {Phys. Rev. B},
  volume = {103},
  issue = {16},
  pages = {L161407},
  numpages = {6},
  year = {2021},
  month = Apr,
  publisher = {American Physical Society},
  doi = {10.1103/PhysRevB.103.L161407},
  url = {https://link.aps.org/doi/10.1103/PhysRevB.103.L161407}
}

@article{Yang2021friedel,
  title = {Friedel oscillations in graphene gapped by breaking $\mathcal{P}$ and $\mathcal{T}$ symmetries: Topological and geometrical signatures of electronic structure},
  author = {Yang, Jin and Shao, Ding-Fu and Zhang, Shu-Hui and Yang, Wen},
  journal = {Phys. Rev. B},
  volume = {104},
  issue = {3},
  pages = {035402},
  numpages = {8},
  year = {2021},
  month = Jul,
  publisher = {American Physical Society},
  doi = {10.1103/PhysRevB.104.035402},
  url = {https://link.aps.org/doi/10.1103/PhysRevB.104.035402}
}

@article{Yang2024wavefronts,
  title={Wavefronts Dislocations of Friedel Oscillations in Graphene: Trigonal Warping Effect},
  author={Yang, Jin and Zhang, Shu-Hui and Yang, Wen},
  journal={physica status solidi (RRL)--Rapid Research Letters},
  volume={18},
  number={4},
  pages={2300378},
  year={2024},
  doi={10.1002/pssr.202300378},
  publisher={Wiley Online Library}
}

@article{Li2025local,
  title={Local magnetic moment oscillation around an Anderson impurity on graphene},
  author={Li, Shuai and Ma, Zhen and Gao, Jin-Hua},
  journal={Science China Physics, Mechanics \& Astronomy},
  volume={68},
  number={1},
  pages={217211},
  year={2025},
  doi={10.1007/s11433-024-2512-3},
  publisher={Springer}
}

@article{He2022,
  title = {Magnetic impurity as a local probe of the $U$(1) quantum spin liquid with spinon Fermi surface},
  author = {He, Wen-Yu and Lee, Patrick A.},
  journal = {Phys. Rev. B},
  volume = {105},
  issue = {19},
  pages = {195156},
  numpages = {15},
  year = {2022},
  month = {May},
  publisher = {American Physical Society},
  doi = {10.1103/PhysRevB.105.195156},
  url = {https://link.aps.org/doi/10.1103/PhysRevB.105.195156}
}

@article{Jahin2025,
  title = {Quasiparticle Interference in Kitaev Quantum Spin Liquids},
  author = {Jahin, Ammar and Zhang, Hao and Hal\'asz, G\'abor B. and Lin, Shi-Zeng},
  journal = {Phys. Rev. Lett.},
  volume = {134},
  issue = {12},
  pages = {126501},
  numpages = {7},
  year = {2025},
  month = {Mar},
  publisher = {American Physical Society},
  doi = {10.1103/PhysRevLett.134.126501},
  url = {https://link.aps.org/doi/10.1103/PhysRevLett.134.126501}
}

@ARTICLE{Ruan2021,
  title     = "Evidence for quantum spin liquid behaviour in single-layer
               1T-TaSe$_2$ from scanning tunnelling microscopy",
  author    = "Ruan, Wei and Chen, Yi and Tang, Shujie and Hwang, Jinwoong and
               Tsai, Hsin-Zon and Lee, Ryan L and Wu, Meng and Ryu, Hyejin and
               Kahn, Salman and Liou, Franklin and Jia, Caihong and Aikawa,
               Andrew and Hwang, Choongyu and Wang, Feng and Choi, Yongseong
               and Louie, Steven G and Lee, Patrick A and Shen, Zhi-Xun and Mo,
               Sung-Kwan and Crommie, Michael F",
  journal   = "Nat. Phys.",
  publisher = "Springer Science and Business Media LLC",
  volume    =  "17",
  number    =  "10",
  pages     = "1154--1161",
  month     =  "Oct",
  year      =  "2021",
  language  = "English",
  url = {https://www.nature.com/articles/s41567-021-01321-0}
}

@article{Novoselov2004electric,
	title={Electric field effect in atomically thin carbon films},
	author={Novoselov, Kostya S and Geim, Andre K and Morozov, Sergei V and Jiang, De-eng and Zhang, Yanshui and Dubonos, Sergey V and Grigorieva, Irina V and Firsov, Alexandr A},
	journal={science},
	volume={306},
	number={5696},
	pages={666--669},
	year={2004},
	url = {https://www.science.org/doi/10.1126/science.1102896},
	publisher={American Association for the Advancement of Science}
}

@article{Zou2018,
  title = {Band structure of twisted bilayer graphene: Emergent symmetries, commensurate approximants, and Wannier obstructions},
  author = {Zou, Liujun and Po, Hoi Chun and Vishwanath, Ashvin and Senthil, T.},
  journal = {Phys. Rev. B},
  volume = {98},
  issue = {8},
  pages = {085435},
  numpages = {14},
  year = {2018},
  month = {Aug},
  publisher = {American Physical Society},
  doi = {10.1103/PhysRevB.98.085435},
  url = {https://link.aps.org/doi/10.1103/PhysRevB.98.085435}
}

@article{Shallcross2008,
	title = {Quantum Interference at the Twist Boundary in Graphene},
	author = {Shallcross, S. and Sharma, S. and Pankratov, O. A.},
	journal = {Phys. Rev. Lett.},
	volume = {101},
	issue = {5},
	pages = {056803},
	numpages = {4},
	year = {2008},
	month = Aug,
	publisher = {American Physical Society},
	doi = {10.1103/PhysRevLett.101.056803},
	url = {https://link.aps.org/doi/10.1103/PhysRevLett.101.056803}
}

@article{Shallcross2010,
	title = {Electronic structure of turbostratic graphene},
	author = {Shallcross, S. and Sharma, S. and Kandelaki, E. and Pankratov, O. A.},
	journal = {Phys. Rev. B},
	volume = {81},
	issue = {16},
	pages = {165105},
	numpages = {15},
	year = {2010},
	month = Apr,
	publisher = {American Physical Society},
	doi = {10.1103/PhysRevB.81.165105},
	url = {https://link.aps.org/doi/10.1103/PhysRevB.81.165105}
}

@article{Mele2010,
	title = {Commensuration and interlayer coherence in twisted bilayer graphene},
	author = {Mele, E. J.},
	journal = {Phys. Rev. B},
	volume = {81},
	issue = {16},
	pages = {161405},
	numpages = {4},
	year = {2010},
	month = Apr,
	publisher = {American Physical Society},
	doi = {10.1103/PhysRevB.81.161405},
	url = {https://link.aps.org/doi/10.1103/PhysRevB.81.161405}
}

@article{Mele2011,
	title = {Band symmetries and singularities in twisted multilayer graphene},
	author = {Mele, E. J.},
	journal = {Phys. Rev. B},
	volume = {84},
	issue = {23},
	pages = {235439},
	numpages = {5},
	year = {2011},
	month = Dec,
	publisher = {American Physical Society},
	doi = {10.1103/PhysRevB.84.235439},
	url = {https://link.aps.org/doi/10.1103/PhysRevB.84.235439}
}

@article{Latil2007,
	title = {Massless fermions in multilayer graphitic systems with misoriented layers: Ab initio calculations and experimental fingerprints},
	author = {Latil, Sylvain and Meunier, Vincent and Henrard, Luc},
	journal = {Phys. Rev. B},
	volume = {76},
	issue = {20},
	pages = {201402},
	numpages = {4},
	year = {2007},
	month = Nov,
	publisher = {American Physical Society},
	doi = {10.1103/PhysRevB.76.201402},
	url = {https://link.aps.org/doi/10.1103/PhysRevB.76.201402}
}

@article{Emtsev2008,
	title = {Interaction, growth, and ordering of epitaxial graphene on SiC{0001} surfaces: A comparative photoelectron spectroscopy study},
	author = {Emtsev, K. V. and Speck, F. and Seyller, Th. and Ley, L. and Riley, J. D.},
	journal = {Phys. Rev. B},
	volume = {77},
	issue = {15},
	pages = {155303},
	numpages = {10},
	year = {2008},
	month = Apr,
	publisher = {American Physical Society},
	doi = {10.1103/PhysRevB.77.155303},
	url = {https://link.aps.org/doi/10.1103/PhysRevB.77.155303}
}

@article{Varchon2008,
	title = {Rotational disorder in few-layer graphene films on $6H\text{\ensuremath{-}}\mathrm{Si}\mathrm{C}(000\text{\ensuremath{-}}1)$: A scanning tunneling microscopy study},
	author = {Varchon, Fran\ifmmode \mbox{\c{c}}\else \c{c}\fi{}ois and Mallet, Pierre and Magaud, Laurence and Veuillen, Jean-Yves},
	journal = {Phys. Rev. B},
	volume = {77},
	issue = {16},
	pages = {165415},
	numpages = {5},
	year = {2008},
	month = Apr,
	publisher = {American Physical Society},
	doi = {10.1103/PhysRevB.77.165415},
	url = {https://link.aps.org/doi/10.1103/PhysRevB.77.165415}
}

@article{Kindermann2011,
	title = {Local sublattice-symmetry breaking in rotationally faulted multilayer graphene},
	author = {Kindermann, M. and First, P. N.},
	journal = {Phys. Rev. B},
	volume = {83},
	issue = {4},
	pages = {045425},
	numpages = {6},
	year = {2011},
	month = Jan,
	publisher = {American Physical Society},
	doi = {10.1103/PhysRevB.83.045425},
	url = {https://link.aps.org/doi/10.1103/PhysRevB.83.045425}
}

@article{Moon2012,
	title = {Energy spectrum and quantum Hall effect in twisted bilayer graphene},
	author = {Moon, Pilkyung and Koshino, Mikito},
	journal = {Phys. Rev. B},
	volume = {85},
	issue = {19},
	pages = {195458},
	numpages = {9},
	year = {2012},
	month = May,
	publisher = {American Physical Society},
	doi = {10.1103/PhysRevB.85.195458},
	url = {https://link.aps.org/doi/10.1103/PhysRevB.85.195458}
}

@article{Li2010observation,
	title={Observation of Van Hove singularities in twisted graphene layers},
	author={Li, Guohong and Luican, A and Lopes dos Santos, JMB and Castro Neto, AH and Reina, A and Kong, J and Andrei, EY},
	journal={Nature physics},
	volume={6},
	number={2},
	pages={109--113},
	year={2010},
	url = {https://doi.org/10.1038/nphys1463},
	publisher={Nature Publishing Group UK London}
}

@article{Morell2010,
	title = {Flat bands in slightly twisted bilayer graphene: Tight-binding calculations},
	author = {Su\'arez Morell, E. and Correa, J. D. and Vargas, P. and Pacheco, M. and Barticevic, Z.},
	journal = {Phys. Rev. B},
	volume = {82},
	issue = {12},
	pages = {121407},
	numpages = {4},
	year = {2010},
	month = Sep,
	publisher = {American Physical Society},
	doi = {10.1103/PhysRevB.82.121407},
	url = {https://link.aps.org/doi/10.1103/PhysRevB.82.121407}
}

@article{Hass2008tbg,
	title = {Why Multilayer Graphene on $4H\mathrm{\text{\ensuremath{-}}}\mathrm{SiC}(000\overline{1})$ Behaves Like a Single Sheet of Graphene},
	author = {Hass, J. and Varchon, F. and Mill\'an-Otoya, J. E. and Sprinkle, M. and Sharma, N. and de Heer, W. A. and Berger, C. and First, P. N. and Magaud, L. and Conrad, E. H.},
	journal = {Phys. Rev. Lett.},
	volume = {100},
	issue = {12},
	pages = {125504},
	numpages = {4},
	year = {2008},
	month = Mar,
	publisher = {American Physical Society},
	doi = {10.1103/PhysRevLett.100.125504},
	url = {https://link.aps.org/doi/10.1103/PhysRevLett.100.125504}
}

@article{Brihuega2012stm,
	title = {Unraveling the Intrinsic and Robust Nature of van Hove Singularities in Twisted Bilayer Graphene by Scanning Tunneling Microscopy and Theoretical Analysis},
	author = {Brihuega, I. and Mallet, P. and Gonz\'alez-Herrero, H. and Trambly de Laissardi\`ere, G. and Ugeda, M. M. and Magaud, L. and G\'omez-Rodr\'{\i}guez, J. M. and Yndur\'ain, F. and Veuillen, J.-Y.},
	journal = {Phys. Rev. Lett.},
	volume = {109},
	issue = {19},
	pages = {196802},
	numpages = {5},
	year = {2012},
	month = Nov,
	publisher = {American Physical Society},
	doi = {10.1103/PhysRevLett.109.196802},
	url = {https://link.aps.org/doi/10.1103/PhysRevLett.109.196802}
}

@article{Buckley1991large,
	title={Large-scale periodic features associated with surface boundaries in scanning tunneling microscope images of graphite},
	author={Buckley, JE and Wragg, JL and White, HW and Bruckdorfer, A and Worcester, DL},
	journal={Journal of Vacuum Science \& Technology B: Microelectronics and Nanometer Structures Processing, Measurement, and Phenomena},
	volume={9},
	number={2},
	pages={1079--1082},
	year={1991},
	url = {https://doi.org/10.1116/1.585264},
	publisher={American Vacuum Society}
}

@article{Yang1992several,
	title={Several large-scale superperiodicities on highly oriented pyrolytic graphite observed by scanning tunneling microscopy},
	author={Yang, X and Bromm, Ch and Geyer, U and Von Minnigerode, G},
	journal={Annalen der Physik},
	volume={504},
	number={1},
	pages={3--10},
	url = {https://doi.org/10.1002/andp.19925040103},
	year={1992},
	publisher={Wiley Online Library}
}

@article{Rong1993,
	title = {Electronic effects in scanning tunneling microscopy: Moir\'e pattern on a graphite surface},
	author = {Rong, Zhao Y. and Kuiper, Pieter},
	journal = {Phys. Rev. B},
	volume = {48},
	issue = {23},
	pages = {17427--17431},
	numpages = {0},
	year = {1993},
	month = Dec,
	publisher = {American Physical Society},
	doi = {10.1103/PhysRevB.48.17427},
	url = {https://link.aps.org/doi/10.1103/PhysRevB.48.17427}
}

@article{Cee1995unusual,
	title={Unusual aspects of superperiodic features on highly oriented pyrolytic graphite},
	author={Cee, Victor J and Patrick, David L and Beebe Jr, Thomas P},
	journal={Surface science},
	volume={329},
	number={1-2},
	pages={141--148},
	year={1995},
	url = {https://doi.org/10.1016/0039-6028(95)00013-5},
	publisher={Elsevier}
}

@article{Bernhardt1998formation,
	title={Formation of superperiodic patterns on highly oriented pyrolytic graphite by manipulation of nanosized graphite sheets with the STM tip},
	author={Bernhardt, TM and Kaiser, B and Rademann, KJSS},
	journal={Surface science},
	volume={408},
	number={1-3},
	pages={86--94},
	year={1998},
	url = {https://doi.org/10.1016/S0039-6028(98)00152-6},
	publisher={Elsevier}
}

@article{Ouseph2000scanning,
	title={Scanning tunneling microscopy observation of dislocations with superlattice structure in graphite},
	author={Ouseph, PJ},
	journal={Applied surface science},
	volume={165},
	number={1},
	pages={38--43},
	url = {https://doi.org/10.1016/S0169-4332(00)00358-5},
	year={2000},
	publisher={Elsevier}
}

@article{Pong2005review,
	title={A review and outlook for an anomaly of scanning tunnelling microscopy (STM): superlattices on graphite},
	author={Pong, Wing-Tat and Durkan, Colm},
	journal={Journal of Physics D: Applied Physics},
	volume={38},
	number={21},
	pages={R329},
	doi = {10.1088/0022-3727/38/21/R01},
	year={2005},
	publisher={IOP Publishing}
}

@ARTICLE{Trambly_de_Laissardiere2010,
  title     = "Localization of dirac electrons in rotated graphene bilayers",
  author    = "Trambly de Laissardi{\`e}re, G and Mayou, D and Magaud, L",
  journal   = "Nano Lett.",
  publisher = "American Chemical Society (ACS)",
  volume    =  10,
  number    =  3,
  pages     = "804--808",
  month     =  {Mar},
  year      =  {2010},
  language  = "English",
  url = {https://pubs.acs.org/doi/full/10.1021/nl902948m}
}

@article{LopesdosSantos2007,
  title = {Graphene Bilayer with a Twist: Electronic Structure},
  author = {Lopes dos Santos, J. M. B. and Peres, N. M. R. and Castro Neto, A. H.},
  journal = {Phys. Rev. Lett.},
  volume = {99},
  issue = {25},
  pages = {256802},
  numpages = {4},
  year = {2007},
  month = {Dec},
  publisher = {American Physical Society},
  doi = {10.1103/PhysRevLett.99.256802},
  url = {https://link.aps.org/doi/10.1103/PhysRevLett.99.256802}
}

@article{Santos2012,
	title = {Continuum model of the twisted graphene bilayer},
	author = {Lopes dos Santos, J. M. B. and Peres, N. M. R. and Castro Neto, A. H.},
	journal = {Phys. Rev. B},
	volume = {86},
	issue = {15},
	pages = {155449},
	numpages = {12},
	year = {2012},
	month = Oct,
	publisher = {American Physical Society},
	doi = {10.1103/PhysRevB.86.155449},
	url = {https://link.aps.org/doi/10.1103/PhysRevB.86.155449}
}

@ARTICLE{Bistritzer2011-mq,
  title     = "Moir{\'e} bands in twisted double-layer graphene",
  author    = "Bistritzer, Rafi and MacDonald, Allan H",
  journal   = "Proc. Natl. Acad. Sci. U. S. A.",
  publisher = "Proceedings of the National Academy of Sciences",
  volume    =  108,
  number    =  30,
  pages     = "12233--12237",
  month     =  jul,
  year      =  2011,
  language  = "English",
  url = {https://www.pnas.org/doi/full/10.1073/pnas.1108174108}
}

@article{Koshino2018,
	title = {Maximally Localized Wannier Orbitals and the Extended Hubbard Model for Twisted Bilayer Graphene},
	author = {Koshino, Mikito and Yuan, Noah F. Q. and Koretsune, Takashi and Ochi, Masayuki and Kuroki, Kazuhiko and Fu, Liang},
	journal = {Phys. Rev. X},
	volume = {8},
	issue = {3},
	pages = {031087},
	numpages = {12},
	year = 2018,
	month = {Sep},
	publisher = {American Physical Society},
	doi = {10.1103/PhysRevX.8.031087},
	url = {https://link.aps.org/doi/10.1103/PhysRevX.8.031087}
}

@article{Kang2018,
	title = {Symmetry, Maximally Localized Wannier States, and a Low-Energy Model for Twisted Bilayer Graphene Narrow Bands},
	author = {Kang, Jian and Vafek, Oskar},
	journal = {Phys. Rev. X},
	volume = {8},
	issue = {3},
	pages = {031088},
	numpages = {9},
	year = {2018},
	month = Sep,
	publisher = {American Physical Society},
	doi = {10.1103/PhysRevX.8.031088},
	url = {https://link.aps.org/doi/10.1103/PhysRevX.8.031088}
}

@article{Ahn2019,
	title = {Failure of Nielsen-Ninomiya Theorem and Fragile Topology in Two-Dimensional Systems with Space-Time Inversion Symmetry: Application to Twisted Bilayer Graphene at Magic Angle},
	author = {Ahn, Junyeong and Park, Sungjoon and Yang, Bohm-Jung},
	journal = {Phys. Rev. X},
	volume = {9},
	issue = {2},
	pages = {021013},
	numpages = {26},
	year = {2019},
	month = Apr,
	publisher = {American Physical Society},
	doi = {10.1103/PhysRevX.9.021013},
	url = {https://link.aps.org/doi/10.1103/PhysRevX.9.021013}
}

@article{Carr2019,
	title = {Derivation of Wannier orbitals and minimal-basis tight-binding Hamiltonians for twisted bilayer graphene: First-principles approach},
	author = {Carr, Stephen and Fang, Shiang and Po, Hoi Chun and Vishwanath, Ashvin and Kaxiras, Efthimios},
	journal = {Phys. Rev. Res.},
	volume = {1},
	issue = {3},
	pages = {033072},
	numpages = {11},
	year = {2019},
	month = Nov,
	publisher = {American Physical Society},
	doi = {10.1103/PhysRevResearch.1.033072},
	url = {https://link.aps.org/doi/10.1103/PhysRevResearch.1.033072}
}

@ARTICLE{Cao2018,
  title     = "Unconventional superconductivity in magic-angle graphene
               superlattices",
  author    = "Cao, Yuan and Fatemi, Valla and Fang, Shiang and Watanabe, Kenji
               and Taniguchi, Takashi and Kaxiras, Efthimios and
               Jarillo-Herrero, Pablo",
  journal   = "Nature",
  publisher = "Springer Science and Business Media LLC",
  volume    =  556,
  number    =  7699,
  pages     = "43--50",
  month     =  apr,
  year      =  2018,
  language  = "English",
  url = {https://www.nature.com/articles/nature26160}
}

@ARTICLE{Lu2019-ga,
  title     = "Superconductors, orbital magnets and correlated states in
               magic-angle bilayer graphene",
  author    = "Lu, Xiaobo and Stepanov, Petr and Yang, Wei and Xie, Ming and
               Aamir, Mohammed Ali and Das, Ipsita and Urgell, Carles and
               Watanabe, Kenji and Taniguchi, Takashi and Zhang, Guangyu and
               Bachtold, Adrian and MacDonald, Allan H and Efetov, Dmitri K",
  journal   = "Nature",
  publisher = "Springer Science and Business Media LLC",
  volume    =  574,
  number    =  7780,
  pages     = "653--657",
  month     =  oct,
  year      =  2019,
  language  = "English",
  url = {https://www.nature.com/articles/s41586-019-1695-0}
}

@ARTICLE{Yankowitz2019,
  title     = "Tuning superconductivity in twisted bilayer graphene",
  author    = "Yankowitz, Matthew and Chen, Shaowen and Polshyn, Hryhoriy and
               Zhang, Yuxuan and Watanabe, K and Taniguchi, T and Graf, David
               and Young, Andrea F and Dean, Cory R",
  journal   = "Science",
  publisher = "American Association for the Advancement of Science (AAAS)",
  volume    =  363,
  number    =  6431,
  pages     = "1059--1064",
  month     =  mar,
  year      =  2019,
  copyright = "http://www.sciencemag.org/about/science-licenses-journal-article-reuse",
  language  = "English",
  url = {https://www.science.org/doi/full/10.1126/science.aav1910}
}

@ARTICLE{Saito2020-zp,
  title     = "Independent superconductors and correlated insulators in twisted
               bilayer graphene",
  author    = "Saito, Yu and Ge, Jingyuan and Watanabe, Kenji and Taniguchi,
               Takashi and Young, Andrea F",
  journal   = "Nat. Phys.",
  publisher = "Springer Science and Business Media LLC",
  volume    =  16,
  number    =  9,
  pages     = "926--930",
  month     =  sep,
  year      =  2020,
  language  = "English",
  url = {https://www.nature.com/articles/s41567-020-0928-3}
}

@ARTICLE{Nuckolls2020-pe,
  title     = "Strongly correlated Chern insulators in magic-angle twisted
               bilayer graphene",
  author    = "Nuckolls, Kevin P and Oh, Myungchul and Wong, Dillon and Lian,
               Biao and Watanabe, Kenji and Taniguchi, Takashi and Bernevig, B
               Andrei and Yazdani, Ali",
  journal   = "Nature",
  publisher = "Springer Science and Business Media LLC",
  volume    =  588,
  number    =  7839,
  pages     = "610--615",
  month     =  dec,
  year      =  2020,
  language  = "English",
  url = {https://www.nature.com/articles/s41586-020-3028-8}
}

@article{Cao2020,
  title = {Strange Metal in Magic-Angle Graphene with near Planckian Dissipation},
  author = {Cao, Yuan and Chowdhury, Debanjan and Rodan-Legrain, Daniel and Rubies-Bigorda, Oriol and Watanabe, Kenji and Taniguchi, Takashi and Senthil, T. and Jarillo-Herrero, Pablo},
  journal = {Phys. Rev. Lett.},
  volume = {124},
  issue = {7},
  pages = {076801},
  numpages = {7},
  year = {2020},
  month = {Feb},
  publisher = {American Physical Society},
  doi = {10.1103/PhysRevLett.124.076801},
  url = {https://link.aps.org/doi/10.1103/PhysRevLett.124.076801}
}

@ARTICLE{Oh2021-xr,
  title     = "Evidence for unconventional superconductivity in twisted bilayer
               graphene",
  author    = "Oh, Myungchul and Nuckolls, Kevin P and Wong, Dillon and Lee,
               Ryan L and Liu, Xiaomeng and Watanabe, Kenji and Taniguchi,
               Takashi and Yazdani, Ali",
  journal   = "Nature",
  publisher = "Springer Science and Business Media LLC",
  volume    =  600,
  number    =  7888,
  pages     = "240--245",
  month     =  dec,
  year      =  2021,
  language  = "English",
  url = {https://www.nature.com/articles/s41586-021-04121-x}
}

@ARTICLE{Jaoui2022-aa,
  title     = "Quantum critical behaviour in magic-angle twisted bilayer
               graphene",
  author    = "Jaoui, Alexandre and Das, Ipsita and Di Battista, Giorgio and
               D{\'\i}ez-M{\'e}rida, Jaime and Lu, Xiaobo and Watanabe, Kenji
               and Taniguchi, Takashi and Ishizuka, Hiroaki and Levitov, Leonid
               and Efetov, Dmitri K",
  journal   = "Nat. Phys.",
  publisher = "Springer Science and Business Media LLC",
  volume    =  18,
  number    =  6,
  pages     = "633--638",
  month     =  jun,
  year      =  2022,
  language  = "English",
  url = {https://www.nature.com/articles/s41567-022-01556-5}
}

@article{Lian2019,
  title = {Twisted Bilayer Graphene: A Phonon-Driven Superconductor},
  author = {Lian, Biao and Wang, Zhijun and Bernevig, B. Andrei},
  journal = {Phys. Rev. Lett.},
  volume = {122},
  issue = {25},
  pages = {257002},
  numpages = {6},
  year = {2019},
  month = {Jun},
  publisher = {American Physical Society},
  doi = {10.1103/PhysRevLett.122.257002},
  url = {https://link.aps.org/doi/10.1103/PhysRevLett.122.257002}
}

@misc{Zhu2025,
      title={Microscopic theory for electron-phonon coupling in twisted bilayer graphene}, 
      author={Ziyan Zhu and Thomas P. Devereaux},
      year={2025},
      eprint={2407.03293},
      archivePrefix={arXiv},
      primaryClass={cond-mat.mes-hall},
      url={https://arxiv.org/abs/2407.03293}, 
}

@ARTICLE{Tian2023,
  title     = "Evidence for Dirac flat band superconductivity enabled by
               quantum geometry",
  author    = "Tian, Haidong and Gao, Xueshi and Zhang, Yuxin and Che, Shi and
               Xu, Tianyi and Cheung, Patrick and Watanabe, Kenji and
               Taniguchi, Takashi and Randeria, Mohit and Zhang, Fan and Lau,
               Chun Ning and Bockrath, Marc W",
  journal   = "Nature",
  publisher = "Springer Science and Business Media LLC",
  volume    =  {614},
  number    =  {7948},
  pages     = "440--444",
  month     =  {feb},
  year      =  {2023},
  language  = "English",
  url = {https://www.nature.com/articles/s41586-022-05576-2}
}

@ARTICLE{Tanaka2025,
  title     = "Superfluid stiffness of magic-angle twisted bilayer graphene",
  author    = "Tanaka, Miuko and Wang, Joel {\^I}-J and Dinh, Thao H and
               Rodan-Legrain, Daniel and Zaman, Sameia and Hays, Max and
               Almanakly, Aziza and Kannan, Bharath and Kim, David K and
               Niedzielski, Bethany M and Serniak, Kyle and Schwartz, Mollie E
               and Watanabe, Kenji and Taniguchi, Takashi and Orlando, Terry P
               and Gustavsson, Simon and Grover, Jeffrey A and Jarillo-Herrero,
               Pablo and Oliver, William D",
  journal   = "Nature",
  publisher = "Springer Science and Business Media LLC",
  volume    =  638,
  number    =  8049,
  pages     = "99--105",
  month     =  feb,
  year      =  2025,
  copyright = "https://www.springernature.com/gp/researchers/text-and-data-mining",
  language  = "English",
  url = {https://www.nature.com/articles/s41586-024-08494-7}
}

@article{Hu2019,
  title = {Geometric and Conventional Contribution to the Superfluid Weight in Twisted Bilayer Graphene},
  author = {Hu, Xiang and Hyart, Timo and Pikulin, Dmitry I. and Rossi, Enrico},
  journal = {Phys. Rev. Lett.},
  volume = {123},
  issue = {23},
  pages = {237002},
  numpages = {6},
  year = {2019},
  month = {Dec},
  publisher = {American Physical Society},
  doi = {10.1103/PhysRevLett.123.237002},
  url = {https://link.aps.org/doi/10.1103/PhysRevLett.123.237002}
}

@article{Julku2020,
  title = {Superfluid weight and Berezinskii-Kosterlitz-Thouless transition temperature of twisted bilayer graphene},
  author = {Julku, A. and Peltonen, T. J. and Liang, L. and Heikkil\"a, T. T. and T\"orm\"a, P.},
  journal = {Phys. Rev. B},
  volume = {101},
  issue = {6},
  pages = {060505},
  numpages = {7},
  year = {2020},
  month = {Feb},
  publisher = {American Physical Society},
  doi = {10.1103/PhysRevB.101.060505},
  url = {https://link.aps.org/doi/10.1103/PhysRevB.101.060505}
}

@article{Xie2020,
  title = {Topology-Bounded Superfluid Weight in Twisted Bilayer Graphene},
  author = {Xie, Fang and Song, Zhida and Lian, Biao and Bernevig, B. Andrei},
  journal = {Phys. Rev. Lett.},
  volume = {124},
  issue = {16},
  pages = {167002},
  numpages = {6},
  year = {2020},
  month = {Apr},
  publisher = {American Physical Society},
  doi = {10.1103/PhysRevLett.124.167002},
  url = {https://link.aps.org/doi/10.1103/PhysRevLett.124.167002}
}

@ARTICLE{Verma2021,
  title     = "Optical spectral weight, phase stiffness, and {T} c bounds for
               trivial and topological flat band superconductors",
  author    = "Verma, Nishchhal and Hazra, Tamaghna and Randeria, Mohit",
  journal   = "Proc. Natl. Acad. Sci. U. S. A.",
  publisher = "Proceedings of the National Academy of Sciences",
  volume    =  118,
  number    =  34,
  pages     = "e2106744118",
  month     =  aug,
  year      =  2021,
  copyright = "https://www.pnas.org/site/aboutpnas/licenses.xhtml",
  language  = "English",
  url = {https://www.pnas.org/doi/full/10.1073/pnas.2106744118}
}

@article{Liu2018,
  title = {Chiral Spin Density Wave and $d+id$ Superconductivity in the Magic-Angle-Twisted Bilayer Graphene},
  author = {Liu, Cheng-Cheng and Zhang, Li-Da and Chen, Wei-Qiang and Yang, Fan},
  journal = {Phys. Rev. Lett.},
  volume = {121},
  issue = {21},
  pages = {217001},
  numpages = {6},
  year = {2018},
  month = {Nov},
  publisher = {American Physical Society},
  doi = {10.1103/PhysRevLett.121.217001},
  url = {https://link.aps.org/doi/10.1103/PhysRevLett.121.217001}
}

@article{Sharma2020,
  title = {Superconductivity from collective excitations in magic-angle twisted bilayer graphene},
  author = {Sharma, Gargee and Trushin, Maxim and Sushkov, Oleg P. and Vignale, Giovanni and Adam, Shaffique},
  journal = {Phys. Rev. Res.},
  volume = {2},
  issue = {2},
  pages = {022040},
  numpages = {5},
  year = {2020},
  month = {May},
  publisher = {American Physical Society},
  doi = {10.1103/PhysRevResearch.2.022040},
  url = {https://link.aps.org/doi/10.1103/PhysRevResearch.2.022040}
}

@ARTICLE{Cea2021,
  title     = "Coulomb interaction, phonons, and superconductivity in twisted
               bilayer graphene",
  author    = "Cea, Tommaso and Guinea, Francisco",
  journal   = "Proc. Natl. Acad. Sci. U. S. A.",
  publisher = "Proceedings of the National Academy of Sciences",
  volume    =  118,
  number    =  32,
  pages     = "e2107874118",
  month     =  aug,
  year      =  2021,
  copyright = "https://creativecommons.org/licenses/by-nc-nd/4.0/",
  language  = "English",
  url = {https://www.pnas.org/doi/full/10.1073/pnas.2107874118}
}

@article{Peng2024,
  title = {Theoretical determination of the effect of a screening gate on plasmon-induced superconductivity in twisted bilayer graphene},
  author = {Peng, Liangtao and Yudhistira, Indra and Vignale, Giovanni and Adam, Shaffique},
  journal = {Phys. Rev. B},
  volume = {109},
  issue = {4},
  pages = {045404},
  numpages = {14},
  year = {2024},
  month = {Jan},
  publisher = {American Physical Society},
  doi = {10.1103/PhysRevB.109.045404},
  url = {https://link.aps.org/doi/10.1103/PhysRevB.109.045404}
}

@ARTICLE{Torma2022,
  title     = "Superconductivity, superfluidity and quantum geometry in twisted
               multilayer systems",
  author    = "T{\"o}rm{\"a}, P{\"a}ivi and Peotta, Sebastiano and Bernevig,
               Bogdan A",
  journal   = "Nat. Rev. Phys.",
  publisher = "Springer Science and Business Media LLC",
  volume    =  4,
  number    =  8,
  pages     = "528--542",
  month     =  jun,
  year      =  2022,
  language  = "English",
  url = {https://www.nature.com/articles/s42254-022-00466-y}
}

@article{Bernevig2021_1,
  title = {{Twisted bilayer graphene. I. Matrix elements, approximations, perturbation theory, and a $k\ifmmode\cdot\else\textperiodcentered\fi{}p$ two-band model}},
  author = {Bernevig, B. Andrei and Song, Zhi-Da and Regnault, Nicolas and Lian, Biao},
  journal = {Phys. Rev. B},
  volume = {103},
  issue = {20},
  pages = {205411},
  numpages = {42},
  year = {2021},
  month = {May},
  publisher = {American Physical Society},
  doi = {10.1103/PhysRevB.103.205411},
  url = {https://link.aps.org/doi/10.1103/PhysRevB.103.205411}
}

@article{Song2021,
  title = {{Twisted bilayer graphene. II. Stable symmetry anomaly}},
  author = {Song, Zhi-Da and Lian, Biao and Regnault, Nicolas and Bernevig, B. Andrei},
  journal = {Phys. Rev. B},
  volume = {103},
  issue = {20},
  pages = {205412},
  numpages = {18},
  year = {2021},
  month = {May},
  publisher = {American Physical Society},
  doi = {10.1103/PhysRevB.103.205412},
  url = {https://link.aps.org/doi/10.1103/PhysRevB.103.205412}
}

@article{Bernevig2021_3,
  title = {{Twisted bilayer graphene. III. Interacting Hamiltonian and exact symmetries}},
  author = {Bernevig, B. Andrei and Song, Zhi-Da and Regnault, Nicolas and Lian, Biao},
  journal = {Phys. Rev. B},
  volume = {103},
  issue = {20},
  pages = {205413},
  numpages = {34},
  year = {2021},
  month = {May},
  publisher = {American Physical Society},
  doi = {10.1103/PhysRevB.103.205413},
  url = {https://link.aps.org/doi/10.1103/PhysRevB.103.205413}
}

@article{Lian2021,
  title = {{Twisted bilayer graphene. IV. Exact insulator ground states and phase diagram}},
  author = {Lian, Biao and Song, Zhi-Da and Regnault, Nicolas and Efetov, Dmitri K. and Yazdani, Ali and Bernevig, B. Andrei},
  journal = {Phys. Rev. B},
  volume = {103},
  issue = {20},
  pages = {205414},
  numpages = {41},
  year = {2021},
  month = {May},
  publisher = {American Physical Society},
  doi = {10.1103/PhysRevB.103.205414},
  url = {https://link.aps.org/doi/10.1103/PhysRevB.103.205414}
}

@article{Bernevig2021_5,
  title = {{Twisted bilayer graphene. V. Exact analytic many-body excitations in Coulomb Hamiltonians: Charge gap, Goldstone modes, and absence of Cooper pairing}},
  author = {Bernevig, B. Andrei and Lian, Biao and Cowsik, Aditya and Xie, Fang and Regnault, Nicolas and Song, Zhi-Da},
  journal = {Phys. Rev. B},
  volume = {103},
  issue = {20},
  pages = {205415},
  numpages = {39},
  year = {2021},
  month = {May},
  publisher = {American Physical Society},
  doi = {10.1103/PhysRevB.103.205415},
  url = {https://link.aps.org/doi/10.1103/PhysRevB.103.205415}
}

@article{Phong2020,
  title = {Obstruction and Interference in Low-Energy Models for Twisted Bilayer Graphene},
  author = {Phong, V\~o Ti\'{e}n and Mele, E. J.},
  journal = {Phys. Rev. Lett.},
  volume = {125},
  issue = {17},
  pages = {176404},
  numpages = {6},
  year = {2020},
  month = {Oct},
  publisher = {American Physical Society},
  doi = {10.1103/PhysRevLett.125.176404},
  url = {https://link.aps.org/doi/10.1103/PhysRevLett.125.176404}
}

@article{Rhodes2025,
  title = {Probing moir\'e electronic structures through quasiparticle interference},
  author = {Rhodes, Luke C. and Houston, Dylan C. and Armitage, Olivia R. and Wahl, Peter},
  journal = {Phys. Rev. B},
  volume = {111},
  issue = {12},
  pages = {L121403},
  numpages = {6},
  year = {2025},
  month = {Mar},
  publisher = {American Physical Society},
  doi = {10.1103/PhysRevB.111.L121403},
  url = {https://link.aps.org/doi/10.1103/PhysRevB.111.L121403}
}

@article{Mesple2025,
  title     = {Experimental evidence of the topological obstruction in twisted
               bilayer graphene},
  author    = {Mesple, F and Mallet, P and Trambly de Laissardi{\`e}re, G and
               Dutreix, C and Lapertot, G and Veuillen, J-Y and Renard, V T},
  journal   = "Nat. Commun.",
  volume    =  16,
  number    =  1,
  pages     = "11478",
  year      =  2025,
  doi = "10.1038/s41467-025-66257-y"
}

@article{Ulman2014,
  title = {Point defects in twisted bilayer graphene: A density functional theory study},
  author = {Ulman, Kanchan and Narasimhan, Shobhana},
  journal = {Phys. Rev. B},
  volume = {89},
  issue = {24},
  pages = {245429},
  numpages = {16},
  year = {2014},
  month = {Jun},
  publisher = {American Physical Society},
  doi = {10.1103/PhysRevB.89.245429},
  url = {https://link.aps.org/doi/10.1103/PhysRevB.89.245429}
}

@article{Lopez-Bezanilla2019,
  title = {Defect-induced magnetism and Yu-Shiba-Rusinov states in twisted bilayer graphene},
  author = {Lopez-Bezanilla, Alejandro and Lado, J. L.},
  journal = {Phys. Rev. Mater.},
  volume = {3},
  issue = {8},
  pages = {084003},
  numpages = {8},
  year = {2019},
  month = {Aug},
  publisher = {American Physical Society},
  doi = {10.1103/PhysRevMaterials.3.084003},
  url = {https://link.aps.org/doi/10.1103/PhysRevMaterials.3.084003}
}

@ARTICLE{Dietrich2023,
  title     = "Vacancies and {Stone--Wales} defects in twisted bilayer graphene
               -- A comparative theoretical study",
  author    = "Dietrich, Fabian and Guevara, Ulises J and Tiutiunnyk, Anton and
               Laroze, David and Cisternas, Eduardo",
  journal   = "FlatChem",
  publisher = "Elsevier BV",
  volume    =  {41},
  pages     = "100541",
  month     =  {sep},
  year      =  {2023},
  language  = "English",
  url = {https://doi.org/10.1016/j.flatc.2023.100541}
}

@article{Chang2024,
  title = {Vacancy-Induced Tunable Kondo Effect in Twisted Bilayer Graphene},
  author = {Chang, Yueqing and Yi, Jinjing and Wu, Ang-Kun and Kugler, Fabian B. and Andrei, Eva Y. and Vanderbilt, David and Kotliar, Gabriel and Pixley, J. H.},
  journal = {Phys. Rev. Lett.},
  volume = {133},
  issue = {12},
  pages = {126503},
  numpages = {9},
  year = {2024},
  month = {Sep},
  publisher = {American Physical Society},
  doi = {10.1103/PhysRevLett.133.126503},
  url = {https://link.aps.org/doi/10.1103/PhysRevLett.133.126503}
}

@article{Weisse2006,
  title = {The kernel polynomial method},
  author = {Wei\ss{}e, Alexander and Wellein, Gerhard and Alvermann, Andreas and Fehske, Holger},
  journal = {Rev. Mod. Phys.},
  volume = {78},
  issue = {1},
  pages = {275--306},
  numpages = {0},
  year = {2006},
  month = {Mar},
  publisher = {American Physical Society},
  doi = {10.1103/RevModPhys.78.275},
  url = {https://link.aps.org/doi/10.1103/RevModPhys.78.275}
}

@article{Le2018,
  title = {Electronic structure and optical properties of twisted bilayer graphene calculated via time evolution of states in real space},
  author = {Le, H. Anh and Do, V. Nam},
  journal = {Phys. Rev. B},
  volume = {97},
  issue = {12},
  pages = {125136},
  numpages = {12},
  year = {2018},
  month = {Mar},
  publisher = {American Physical Society},
  doi = {10.1103/PhysRevB.97.125136},
  url = {https://link.aps.org/doi/10.1103/PhysRevB.97.125136}
}

@article{Do_2019,
  title = {{Time-evolution patterns of electrons in twisted bilayer graphene}},
  author = {Do, V. Nam and Le, H. Anh and Bercioux, D.},
  journal = {Phys. Rev. B},
  volume = {99},
  issue = {16},
  pages = {165127},
  numpages = {14},
  year = {2019},
  month = {Apr},
  publisher = {American Physical Society},
  doi = {10.1103/PhysRevB.99.165127},
  url = {https://link.aps.org/doi/10.1103/PhysRevB.99.165127}
}

@article{Do_2021,
  title = {{Optical Hall response of bilayer graphene: Manifestation of chiral hybridized states in broken mirror symmetry lattices}},
  author = {Do, V. Nam and Le, H. Anh and Nguyen, V. Duy and Bercioux, D.},
  journal = {Phys. Rev. Res.},
  volume = {2},
  issue = {4},
  pages = {043281},
  numpages = {13},
  year = {2020},
  month = {Nov},
  publisher = {American Physical Society},
  doi = {10.1103/PhysRevResearch.2.043281},
  url = {https://link.aps.org/doi/10.1103/PhysRevResearch.2.043281}
}

@misc{pybinding,
  author       = {Dean Moldovan and Mi\u{a} Andelkovi\'{c} and Francois Peeters},
  title        = {{pybinding v0.9.5: a Python package for tight-binding calculations}},
  month        = aug,
  year         = 2020,
  publisher    = {Zenodo},
  version      = {v0.9.5},
  doi          = {10.5281/zenodo.4010216},
  url          = {https://doi.org/10.5281/zenodo.4010216},
}

@ARTICLE{Hong2021-pn,
  title     = "General, strong impurity-strength dependence of quasiparticle
               interference",
  author    = "Hong, Seung-Ju and Lihm, Jae-Mo and Park, Cheol-Hwan",
  journal   = "J. Phys. Chem. C Nanomater. Interfaces",
  publisher = "American Chemical Society (ACS)",
  volume    =  125,
  number    =  13,
  pages     = "7488--7494",
  month     =  {Apr},
  year      =  {2021},
  copyright = "https://creativecommons.org/licenses/by-nc-nd/4.0/",
  language  = "English",
  doi = {10.1021/acs.jpcc.1c01410}
}

@article{Buifmmode1996,
  title = {Quantum copying: Beyond the no-cloning theorem},
  author = {Bu\ifmmode \check{z}\else \v{z}\fi{}ek, V. and Hillery, M.},
  journal = {Phys. Rev. A},
  volume = {54},
  issue = {3},
  pages = {1844--1852},
  numpages = {0},
  year = {1996},
  month = {Sep},
  publisher = {American Physical Society},
  doi = {10.1103/PhysRevA.54.1844},
  url = {https://link.aps.org/doi/10.1103/PhysRevA.54.1844}
}

@ARTICLE{Rhim2020-rj,
  title     = "Quantum distance and anomalous Landau levels of flat bands",
  author    = "Rhim, Jun-Won and Kim, Kyoo and Yang, Bohm-Jung",
  journal   = "Nature",
  publisher = "Springer Science and Business Media LLC",
  volume    =  584,
  number    =  7819,
  pages     = "59--63",
  month     =  aug,
  year      =  2020,
  language  = "English",
  doi = {10.1038/s41586-020-2540-1}
}

@article{Abouelkomsan2023,
  title = {Quantum metric induced phases in Moir\'e materials},
  author = {Abouelkomsan, Ahmed and Yang, Kang and Bergholtz, Emil J.},
  journal = {Phys. Rev. Res.},
  volume = {5},
  issue = {1},
  pages = {L012015},
  numpages = {8},
  year = {2023},
  month = {Feb},
  publisher = {American Physical Society},
  doi = {10.1103/PhysRevResearch.5.L012015},
  url = {https://link.aps.org/doi/10.1103/PhysRevResearch.5.L012015}
}

@BOOK{Economou2006-te,
	title     = "Green's Functions in Quantum Physics",
	author    = "Economou, Eleftherios N",
	publisher = "Springer",
	series    = "Springer Series in Solid-State Sciences",
	edition   =  3,
	month     =  jun,
	year      =  2006,
	address   = "Berlin, Germany",
	doi = "10.1007/3-540-28841-4"
}

@article{Nam2017,
  title = {Lattice relaxation and energy band modulation in twisted bilayer graphene},
  author = {Nam, Nguyen N. T. and Koshino, Mikito},
  journal = {Phys. Rev. B},
  volume = {96},
  issue = {7},
  pages = {075311},
  numpages = {12},
  year = {2017},
  month = Aug,
  publisher = {American Physical Society},
  doi = {10.1103/PhysRevB.96.075311},
  url = {https://link.aps.org/doi/10.1103/PhysRevB.96.075311}
}

@article{Queiroz2018,
  title = {Selection Rules for Quasiparticle Interference with Internal Nonsymmorphic Symmetries},
  author = {Queiroz, Raquel and Stern, Ady},
  journal = {Phys. Rev. Lett.},
  volume = {121},
  issue = {17},
  pages = {176401},
  numpages = {6},
  year = {2018},
  month = Oct,
  publisher = {American Physical Society},
  doi = {10.1103/PhysRevLett.121.176401},
  url = {https://link.aps.org/doi/10.1103/PhysRevLett.121.176401}
}

@article{Ferreira2015,
  title = {Critical Delocalization of Chiral Zero Energy Modes in Graphene},
  author = {Ferreira, Aires and Mucciolo, Eduardo R.},
  journal = {Phys. Rev. Lett.},
  volume = {115},
  issue = {10},
  pages = {106601},
  numpages = {5},
  year = {2015},
  month = Aug,
  publisher = {American Physical Society},
  doi = {10.1103/PhysRevLett.115.106601},
  url = {https://link.aps.org/doi/10.1103/PhysRevLett.115.106601}
}

@article{Joao2020kite,
  title={KITE: high-performance accurate modelling of electronic structure and response functions of large molecules, disordered crystals and heterostructures},
  author={Jo{\~a}o, Sim{\~a}o M and Andelkovi{\'c}, Mi{\v{s}}a and Covaci, Lucian and Rappoport, Tatiana G and Lopes, Jo{\~a}o MVP and Ferreira, Aires},
  journal={Royal Society Open Science},
  volume={7},
  number={2},
  year={2020},
  doi={10.1098/rsos.191809},
  publisher={The Royal Society}
}

@article{Braun2014,
  title = {Numerical evaluation of Green's functions based on the Chebyshev expansion},
  author = {Braun, A. and Schmitteckert, P.},
  journal = {Phys. Rev. B},
  volume = {90},
  issue = {16},
  pages = {165112},
  numpages = {7},
  year = {2014},
  month = Oct,
  publisher = {American Physical Society},
  doi = {10.1103/PhysRevB.90.165112},
  url = {https://link.aps.org/doi/10.1103/PhysRevB.90.165112}
}

@article{Zhang2019local,
  title = {Local and global patterns in quasiparticle interference: A reduced response function approach},
  author = {Zhang, Dan-Bo and Han, Qiang and Wang, Z. D.},
  journal = {Phys. Rev. B},
  volume = {100},
  issue = {20},
  pages = {205112},
  numpages = {10},
  year = {2019},
  month = Nov,
  publisher = {American Physical Society},
  doi = {10.1103/PhysRevB.100.205112},
  url = {https://link.aps.org/doi/10.1103/PhysRevB.100.205112}
}

@article{Hermann2012periodic,
  title={Periodic overlayers and moir{\'e} patterns: theoretical studies of geometric properties},
  author={Hermann, Klaus},
  journal={Journal of Physics: Condensed Matter},
  volume={24},
  number={31},
  pages={314210},
  year={2012},
  doi = {10.1088/0953-8984/24/31/314210},
  publisher={IOP Publishing}
}

@article{Liu2024visualizing,
	title={Visualizing a single wavefront dislocation induced by orbital angular momentum in graphene},
	author={Liu, Yi-Wen and Zhuang, Yu-Chen and Ren, Ya-Ning and Yan, Chao and Zhou, Xiao-Feng and Yang, Qian and Sun, Qing-Feng and He, Lin},
	journal={Nature Communications},
	volume={15},
	number={1},
	pages={3546},
	year={2024},
	doi={10.1038/s41467-024-47756-w},
	publisher={Nature Publishing Group UK London}
}

@article{Zhang2020localBerry,
	title = {Local Berry Phase Signatures of Bilayer Graphene in Intervalley Quantum Interference},
	author = {Zhang, Yu and Su, Ying and He, Lin},
	journal = {Phys. Rev. Lett.},
	volume = {125},
	issue = {11},
	pages = {116804},
	numpages = {6},
	year = {2020},
	month = Sep,
	publisher = {American Physical Society},
	doi = {10.1103/PhysRevLett.125.116804},
	url = {https://link.aps.org/doi/10.1103/PhysRevLett.125.116804}
}

@article{Ren2026,
	title = {Unconventional intervalley scattering around an anisotropic atomic collapse potential},
	author = {Ren, Hui-Ying and Zhuang, Yu-Chen and Zhao, Wen-Xin and Ren, Ya-Ning and Sun, Qing-Feng and He, Lin},
	journal = {Phys. Rev. B},
	volume = {113},
	issue = {8},
	pages = {L081405},
	numpages = {8},
	year = {2026},
	month = {Feb},
	publisher = {American Physical Society},
	doi = {10.1103/sqb6-3s4c},
	url = {https://link.aps.org/doi/10.1103/sqb6-3s4c}
}

@ARTICLE{Goncalves2022-eg,
	title     = "Incommensurability-induced sub-ballistic narrow-band-states in
	twisted bilayer graphene",
	author    = "Gon{\c c}alves, Miguel and Olyaei, Hadi Z and Amorim, Bruno and Mondaini, Rubem and Ribeiro, Pedro and Castro, Eduardo V",
	journal   = "2D Mater.",
	publisher = "IOP Publishing",
	volume    =  9,
	number    =  1,
	pages     = "011001",
	month     =  jan,
	year      =  2022,
	doi = {10.1088/2053-1583/ac3259},
	copyright = "https://iopscience.iop.org/page/copyright"
}
	
	\pagebreak
	\onecolumngrid
	\pagebreak
	
	\section*{Appendix}
	\subsection*{Reduction in Size of Commensurate Supercell}
	In this section, we present a simplified description of the commensurate TBG lattice when $\lambda=3$. We start with the basic definitions of superlattice in a commensurate structure. As schematically illustrated in Fig.~\ref{fig:1lattice}(b), the primitive superlattice vectors are given by Eq.~\eqref{Eq: coincident vector}. If we rotate $\bm{L}_1$ by $60^\circ$ to obtain $\bm{L}_2$, we arrive at
	\begin{align}
		\bm{L}_2 = e^{i\frac{\pi}{3}}\bm{L}_1 = m\bm{a}_2^{(1)} + n\left(\bm{a}_2^{(1)}-\bm{a}_1^{(1)}\right) = -n\bm{a}_1^{(1)} + (m+n)\bm{a}_2^{(1)}.
	\end{align}
	The number of graphene's unit cells inside the supercell is given by
	\begin{align}
		N_{\text{uc}} = 2\frac{|\bm{L}_1\times\bm{L}_2|}{\left|\bm{a}_1^{(1)}\times\bm{a}_2^{(1)}\right|} = 2\det\begin{pmatrix}
			m & n \\ -n & m+n
		\end{pmatrix} = 2(m^2 + n^2 + mn),
	\end{align}
	corresponding to $4(m^2 + n^2 + mn)$ carbon atoms per supercell. However, the relation~\eqref{Eq: coincident vector} does not hold when $\lambda=3$.
	Although the vectors $\bm{L}_1$ and $\bm{L}_2$ still define a hexagonal supercell in this case, the supercell constructed from these vectors are not the primitive supercell. To see this, we check if there are any other coincidence lattice points hidden strictly inside the area spanned by $\bm{L}_1$ and $\bm{L}_2$.
	
	For a hexagonal lattice, the only possible internal lattice points that maintain the symmetry lie at the $1/3$ or $2/3$ positions along the long diagonal of the unit cell. We therefore consider the vector at the $1/3$ position first
	\begin{equation}
		\bm{L}_1' = \frac{1}{3}(\bm{L}_1 + \bm{L}_2) = \frac{1}{3} \left[ (m-n)\bm{a}_1^{(1)} + (m+2n)\bm{a}_2^{(1)} \right].
	\end{equation}
	For $\bm{L}_1'$ to be a lattice point of the crystal, its coefficients in the basis $\left\{\bm{a}_1^{(1)},\bm{a}_2^{(1)}\right\}$ must be integers.
	We note that $m+2n = (m-n) + 3n$. Because $3m$ is always divisible by 3, the entire vector $\bm{L}_1'$ consists of integers if and only if $(m-n)$ is divisible by 3. This leads to two distinct physical cases governed by the value of $\gcd(m-n, 3)$:
	\begin{itemize}
		\item \textbf{Case 1}: $\gcd(m-n, 3) = 1$\\
		The coefficients of $\bm{L}_1'$ are fractions and thus there is no smaller coincidence site. The primitive superlattice vector is defined by $\bm{L}_1$, whose length is $L = a\sqrt{m^2 + mn + n^2}$.
		\item \textbf{Case 2}: $\gcd(m-n, 3) = 3$\\
		The coefficients of $\bm{L}_1'$ are integers, which means that the actual repeating primitive superlattice vector has a magnitude smaller than $\bm{L}_1$. The new superlattice constant is the length of $\bm{L}_1'$
		\begin{align*}
			L = \frac{1}{3}\sqrt{|\bm{L}_1 + \bm{L}_2|^2} &= \frac{a}{3}\sqrt{(m-n)^2 + (m-n)(m+2n) + (m+2n)^2} \nonumber\\
			&= \frac{a}{3}\sqrt{3(m^2 + mn + n^2)} = \frac{a\sqrt{m^2 + mn + n^2}}{\sqrt{3}}.
		\end{align*}
	\end{itemize}
	By combining the two cases, we obtain the unified equation for the superlattice constant $L$ as 
	\begin{equation}
		L = \frac{a \sqrt{m^2 + mn + n^2}}{\sqrt{\lambda}}.
	\end{equation}
	Here, we note that $\bm{L}_1'$ is rotated by $30^{\circ}$ from vector $\bm{L}_1$. This rotation of the primitive superlattice vectors gives rise to a different folding mechanisms of the Dirac cones, as mentioned in the main text and sketched in Fig.~\ref{fig:3emergence}(a). The second primitive superlattice vector is then given by
	\begin{align}
		\bm{L}_2' = e^{i\frac{\pi}{3}}\bm{L}_1' &= \frac{1}{3} \left[(m-n)\bm{a}_2^{(1)} + (m+2n)\left(\bm{a}_2^{(1)}-\bm{a}_1^{(1)}\right)\right]\nonumber\\
		&= \frac{1}{3} \left[-(m+2n)\bm{a}_1^{(1)} + (2m+n)\bm{a}_2^{(1)}\right].
	\end{align}
	\subsection*{Band Structure}
	In this section, we comment on the band structure of TBG for some twist angles. The band structure is considered for only twist angles satisfying the Eq.~\eqref{Eq: twist angle}. This ensures that translation symmetry is present and we can identify a supercell and the associated primitive superlattice vectors $(\bm{L}_1,\bm{L}_2)$. It is worth noting that another equivalent relation, proposed by Lopes dos Santos \textit{et al.}~\cite{Santos2012}, is $\cos\theta=\frac{3m^2+3mr+r^2/2}{3m^2+3mr+r^2}$, which can be transformed back to Eq.~\eqref{Eq: twist angle} via the substitution $r\rightarrow n-m$. The band folding mechanisms are different for different values of $r$, as commented in the main text and noted by Refs.~\cite{Mele2010} and ~\cite{Zou2018}.
	
\begin{figure}[h!]
		\includegraphics[width=\linewidth]{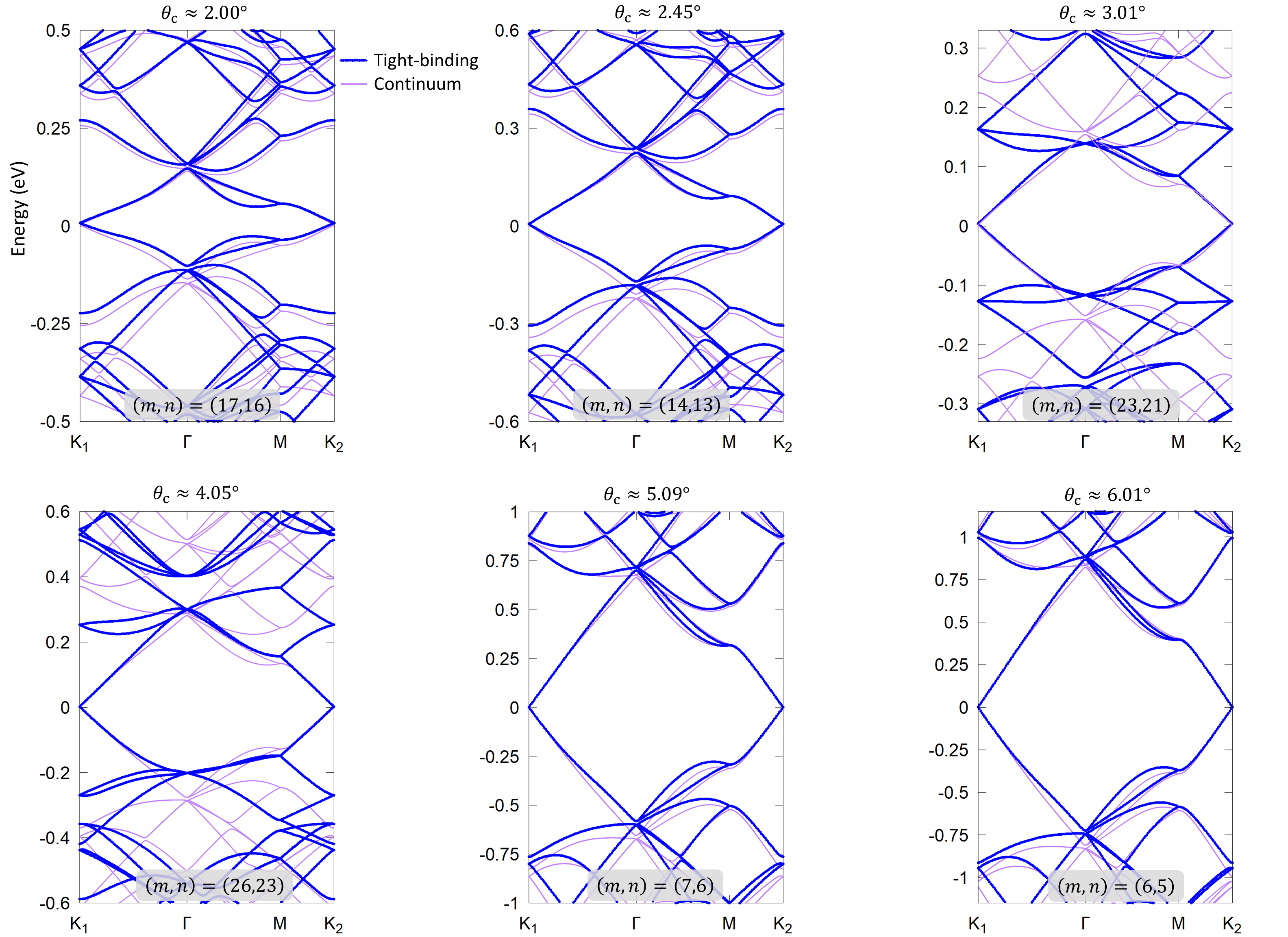}
		\caption{\label{Sfig:bands} \textbf{Band structures}. Band structures of TBG for different twist angles, obtained from tight-binding and continuum model. Due to the difference in size of the superlattice Brillouin zone and the moir\'{e} Brillouin zone, the energies of the continuum Hamiltonian are multiplied by $|m-n|/\lambda$.}
	\end{figure}

	Using the tight-binding and continuum models described in the main text, we present the band structure of TBG for several twist angles in Fig.~\ref{Sfig:bands}. The superlattice Brillouin zone is assumed to be a hexagon defined by the reciprocal lattice vectors $\bm{G}_1$ and $\bm{G}_2$: $|\bm{G}_1|=|\bm{G}_2|=\frac{4\pi}{\sqrt{3}|\bm{L}_1|}$ and $\angle(\bm{G}_1,\bm{L}_1)=\frac{\pi}{6}$. Meanwhile, the moir\'{e} Brillouin zone is defined by the reciprocal moir\'{e} lattice vectors, whose lengths are $\sqrt{3}|\Delta\bm{K}|$. We see that the two models agree well with each other, especially for the conduction bands, when $m-n=1$ since the superlattice Brillouin zone coincides with the moir\'{e} Brillouin zone in this case. The deviation between the models in the valence bands arises from the pronounced particle-hole asymmetry of the tight-binding band structure. On the other hand, for $m-n=2$ or $3$, the bands fold differently, leading to the difference between the band structures of the two models.
	
	\pagebreak
	
	\subsection*{Form Factor}
\begin{figure}[h!]
		\includegraphics[width=\linewidth]{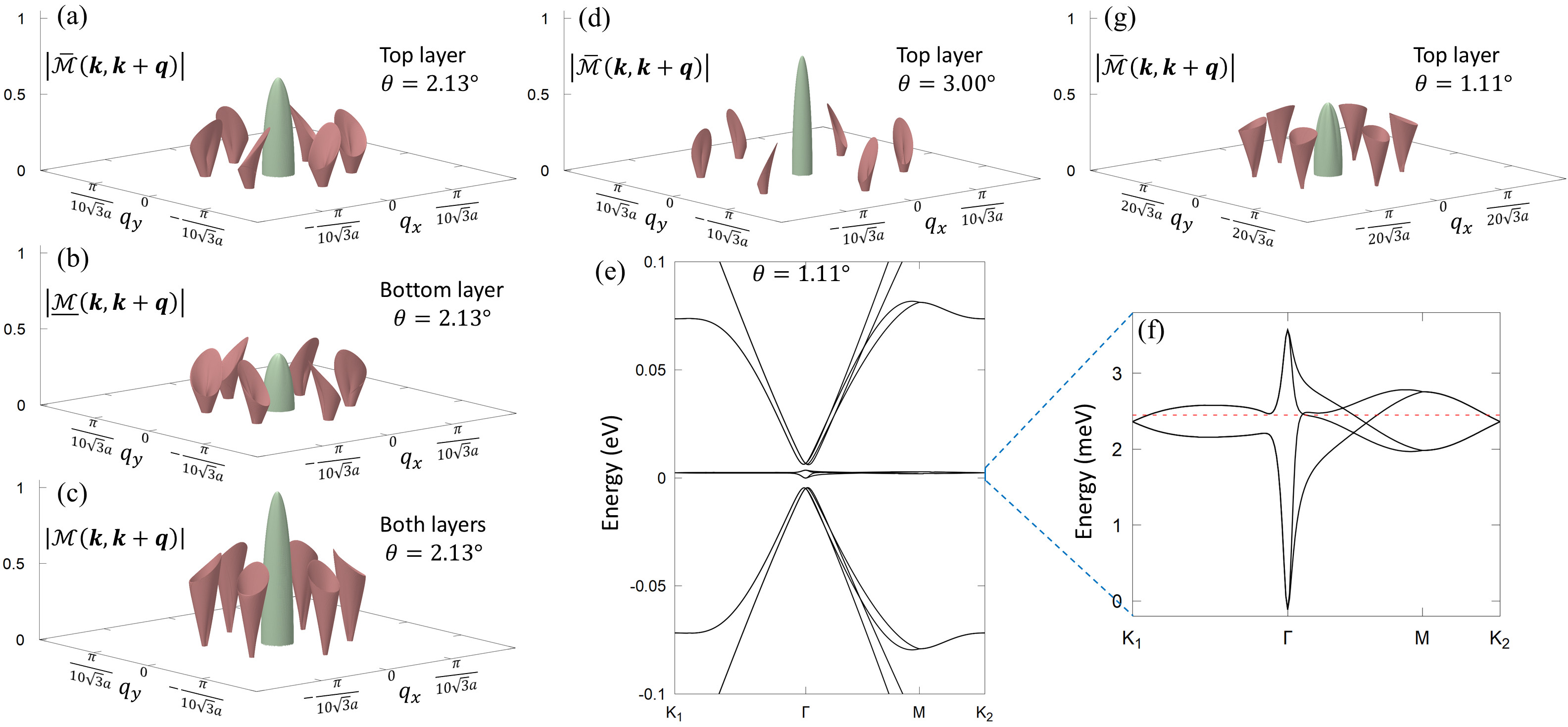}
		\caption{\label{Sfig:overlap} \textbf{Intravalley Form Factor}. (a), (b), (c) The form factor norm between states at $E_\text{F}=20$~meV, projected onto the top layer (a), bottom layer (b), and both layers (c). The twist angle is $2.13^\circ$. (d) The form factor between states at $E_\text{F}=20$~meV, projected onto the top layer, for $\theta=3.00^\circ$. (e) Band structure of the continuum Hamiltonian when $\theta=1.1^\circ$. (f) Magnified picture of the flat bands. The red dashed line indicates the Fermi energy $E_\text{F}=2.45$~meV, at which the form factor of states at the Dirac cones is plotted in (g).}
	\end{figure}

	In the main text, to compare with the FT-LDOS, the form factor norm is presented as the SLFFB with the constraint that only back-scattering processes are considered, i.e., $\left|\bar{\mathcal{M}}(\bm{k},\bm{k}+\bm{q})\right|\delta_{\bm{k}-\bm{K}_j,\bm{K}_l-\bm{k}-\bm{q}}$, for all interference mechanisms except the intravalley intralayer one. Here, the original form factor norm for $\theta=2.13^\circ$ is shown in Figs.~\ref{Sfig:overlap}(a-c). In particular, the form factor norm of eigenstates projected onto the top layer, $\bar{\mathcal{M}}(\bm{k},\bm{k}+\bm{q}) = \sum_{\alpha{(z>0)}}C_{\alpha(\bm{k}+\bm{q})}^{\dagger}C_{\alpha\bm{k}}$, is shown in Fig.~\ref{Sfig:overlap}(a), displaying a chiral structure. The form factor norm of eigenstates projected onto the bottom layer, $\underline{\mathcal{M}}(\bm{k},\bm{k}+\bm{q}) = \sum_{\alpha{(z<0)}}C_{\alpha(\bm{k}+\bm{q})}^{\dagger}C_{\alpha\bm{k}}$, is shown in Fig.~\ref{Sfig:overlap}(b) with the chirality being reversed. Meanwhile, the form factor norm of complete eigenstates, $\mathcal{M}(\bm{k},\bm{k}+\bm{q}) = \sum_{\alpha}C_{\alpha(\bm{k}+\bm{q})}^{\dagger}C_{\alpha\bm{k}}$, is shown in Fig.~\ref{Sfig:overlap}(c), displaying no chiral structure. Additionally, $\left|\bar{\mathcal{M}}(\bm{k},\bm{k}+\bm{q})\right|$ for twist angle $\theta=3.00^\circ$ is shown in Fig.~\ref{Sfig:overlap}(d), which displays a similar chiral structure.
	
	We consider the form factor norm at the magic angle. In our continuum model, we now choose $u=105$~meV and $w=110$~meV so that the four central bands are well separated from other bands, mimicking the effect of lattice relaxation. The magic angle is determined from the effective two-band model presented in the main text [Eq.~\eqref{Eq: two-band}] by identifying the condition of a vanishing group velocity, $1-3w^2/(\hbar v_\text{D}|\boldsymbol{\kappa}|)^2=0$. The band structure at the magic angle $\theta=1.11^\circ$ is depicted in Fig.~\ref{Sfig:overlap}(e), where the four central bands become nearly flat. These bands are magnified in Fig.~\ref{Sfig:overlap}(f), showing a bandwidth around 4~meV. Taking into account only states close to the $\bm{K}$ points on the Fermi surface at $E_\text{F}\approx2.45$~meV, we compute the form factor norm of these states projected on the top layer, which can be seen in Fig.~\ref{Sfig:overlap}(g). For this twist angle, the chiral structure observed in Figs.~\ref{Sfig:overlap}(a) and~\ref{Sfig:overlap}(d) is no longer apparent.
	
	\pagebreak
	
	\subsection*{Gated Twisted Bilayer Graphene}
	Although the single-layer form factor of TBG is captured well by the QPI patterns, the quantity that determines the quantum geometry of the Bloch states of the entire system is the form factor, which involves the electronic states in both layers of TBG. The norm of the form factor, however, does not exhibit a chiral structure since the opposite chiralities of both layers cancel each other. In order to make this quantity chiral, we may apply a gate bias to the TBG since the Fermi surface of one chirality shrinks while the other enlarges, resulting a net nonzero chirality. An example that would have a chiral structure accordingly in this case is the Hilbert–Schmidt quantum distance between Bloch states, $d(\mathbf{k},\mathbf{k}+\mathbf{q}) = \sqrt{1 - \left|\mathcal{M}(\mathbf{k},\mathbf{k}+\mathbf{q})\right|^2}$, whose infinitesimal change when $|\mathbf{q}|\rightarrow0$ gives the quantum metric of electrons~\cite{Buifmmode1996,Rhim2020-rj,Abouelkomsan2023}.
	\begin{figure}[h!]
		\includegraphics[width=\linewidth]{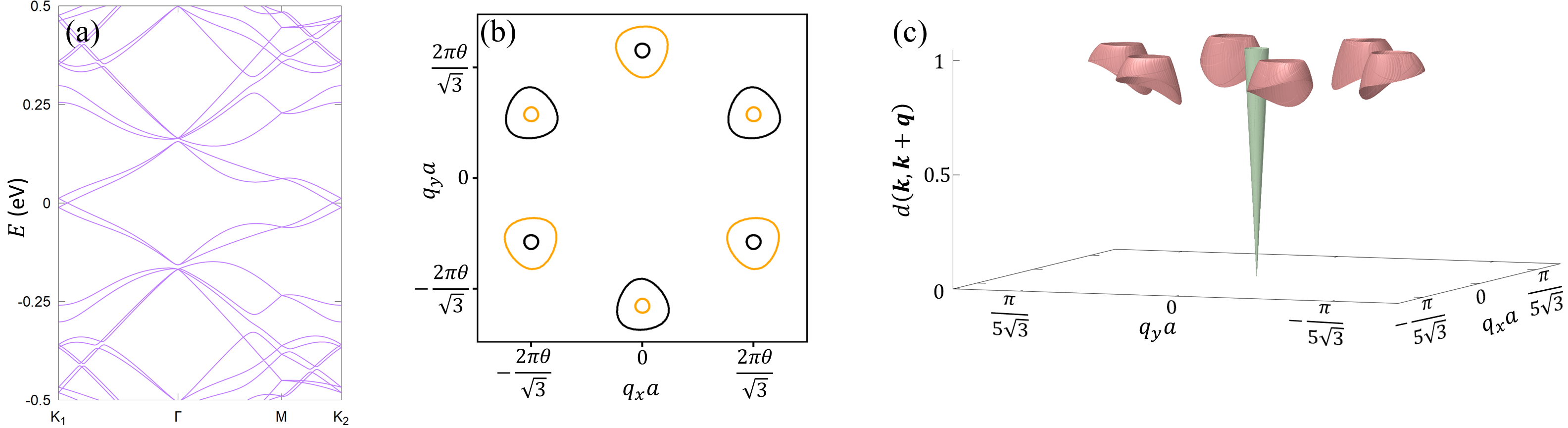}
		\caption{\label{Sfig:qdistance} \textbf{Quantum distance}. The band structure (a), the Fermi surface (b), and the quantum distance between the Bloch states (c) of TBG at $E_{\text{F}}=20$~meV for a twist angle $\theta=2.13^\circ$ and gate bias $V_{\text{G}}=80$~meV. In (b), the black and orange curves are the Fermi contours at the $K$ and $K'$ valleys, respectively.}
	\end{figure}
	
	In the presence of a gate bias, the Dirac cones of the continuum Hamiltonian Eq.~\eqref{Eq: continuum} are now given by
	%
	%
	\begin{equation}
		h_{\pm}(\mathbf{k}) = \frac{\sqrt{3}t_0}{2}\begin{pmatrix}
			\pm V_\text{g}/2 & |\mathbf{k}|e^{i(\varphi_{\mathbf{k}}\pm\theta/2)}\\
			|\mathbf{k}|e^{-i(\varphi_{\mathbf{k}}\pm\theta/2)} & \pm V_\text{g}/2
		\end{pmatrix}
	\end{equation}
	%
	%
	with $V_\text{g}$ being the gate bias. The band structure for $V_\text{g}=80$~meV and $\theta=2.13^\circ$ is shown in Fig.~\ref{Sfig:qdistance}(a), where the valley degeneracy of the Dirac cones is lifted. Interestingly, the Fermi surface at an energy slightly above the Dirac points exhibits a chiral structure where the trigonal warping contours are all rotated anticlockwise --- see Fig.~\ref{Sfig:qdistance}(b). The quantum distance is shown in Fig.~\ref{Sfig:qdistance}(c). As expected, the quantum distance between Bloch states at the same valley ($K$ or $K'$) of different layers exhibit a chiral structure. Its minimum value is 0 when $\mathbf{q}=0$ and its maximum value is 1 occurring between states on the same valley as well as on adjacent ones. 
	
	\pagebreak
	
	\subsection*{Additional Data of FT-LDOS}
	\begin{figure}[h!]
		\includegraphics[width=\linewidth]{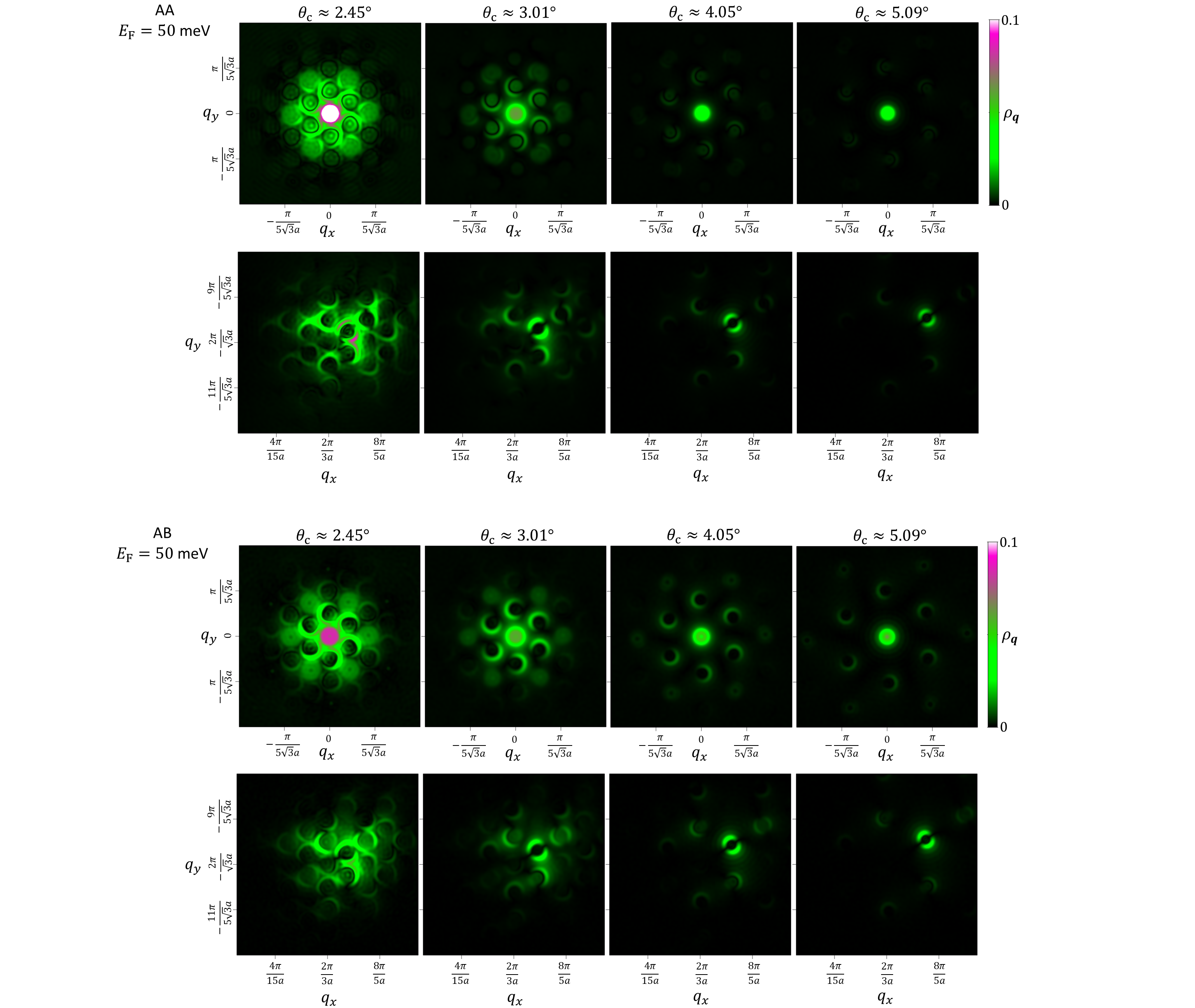}
		\caption{\label{Sfig:angles} \textbf{FT-LDOS for different commensurate twist angles}. The on-site defect ($\varepsilon_0=0.8|t|$) is located at the center of either the AA-stacking region (upper panel) or the AB-stacking region (lower panel) while the Fermi energy is fixed at $E_\text{F}=50$~meV. In both panels, the interlayer interference signals decrease in strength for larger twist angles as the coupling between the two layers becomes weaker and the DOS decreases. The intensity in the upper panel is stronger than the lower one since the wave functions of states at the Dirac cones localize more at the AA-stacking region. The interlayer signals are composed of circles that are either filled or empty, which may indicate different decay laws of the Friedel oscillations~\cite{Pereg-Barnea2008}.}
	\end{figure}
	
	\pagebreak
	
	\begin{figure}
		\includegraphics[width=\linewidth]{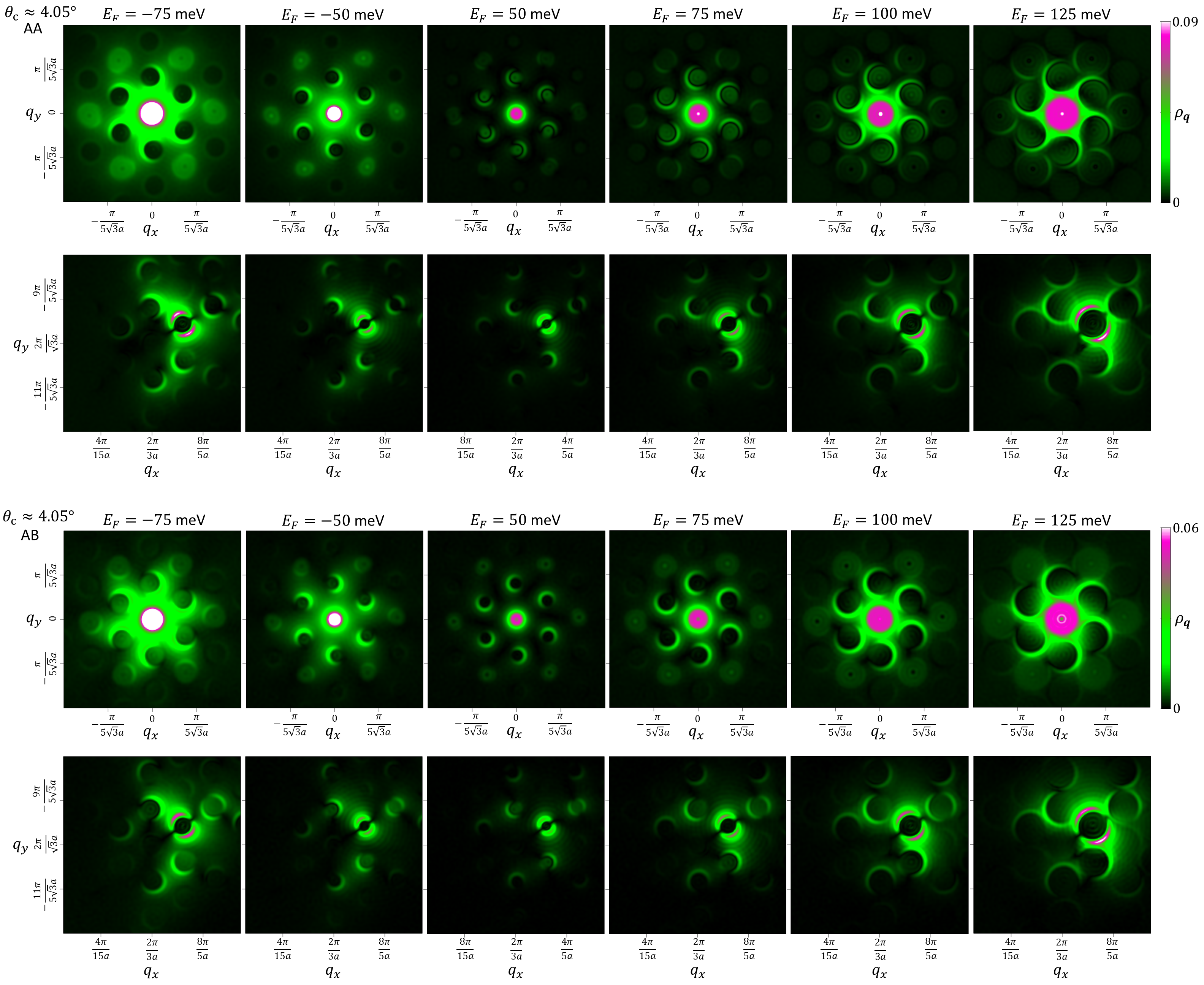}
		\caption{\label{Sfig:fermi} \textbf{FT-LDOS for different Fermi energies}. The on-site defect ($\varepsilon_0=0.8|t|$) is located at the center of either the AA-stacking region (upper panel) or the AB-stacking region (lower panel) when the twist angle is $\theta_\text{c}\approx4.05^\circ$. Both panels show clearly the reversal of chirality and the asymmetry in intensity between the bands above and below the charge neutrality point. Similar to Fig.~\ref{Sfig:angles}, the intensity is stronger when the defect is placed at the center of the AA-stacking region. It should be noted that the chirality reversal in this case differs from that observed when going from the top to the bottom layer, i.e., Fig.~\ref{fig:2qpi}(b), since the positions of circular signals in the intervalley interference remain unchanged.}
	\end{figure}
	
	\pagebreak
	
	\begin{figure}
		\includegraphics[width=\linewidth]{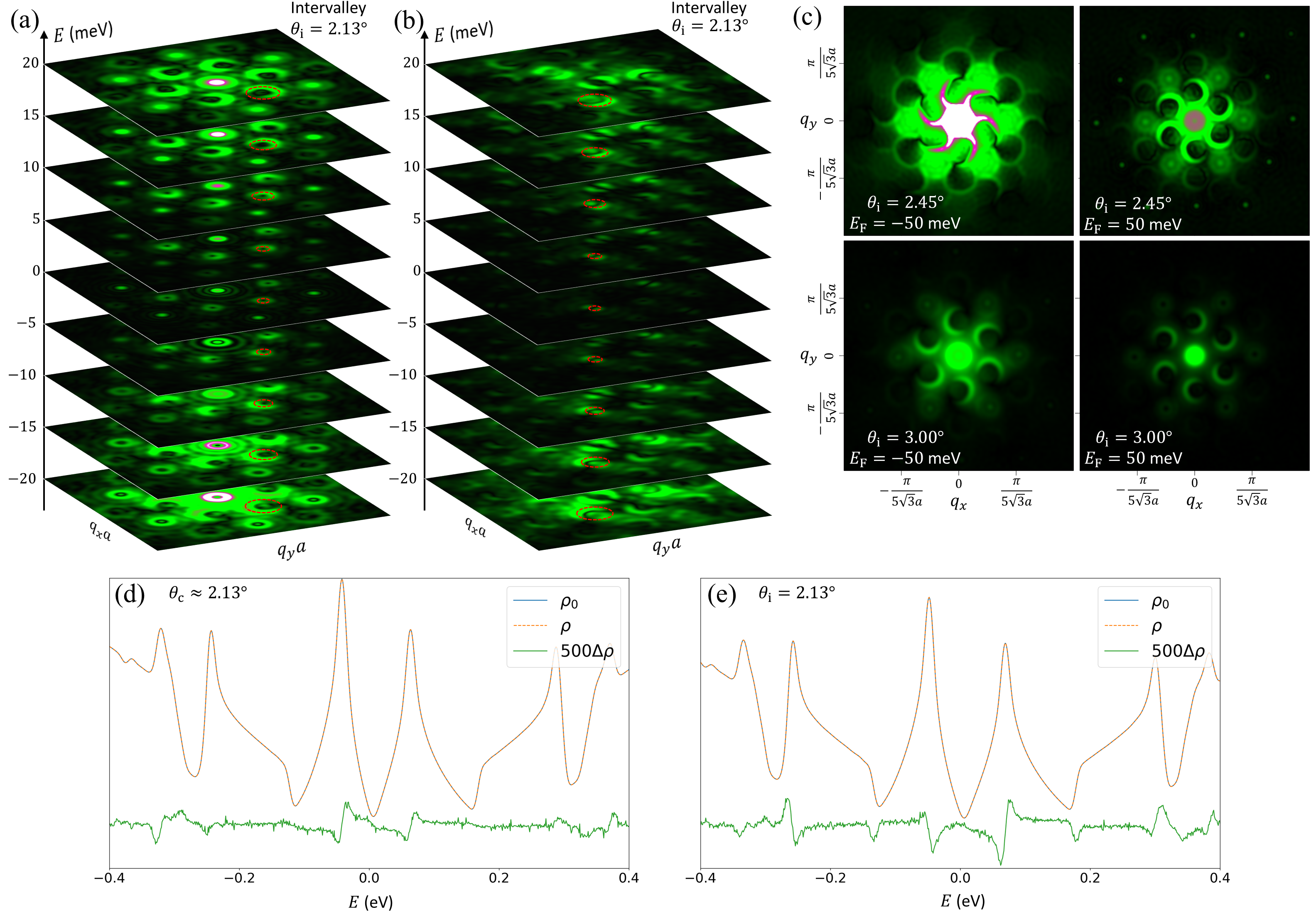}
		\caption{\label{Sfig:incom} \textbf{Dirac cones in incommensurate lattice}. The FT-LDOS of intravalley (a) and intervalley (b) interference for an incommensurate lattice with twist angle $\theta_\text{i}=2.13^\circ$. The red dashed circles have radii that are twice the Fermi momentum, guiding the eyes how the Dirac cones exist even in incommensurate TBG. The circles at $E=0$~meV have nonzero radii due to the finite lifetime of the quasiparticles ($\eta=5$~meV). (c) The FT-LDOS of intravalley signals for two incommensurate lattices with twist angles $\theta_\text{i}=2.45^\circ$ (top row) and $\theta_\text{i}=3.00^\circ$ (bottom row) at two energies $E_\text{F}=-50$~meV (left column) and $E_\text{F}=50$~meV (right column). For larger twist angles, the signals agree better with those of their approximate commensurate lattice. The existence of Dirac cones in incommensurate TBG verifies the emergent symmetries for this range of twist angles~\cite{Zou2018}. However, for smaller twist angles, especially the magic angle, further investigations are needed since there is difference in transport between commensurate and incommensurate TBG~\cite{Goncalves2022-eg}. (d,e) The DOS of TBG and its variation due to the defect when the twist angle is (d) commensurate and (e) incommensurate. At the charge neutrality point, while $\Delta\rho(\omega)$ resembles an odd function in energy for the commensurate lattice, it is similar to an even function in energy for the incommensurate one. The origin of this deviation remains an open question.}
	\end{figure}

	\begin{figure}
		\includegraphics[width=\linewidth]{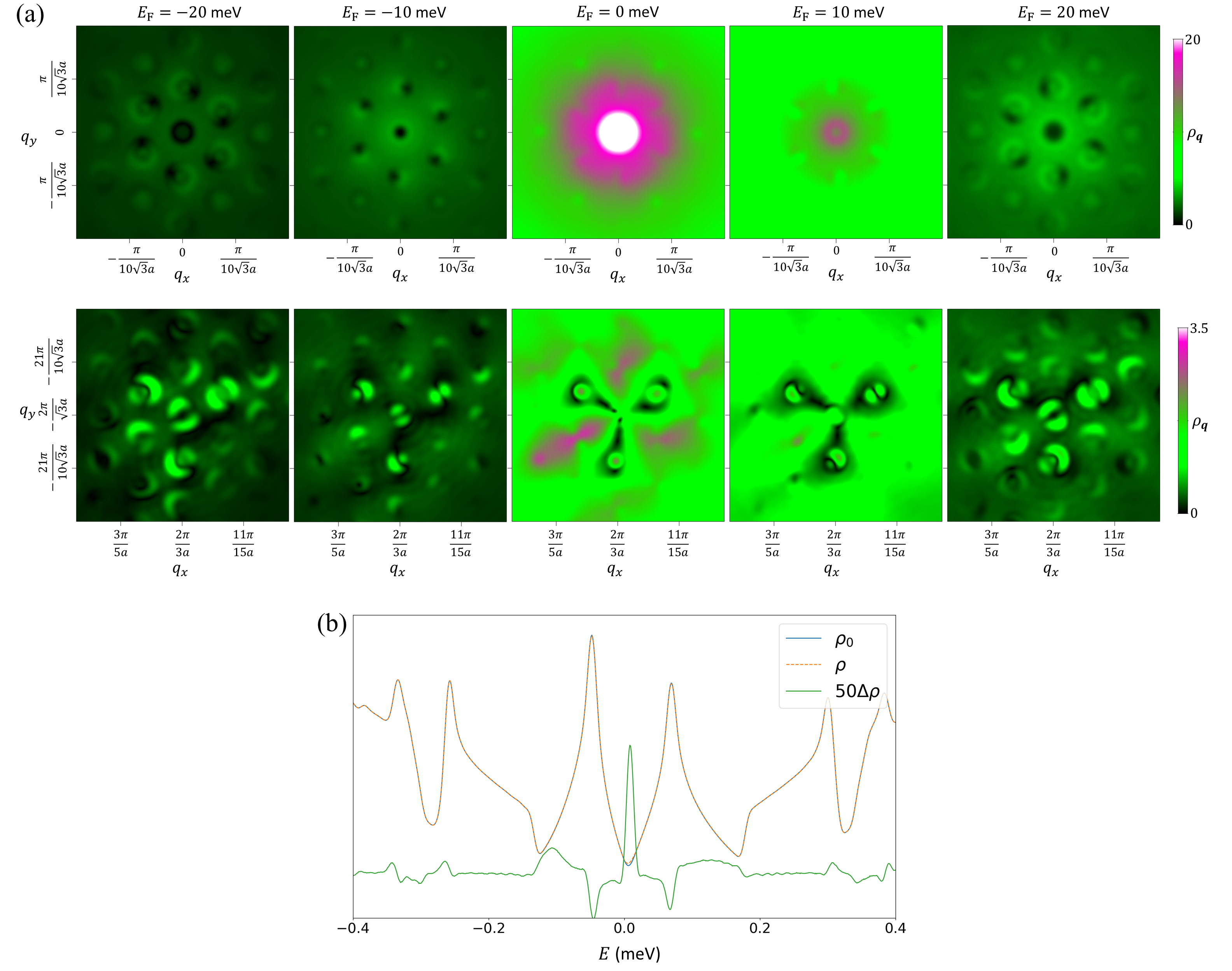}
		\caption{\label{Sfig:vacancy} \textbf{Defect as a vacancy in an incommensurate lattice}. (a) FT-LDOS variation due to a vacancy for different Fermi energies. The vacancy is located at the center of the AB-stacking region with the twist angle being $\theta_\text{i}=2.13^\circ$. The top and bottom rows show the intravalley and intervalley interference, respectively. While the intervalley interference resembles that of on-site defect, the chirality of intravalley interlayer interference is reversed. The signals are strongly suppressed for energies $E_\text{F}$ close to the Dirac points. (b) The DOS of TBG and its variation due to the vacancy. At the charge neutrality point, $\Delta\rho(\omega)$ is analogous to an even function in energy. Another way to see this picture is that the DOS variation at the two van Hove singularities behave similarly. We note that the spontaneous time-reversal symmetry breaking due to the vacancy~\cite{Dietrich2023,Chang2024} is still neglected here.}
	\end{figure}
	
	\begin{figure}
		\includegraphics[width=\linewidth]{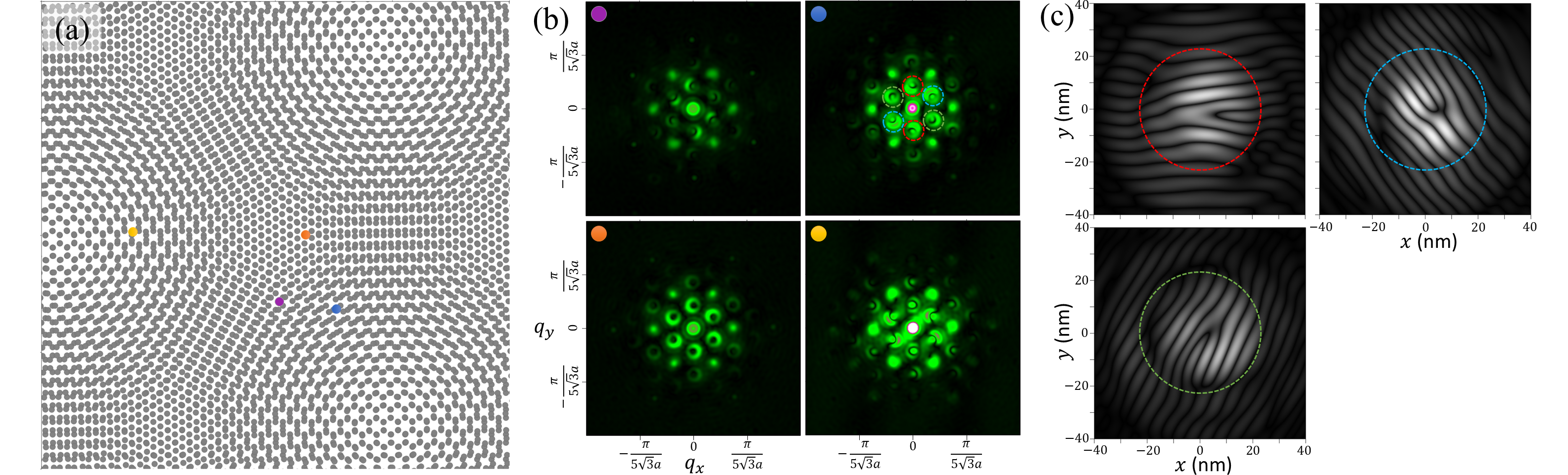}
		\caption{\label{Sfig:C3} \textbf{Quasiparticle interference when the defect position breaks $C_3$ symmetry}. (a) TBG lattice for $\theta_\text{c}\approx2.13^\circ$ with four cases of defect's position, denoted by the four colored dots. (b) The intravalley signals of FT-LDOS associated with the four cases depicted in (a) at the energy $E_{\text{F}}=15$~meV. The signal intensity increases as the defect is placed closer to the AA-stacking region, which is expected since the wave functions of the Dirac cone localize here. When the defect is distant from the center of either AA- or AB-stacking region, the $C_3$-rotational symmetry over the moir\'{e} length scale is broken, but a $C_{2z}$-rotational symmetry remains intact. The latter symmetry results from the reciprocal scattering and interference of the eigenstates. (c) Fourier-filtered LDOS of selected signals depicted in the top right panel of (b). The dislocations near the center still have two additional wavefronts, similar to Figs.~\ref{fig:4dislocation}(c,f). However, if we consider a circle centered at the defect, the number of additional wavefronts varies against the magnitude of its radius. We note that here only the LDOS of layer~1 is considered.}
	\end{figure}
	
\end{document}